\newif\ifconfver
\confverfalse      

\newif\ifcutshort      
\cutshorttrue

\newif\ifcutshortlvltwo  
\cutshortlvltwofalse

\ifconfver
    \documentclass[10pt,twocolumn,twoside]{IEEEtran}
\else
    \documentclass[12pt,draftcls,onecolumn,twoside]{IEEEtran}
\fi

\usepackage{array}
\usepackage{cite}
\usepackage{calc}
\usepackage{caption}
\usepackage{graphicx}
\usepackage{psfrag}
\usepackage[tight,footnotesize]{subfigure}
\usepackage{stfloats}
\usepackage{url}
\usepackage{subfig}
\usepackage{longtable}
\usepackage{amsmath,amsfonts,amssymb,amsbsy,bm,paralist,theorem,ifthen,color}
\usepackage{algorithmic,algorithm}

\newcommand\xb{\ensuremath{{\bm x}}}

\newcommand\yb{\ensuremath{{\bm y}}}

\newcommand\ub{\ensuremath{{\bm u}}}

\newcommand\Ab{\ensuremath{{\bm A}}}

\newcommand\bb{\ensuremath{{\bm b}}}

\newcommand\db{\ensuremath{{\bm d}}}

\newcommand\Fc{\ensuremath{\mathcal{F}}}
\newcommand\fb{\ensuremath{{\bm f}}}

\newcommand\gb{\ensuremath{{\bm g}}}

\newcommand\pb{\ensuremath{{\bm p}}}

\newcommand\Dc{\ensuremath{{\mathcal{D}}}}

\newcommand\Lc{\ensuremath{{\mathcal{L}}}}
\newcommand\Pc{\ensuremath{{\mathcal{P}}}}

\newcommand\Wb{\ensuremath{{\bm W}}}

\newcommand\zb{\ensuremath{{\bm z}}}
\newcommand\alphab{\ensuremath{{\bm \alpha}}}
\newcommand\betab{\ensuremath{{\bm \beta}}}

\newcommand\lambdab{\ensuremath{{\bm \lambda}}}

\newcommand\zerob{\ensuremath{{\bm 0}}}
\newcommand\etab{\ensuremath{{\bm \eta}}}

\newcommand\psib{\ensuremath{{\bm \psi}}}
\newcommand\Psib{\ensuremath{{\bf \Psi}}}

\newcommand\Xc{\ensuremath{{\mathcal{X}}}}

\newtheorem{Lemma}{Lemma}
\newtheorem{Proposition}{Proposition}
\newtheorem{Theorem}{Theorem}

\newtheorem{Assumption}{Assumption}

\ifconfver
\def\blue{\color{black}}
\else
\def\blue{\color{black}}
\fi

\begin{document}
\bibliographystyle{IEEEtran}
\title{Distributed Constrained Optimization by Consensus-Based Primal-Dual Perturbation Method}

\ifconfver \else {\linespread{1.1} \rm \fi

\author{\vspace{0.0cm}Tsung-Hui Chang$^\star$, \emph{Member, IEEE}, Angelia Nedi\'{c}$^\dag$, \emph{Member, IEEE}, and \\Anna Scaglione$^\ddag$, \emph{Fellow, IEEE} \\
\thanks{
The work of Tsung-Hui Chang is supported by National Science Council, Taiwan (R.O.C.), by Grant
NSC 102-2221-E-011-005-MY3. The work of Angelia Nedi\'{c} is supported by  the NSF grants CMMI 07-42538
and CCF 11-11342.
The work of Anna Scaglione is supported
by NSF CCF-1011811. }
\thanks{$^\star$Tsung-Hui Chang is the corresponding author. Address:
Department of Electronic and Computer Engineering, National Taiwan University of Science and Technology, Taipei 10607, Taiwan, (R.O.C.). E-mail:
tsunghui.chang@ieee.org. }\thanks{$^\dag$Angelia Nedi\'{c} is with
Department of Industrial and Enterprise Systems Engineering, University of Illinois at Urbana-Champaign, Urbana, IL 61801, USA. E-mail: angelia@illinois.edu.}\thanks{$^\ddag$Anna Scaglione is with Department of Electrical and Computer Engineering, University of California, Davis, CA 95616, USA. E-mail: ascaglione@ucdavis.edu.}

}

 \markboth{Submitted to IEEE TRANSACTIONS ON Automatic Control, Sept. 2012, revised April 2013 and Nov. 2013}{Submitted to IEEE TRANSACTIONS ON Automatic Control, Sept. 2012, revised April 2013 and Nov. 2013} \maketitle

\ifconfver

\else

\begin{center}
 Revised, Nov. 2013
\end{center}
\fi
\vspace{0.05cm}
\begin{abstract}
Various distributed optimization methods have been developed for solving problems which have simple local constraint sets and
whose objective function is the sum of local cost functions of distributed agents in a network. Motivated by emerging applications in smart grid and distributed sparse regression, this paper studies distributed optimization methods for solving general problems which have a coupled global cost function and have inequality constraints.
We consider a network scenario where each agent has no global knowledge and can access only its local mapping and constraint functions. To solve this problem in a distributed manner, we propose
a consensus-based distributed primal-dual perturbation (PDP) algorithm. In the algorithm, agents employ the average consensus technique to estimate the global cost and constraint functions via exchanging messages with neighbors, and meanwhile use a local primal-dual perturbed subgradient method to approach a global optimum. The proposed PDP method not only can handle smooth inequality constraints but also non-smooth constraints such as some sparsity promoting constraints arising in sparse optimization. We prove that the proposed PDP algorithm converges to an optimal primal-dual solution of the original problem, under standard problem and network assumptions. Numerical results illustrating the performance of the proposed algorithm for a distributed
demand response control problem in smart grid are also presented.
\end{abstract}

\ifconfver \else \IEEEpeerreviewmaketitle} \fi

\ifconfver \else
\newpage
\fi

\vspace{-0.3cm}
\section{Introduction}\label{sec: intro}
Distributed optimization methods are becoming popular options for solving several engineering problems, including parameter estimation, detection and localization problems in sensor networks \cite{Lesser2003,Rabbat2004}, resource allocation problems in peer-to-peer/multi-cellular communication networks \cite{Chiang2008,ChaoTSP2012}, and {distributed learning and regression problems in control \cite{Belomestny2010} and machine learning \cite{Hastie2001Book,Elad2009Book,Yamada2009}}, to name a few.
In these applications, rather than pooling together all the relevant parameters that define the optimization problem, distributed agents, which have access to a local subset of such parameters, collaborate with each other to minimize {a global cost function}, subject to local variable constraints.
Specifically, since it is not always efficient for the agents to exchange across the network the local cost and constraint functions, owing to the large size of network, time-varying network topology, energy constraints and/or privacy issues, distributed optimization methods that utilize only local information and messages exchanged between connecting neighbors have been of great interest; see \cite{Johansson08,Angelia2009_multiagent,Angelia2010,Lobel2011,Ram2010_stoc,Chen_Sayed2012,MZhu2012,Yuan2011} and references therein.

{\bf Contributions:} Different from the existing works \cite{Johansson08,Angelia2009_multiagent,Angelia2010,Lobel2011,Ram2010_stoc,Chen_Sayed2012} where the local variable constraints are usually simple (in the sense that they can be handled via simple projection) and independent among agents, in this paper, we consider a problem formulation that has a general set of convex inequality constraints that couple all the agents' optimization variables. In addition, {similar to}~\cite{Ram2012}, the considered problem has a global (non-separable) convex cost function that is a function of the sum of local mapping functions of the local optimization variabless. {Such a problem formulation appears, for example, in the classical regression problems which have a wide range of applications. In addition, the considered formulation also arises in the demand response control and power flow control problems in the emerging smart grid systems \cite{ChangPES2012,LiLow2011PES,Juan2009}. More discussions about applications are presented in Section \ref{subsec: applications}.
}

In this paper, we assume that each agent knows only the local mapping function and local constraint function. To solve this problem in a distributed fashion, in this paper, we develop a novel \emph{distributed consensus-based
primal-dual perturbation (PDP)} algorithm, which combines the ideas of the primal-dual perturbed (sub-)gradient method \cite{Kallio1994,Kallio1999} and the average consensus techniques \cite{Olshevsky2006,Xiao2003,Angelia2009_multiagent}.
In each iteration of the proposed algorithm, agents exchange their local estimates of the global cost and constraint functions with their neighbors, followed by performing one-step of primal-dual variable (sub-)gradient update. {Instead of using the primal-dual iterates {computed at the preceding iteration} as in most of the existing primal-dual subgradient based methods \cite{MZhu2012,Yuan2011}, the (sub-)gradients in the proposed distributed PDP algorithm} are computed based on some perturbation points which can be efficiently computed using the messages exchanged from neighbors. In particular, we provide two efficient ways to compute the perturbation points that can respectively handle the smooth and non-smooth constraint functions. More importantly, we build convergence analysis results showing that the proposed distributed PDP algorithm ensures a strong convergence of the local primal-dual iterates to a global optimal primal-dual solution of the considered problem. {The proposed algorithm is applied to a distributed sparse regression problem and a distributed demand response control problem in smart grid. Numerical results for the two applications are presented to demonstrate the effectiveness of the proposed algorithm.}

{\bf Related works:} {Distributed dual subgradient method (e.g., dual decomposition) \cite{YangJohansson2010} is a popular approach to
solving a problem with coupled inequality constraints in a distributed manner. However, given the dual variables, this method requires the agents to globally solve the local subproblems, which may require considerable computational efforts if the local cost and constraint functions have some complex structure.
} Consensus-based distributed primal-dual (PD) subgradient methods have been developed recently in \cite{MZhu2012,Yuan2011} for solving a problem with an objective function which is the sum of local convex cost functions, and with global convex inequality constraints. In addition to having a different cost function from our problem formulation, the works in \cite{MZhu2012,Yuan2011} assumed that all the agents in the network have global knowledge of the inequality constraint function; the two are in sharp contrast to our formulation where a non-separable objective function is considered and each agent can access only its local constraint function.
{Moreover, these works adopted the conventional PD subgradient updates \cite{Uzawa1958,Angelia2009_saddle} without perturbation. Numerical results will show that these methods do not perform as well as the proposed algorithm with perturbation.}
Another recent development is the Bregman-distance based PD subgradient method proposed in \cite{Srivastava12_minmax} for solving an epigraph formulation of a min-max problem. The method in \cite{Srivastava12_minmax}, however, assumes that the Lagrangian function has a unique saddle point, in order to guarantee the convergence of the primal-dual iterates. In contrast, our proposed algorithm, which uses the perturbed subgradients, does not require such assumption.

{\bf Synopsis:} Section \ref{sec: model and prob} presents the problem formulation, {applications,} and a brief review of the centralized PD subgradient methods. Section \ref{sec: proposed} presents the proposed distributed consensus-based PDP algorithm. The assumptions and convergence analysis results are given in Section \ref{sec: analysis}. Numerical results are presented in Section \ref{sec: simulation}. Finally, the conclusions and discussion of future extensions are drawn in Section \ref{sec: conclusion}.

\vspace{-0.4cm}
\section{Problem Formulation, Applications and Brief Review}\label{sec: model and prob}

\subsection{Problem Formulation} \label{subsec: formualtion}
We consider a network with $N$ agents, denoted by $\mathcal{V}=\{1,\ldots,N\}$.  We assume that, for all $i=1,\ldots,N$, each agent $i$ has a local decision variable\footnote{Here, without loss of generality, we assume that all the agents have the same variable dimension $K$. The proposed algorithm and analysis can be easily generalized to the case with different variable dimensions.} $\xb_i \in \mathbb{R}^K$, a local constraint set $\mathcal{X}_i \subseteq \mathbb{R}^{K}$, and a local mapping function {$\fb_i:\mathbb{R}^{K} \rightarrow \mathbb{R}^M$}, in which
$\fb_i=\left(f_{i1},\ldots,f_{iM}\right)^T$ with each
$f_{im}:\mathbb{R}^{K} \rightarrow \mathbb{R}$ being continuous.
The network cost function is given by
\begin{align}\label{eq: network utility}
\bar\Fc(\xb_1,\ldots,\xb_N)\triangleq \Fc\left(\sum_{i=1}^N \fb_i(\xb_i)\right),
\end{align}
where {$\Fc:\mathbb{R}^M\rightarrow \mathbb{R}$} and  $\bar\Fc:\mathbb{R}^{NK}\rightarrow \mathbb{R}$ are continuous.
In addition, the agents are subject to a global inequality constraint $\sum_{i=1}^N \gb_i(\xb_i)\preceq \zerob$, where $\gb_i:\mathbb{R}^{K} \rightarrow \mathbb{R}^P$ are continuous {mappings} for all $i=1,\ldots,N$; {specifically, $\gb_i=\left(g_{i1},\ldots,g_{iP}\right)^T$, with each
$g_{ip}:\mathbb{R}^{K} \rightarrow \mathbb{R}$ being continuous. The vector inequality
$\sum_{i=1}^N \gb_i(\xb_i)\preceq \zerob$ is understood coordinate-wise.}

We assume that each agent $i$ can access $\Fc(\cdot)$, $\fb_i(\cdot)$, $\gb_i(\cdot)$ and $\mathcal{X}_i$ only, for all $i=1,\ldots,N$. Under this local knowledge constraint, the agents seek to cooperate with each other to minimize the total network cost $\bar\Fc(\xb_1,\ldots,\xb_N)$ (or maximize the network utility $-\bar\Fc(\xb_1,\ldots,\xb_N)$). Mathematically, the optimization problem can be formulated as follows
\begin{align}\label{problem}
  \min_{\substack{\xb_i\in \mathcal{X}_i, \\ i=1,\ldots,N}}~ \bar\Fc(\xb_1,\ldots,\xb_N)~~
  \text{s.t.}~  \sum_{i=1}^N \gb_i(\xb_i) \preceq \zerob. 
\end{align}
The goal of this paper is to develop a \emph{distributed} algorithm for solving \eqref{problem}
with each agent communicating with {their neighbors only}.

\subsection{\blue Application to Smart Grid Control} \label{subsec: applications}
{\blue In this subsection, we discuss some smart grid control problems where the problem formulation \eqref{problem} may arise.}
Consider a power grid system where a retailer (e.g., the utility company) bids electricity from the power market and serves a residential/industrial neighborhood with $N$ customers. In addition to paying for its market bid, the retailer has to pay additional cost if there is a deviation between the bid purchased in earlier market settlements and the real-time
aggregate load of the customers. Any demand excess or shortfall results in a cost for the retailer that mirrors the effort to maintain the power balance. In the smart grid, thanks to the advances in communication and sensory technologies, it is envisioned that the retailer can observe the load of customers and can even control the power usage of some of the appliances (e.g., controlling the charging rate of electrical vehicles and turning ON/OFF air conditioning systems), which is known as the demand side management (DSM); see \cite{dsm2} for a recent review.

We let $p_t$, $t=1,\ldots,T$, be the power bids over a time horizon of length $T$, and
let $\psi_{i,t}(\xb_i)$, $t=1,\ldots,T$, be the load profile of customer $i$,
where {\blue $\xb_i \in \mathbb{R}^K$ contains some control variables.}
The structures of $\psi_{i,t}$ and $\xb_i$ depend on the appliance load model.
As mentioned, the retailer aims to minimize the cost caused by power imbalance, e.g., \cite{dsm2,LiLow2011PES,ChangPES2012}
\ifconfver
\begin{align}\label{DSM problem}
 \min_{\xb_1\in \mathcal{X}_1,\ldots,\xb_N\in \mathcal{X}_N}~ &C_{\rm p}\bigg[\bigg(\sum_{i=1}^N\psib_i(\xb_i)-\pb\bigg)^+\bigg] \notag \\
 &~~~~~~+ C_{\rm s}\bigg[\bigg(\pb-\sum_{i=1}^N\psib_i(\xb_i)\bigg)^+\bigg]
\end{align}
\else
\begin{align}\label{DSM problem}
 \min_{\xb_1\in \mathcal{X}_1,\ldots,\xb_N\in \mathcal{X}_N}~C_{\rm p}
 \bigg[\bigg(\sum_{i=1}^N\psib_i(\xb_i)-\pb\bigg)^+\bigg] + C_{\rm s}\bigg[\bigg(\pb-\sum_{i=1}^N\psib_i(\xb_i)\bigg)^+\bigg]
\end{align}
\fi
where $(x)^+=\max\{x,0\}$, $\mathcal{X}_i$ denotes the local control constraint set and $C_{\rm p},C_{\rm s}:\mathbb{R}^{T} \rightarrow \mathbb{R}$ denote the cost functions due to insufficient and excessive power bids, respectively. Moreover, {\blue let $\pb=(p_1,\ldots,p_T)^T$} and
$\psib_i=(\psi_{i,1},\ldots,\psi_{i,T})^T.$
By defining $\zb=(\sum_{i=1}^N\psib_i(\xb_i)-\pb)^+$ and assuming that $C_{\rm p}$ is monotonically increasing, one can write \eqref{DSM problem} as
\begin{align}\label{DSM problem 2}
 \min_{\xb_1\in \mathcal{X}_1,\ldots,\xb_N\in \mathcal{X}_N}~ &C_{\rm p}[\zb]
 + C_{\rm s}\bigg[\zb-\sum_{i=1}^N\psib_i(\xb_i)+\pb\bigg] \\
 {\rm s.t.}~&~  \sum_{i=1}^N\psib_i(\xb_i)-\pb-\zb \preceq \zerob, \notag
\end{align}
which belongs to the considered formulation in \eqref{problem}.
Similar problem formulations also arise in the microgrid control problems \cite{Juan2009,GuanTSG2010} where the microgrid controller requires not only to control the loads but also to control the local power generation and local power storage (i.e., power flow control), in order to maintain power balance within the microgrid; see \cite{GuanTSG2010} for detailed formulations.
Distributed control methods are appealing to the smart grid application since
all the agents are identical and failure of one agent would not have significant impact on the performance of the whole system \cite{Cai2010}. Besides, it also
spares the retailer/microgrid controller from the task of collecting {\blue real-time information of customers}, which not only infringes on the customer's privacy but also is not easy for a large scale neighborhood. In Section \ref{sec: simulation}, the proposed distributed algorithm will be applied to a DSM problem as in \eqref{DSM problem 2}.

{\blue In addition to the smart grid applications, problem \eqref{problem} incorporates the important regression problems
which widely appear in control \cite{Belomestny2010}, machine learning \cite{Hastie2001Book,Elad2009Book}, data mining \cite{Hershberger2001,Kargupta1999} and imaging processing \cite{Elad2009Book} applications.
Formulation \eqref{problem} also encompasses the network flow control problems \cite{BK:Bersekas_netowork}; see
\cite{Madan2006} for an example which considered maximizing the network lifetime. The proposed distributed algorithm therefore can be applied to these problem as well. For example, in \cite{Changglobalsip13}, we have shown how the proposed distributed algorithm can be applied to handle a distributed sparse regression problem. }

\vspace{-0.4cm}
\subsection{\blue Centralized PD Subgradient Method}

Let us consider the following {Lagrange} dual problem of \eqref{problem}:
\begin{align}\label{dual}
  \max_{\substack{\lambdab\succeq \zerob}}
  \Bigg\{\min_{\substack{\xb\in \Xc}}~ \Lc(\xb,\lambdab)\Bigg\}, 
\end{align}
where $\xb=(\xb_1^T,\ldots,\xb_N^T)^T$,  $\Xc=\Xc_1\times\cdots\times \Xc_N$,
$\lambdab \in {\mathbb{R}^{P}_+}$ (i.e., the non-negative orthant in $\mathbb{R}^{P}$) is the dual variable associated with
the inequality constraint $\sum_{i=1}^N \gb_i(\xb_i) \preceq \zerob$, 
and $\Lc:\mathbb{R}^{NK}\times {\mathbb{R}^{P}_+}\rightarrow \mathbb{R}$ is the Lagrangian function, given by
\begin{align}\label{lagrangian}
\Lc(\xb,\lambdab)=
\bar\Fc(\xb_1,\ldots,\xb_N)
+\lambdab^T\left(\sum_{i=1}^N\gb_i(\xb_i)\right). 
\end{align}
{\blue Throughout the paper, we assume that problem \eqref{problem} is convex, i.e., $\Xc$ is closed and convex, $\bar \Fc(\xb)$ is convex in $\xb$ and each $\gb_i(\xb_i)$ is convex in $\xb_i$.
We also assume that the Slater condition holds,
i.e., there is an $(\bar\xb_1,\ldots,\bar\xb_N)$ that lies in the relative interior of
  $\Xc_1\times\cdots\times\Xc_N$ such that
 $ \sum_{i=1}^N\gb_i(\bar\xb_i) \prec \zerob. $
Hence, the strong duality holds for problem \eqref{problem} \cite{BK:BoydV04},
problem \eqref{problem} can be handled by solving its dual \eqref{dual}.}
{\blue A classical approach  is the dual subgradient method \cite{BK:Bertsekas2003_analysis}.
One limitation of such method is that the inner problem $\min_{\substack{\xb\in \Xc}}~ \Lc(\xb,\lambdab)$ needs to be globally solved at each iteration, which, however, is not always easy, especially when $\fb_i(\xb_i)$ and $\gb_i(\xb_i)$ are complex or when the problem is large scale.}
%
Another approach is the primal-dual (PD) subgradient method \cite{Uzawa1958,Zabotin88} {\blue which handles the inner problem inexactly.}
More precisely, at iteration $k$, the PD subgradient method performs
\begin{subequations}\label{eq: primal-dual central}
\begin{align}
    \xb^{(k)}&=\Pc_{\Xc}(\xb^{(k-1)}-a_k ~ \Lc_{\xb}(\xb^{(k-1)},\lambdab^{(k-1)})), \label{eq: primal-dual central a}\\
    \lambdab^{(k)}&=(\lambdab^{(k-1)}+a_k ~ \Lc_{\lambdab}(\xb^{(k-1)},\lambdab^{(k-1)}))^+,\label{eq: primal-dual central b} 
\end{align}
\end{subequations}
where $\Pc_{\Xc}: \mathbb{R}^{NK} \rightarrow \Xc$ is a projection function, 
$a_k>0$ is a step size,
and
\vspace{-0.0cm}
\ifconfver
\begin{subequations}\label{eq: subgradients central}
\begin{align}
  &\Lc_{\xb}(\xb^{(k)},\lambdab^{(k)})
  \!\triangleq\!\!
\begin{bmatrix}
  \Lc_{\xb_1}(\xb^{(k)},\lambdab^{(k)}) \\
  \vdots \\
  \Lc_{\xb_N}(\xb^{(k)},\lambdab^{(k)})
\end{bmatrix}\!\!\notag \\
 &=\!\!
  \begin{bmatrix}
    \nabla \fb_1^T(\xb_1^{(k)})\nabla \Fc(\sum_{i=1}^N \fb_i(\xb_i^{(k)})) + {\nabla  \gb^T_1(\xb_1^{(k)}) \lambdab^{(k)} }\cr
    \vdots \\
     \nabla \fb_N^T(\xb_N^{(k)})\nabla \Fc(\sum_{i=1}^N \fb_i(\xb_i^{(k)}))+ {\nabla  \gb^T_N(\xb_N^{(k)}) \lambdab^{(k)} }\cr
  \end{bmatrix},
  \label{eq: subgradients central a}\\
  &\Lc_{\lambdab}(\xb^{(k)},\lambdab^{(k)})\triangleq \sum_{i=1}^N \gb_i(\xb_i^{(k)}),
    \label{eq: subgradients central b} 
\end{align}
\end{subequations}
\else
\begin{subequations}\label{eq: subgradients central}
\begin{align}
\!\!\!\!\!\! \Lc_{\xb}(\xb^{(k)},\lambdab^{(k)})&
  \!\triangleq\!\!
\begin{bmatrix}
  \Lc_{\xb_1}(\xb^{(k)},\lambdab^{(k)}) \\
  \vdots \\
  \Lc_{\xb_N}(\xb^{(k)},\lambdab^{(k)})
\end{bmatrix}\!\!=\!\!
  \begin{bmatrix}
    \nabla \fb_1^T(\xb_1^{(k)})\nabla \Fc(\sum_{i=1}^N \fb_i(\xb_i^{(k)})) + {\nabla  \gb^T_1(\xb_1^{(k)}) \lambdab^{(k)} }\cr
    \vdots \\
     \nabla \fb_N^T(\xb_N^{(k)})\nabla \Fc(\sum_{i=1}^N \fb_i(\xb_i^{(k)}))+ {\nabla  \gb^T_N(\xb_N^{(k)}) \lambdab^{(k)} }\cr
  \end{bmatrix},
  \label{eq: subgradients central a}\\
   \Lc_{\lambdab}(\xb^{(k)},\lambdab^{(k)})&\triangleq \sum_{i=1}^N \gb_i(\xb_i^{(k)}),
    \label{eq: subgradients central b} 
\end{align}
\end{subequations}
\fi
represent the subgradients of $\Lc$ at $(\xb^{(k)},\lambdab^{(k)})$ with
{respect to} $\xb$ and $\lambdab$, respectively.
Each $\nabla  \gb_i(\xb_i^{(k)})$ is a $P\times K$ Jacobian matrix with rows equal to the
subgradients $\nabla g^T_{ip}(\xb_i)$, $p=1,\ldots,P$ (gradients if they are continuously differentiable), and each $\nabla  \fb_i(\xb_i^{(k)})$ is a $M\times K$ Jacobian matrix with rows containing the
gradients $\nabla f^T_{im}(\xb_i)$, $m=1,\ldots,M.$

{The idea behind the PD subgradient method lies in the well-known saddle-point relation:}

\begin{Theorem} {\rm (Saddle-Point Theorem) \cite{BK:BoydV04}} \label{thm: saddle point thm}
The {point} $(\xb^\star,\lambdab^\star)
\in \Xc\times\mathbb{R}^P_+$ is a primal-dual solution pair of
problems \eqref{problem} and \eqref{dual} if and only if {there holds:}
\begin{align}
    \Lc(\xb^\star,\lambdab)
        \leq \Lc(\xb^\star,\lambdab^\star) \leq
        \Lc(\xb,\lambdab^\star)~\forall \xb\in \Xc,~\lambdab\succeq \zerob.
\end{align} 
\end{Theorem}
{According to Theorem \ref{thm: saddle point thm}, if the PD subgradient method converges to a saddle point of the Lagrangian function~\eqref{lagrangian}, then it solves the original problem~\eqref{problem}.} Convergence properties of the PD method in \eqref{eq: primal-dual central} have been studied extensively; see, for example,
\cite{Uzawa1958,Zabotin88,Angelia2009_saddle}.
{In such methods, typically a subsequence of the sequence
$(\xb^{(k)},\lambdab^{(k)})$ converges to a saddle point of the Lagrangian function in \eqref{lagrangian}. To ensure the convergence of the whole sequence $(\xb^{(k)},\lambdab^{(k)})$, it is often assumed that the Lagrangian function is strictly convex in $\xb$ and strictly concave in $\lambdab$, which does not hold in general however.
}

One of the approaches to circumventing this condition is the primal-dual perturbed (PDP) subgradient method in \cite{Kallio1994,Kallio1999}. Specifically, \cite{Kallio1994} suggests to update $\xb^{(k-1)}$ and $\lambdab^{(k-1)}$ based on some perturbation points, denoted by $\hat\alphab^{(k)}$ and $\hat\betab^{(k)}$, respectively. The PDP updates are
\begin{subequations}\label{eq: primal-dual central perturbed}
\begin{align}
    \xb^{(k)}&=\Pc_{\Xc}(\xb^{(k-1)}-a_k ~ \Lc_{\xb}(\xb^{(k-1)},\hat\betab^{(k)} )), \label{eq: primal-dual central a2}\\
    \lambdab^{(k)}&=(\lambdab^{(k-1)}+a_k ~ \Lc_{\lambdab}(\hat\alphab^{(k)},\lambdab^{(k-1)}))^+\label{eq: primal-dual central b2}. 
\end{align}
\end{subequations}
Note that, in \eqref{eq: primal-dual central a2}, we have replaced $\lambdab^{(k-1)}$ by $\hat\betab^{(k)}$, and, in \eqref{eq: primal-dual central b2}, replaced $\xb^{(k-1)}$ by $\hat\alphab^{(k)}$, and thus $\Lc_{\xb}(\xb^{(k-1)},\hat\betab^{(k)} )$ and $\Lc_{\lambdab}(\hat\alphab^{(k)},\lambdab^{(k-1)})$ are perturbed subgradients.
It was shown in  \cite{Kallio1994} that, with carefully chosen
$(\hat\alphab^{(k)},\hat\betab^{(k)})$ and the
step size $a_k$, the primal-dual iterates in
\eqref{eq: primal-dual central perturbed} converge to a saddle point of \eqref{dual}, without any strict convexity and concavity assumptions on $\Lc$.

There are several ways to generate the perturbation points $\hat\alphab^{(k)}$ and $\hat\betab^{(k)}$. Our interests lie specifically on those that are computationally as efficient as the PD subgradient updates in \eqref{eq: primal-dual central perturbed}. Depending on the smoothness of $\{g_{ip}\}_{p=1}^P$, we consider the following two methods:

{\bf Gradient Perturbation Points:} A simple approach to computing the perturbation points is using the conventional gradient updates {exactly as} in \eqref{eq: primal-dual central}, i.e.,
\begin{subequations}\label{eq: gradient perturbation}
\begin{align}
    \hat\alphab^{(k)}&=\Pc_{\Xc}(\xb^{(k-1)}-\rho_1 ~ \Lc_{\xb}(\xb^{(k-1)},\lambdab^{(k-1)})), \label{eq: gradient perturbation a3}\\
    \hat\betab^{(k)}&=(\lambdab^{(k-1)}+\rho_2 ~ \Lc_{\lambdab}(\xb^{(k-1)},\lambdab^{(k-1)}))^+\label{eq: gradient perturbation b3} 
    \end{align}
\end{subequations} where $\rho_1>0$ and $\rho_2>0$ are constants. {
The PDP subgradient method thus combines \eqref{eq: primal-dual central perturbed} and \eqref{eq: gradient perturbation}, which involve two primal and dual subgradient updates.} Even though the updates are relatively simple, this method requires smooth constraint functions $g_{ip}$, $p=1,\ldots,P$.
%
%

{\blue In cases {when} $g_{ip}$, $p=1,\ldots,P,$ are non-smooth, we propose to use the following proximal perturbation point {approach},
which is novel and has not appeared in earlier works \cite{Kallio1994,Kallio1999}.}

{\bf Proximal Perturbation Points:} When $g_{ip}$, $p=1,\ldots,P,$ are non-smooth, we compute the perturbation point $\hat\alphab^{(k)}$ by the following proximal gradient {update}\footnote{If not mentioned specifically, the norm function $\|\cdot\|$ stands for the Euclidian norm.} \cite{Nesterov2005}:
\ifconfver
\begin{align}
\!    \hat\alphab^{(k)}&=
    \arg\min_{\alphab \in \Xc}
    \bigg\{\sum_{i=1}^N\gb_i^T(\alphab_i)\lambdab^{(k-1)}\notag \\
    &~~~+\frac{1}{2\rho_1} \bigg\|\alphab- ( \xb^{(k-1)} -\rho_1 \nabla \bar\Fc(\xb^{(k-1)}) )\bigg\|^2 \bigg\} \label{eq: proximal perturbation point}
 \end{align}
 \else
\begin{align}
    \hat\alphab^{(k)}=&
    \arg\min_{\alphab \in \Xc}
    \bigg\{\sum_{i=1}^N\gb_i^T(\alphab_i)\lambdab^{(k-1)}+
    \frac{1}{2\rho_1} \bigg\|\alphab- ( \xb^{(k-1)} -\rho_1 \nabla \bar\Fc(\xb^{(k-1)}) )\bigg\|^2 \bigg\}, \label{eq: proximal perturbation point}
 \end{align}
 \fi
where $\alphab=(\alphab_1^T,\ldots,\alphab_N^T)^T$ and
\begin{align}
  \nabla \bar\Fc(\xb^{(k-1)})=
  \begin{bmatrix}
     \nabla \fb_1^T(\xb_1^{(k-1)}) {\blue\nabla  \Fc(\sum_{i=1}^N \fb_i(\xb_i^{(k-1)}))} \\
     \vdots\\
     \nabla \fb_N^T(\xb_N^{(k-1)}) {\blue\nabla \Fc(\sum_{i=1}^N \fb_i(\xb_i^{(k-1)}))}
  \end{bmatrix}.
\end{align}
{It is worthwhile to note that, when $g_{ip},$ $p=1,\ldots,P$, are some \emph{sparsity promoting functions} (e.g., the 1-norm, 2-norm and the nuclear norm) that often arise in sparse optimization problems \cite{Elad2009Book,Mateos2010,Mota2012}, the proximal perturbation point in \eqref{eq: proximal perturbation point}
can be solved very efficiently and {may even have closed-form} solutions.
For example, if $\gb_i(\alphab_i)=\|\alphab_i\|_1$ for all $i$ $(P=1)$, and $\Xc=\mathbb{R}^{KN}$, \eqref{eq: proximal perturbation point} has a closed-form solution known as the soft thresholding operator \cite{Elad2009Book}:
\begin{align}\label{soft thresholding}
    \hat\alphab^{(k)} = ( \bb - \lambda^{(k-1)}\rho_1\mathbf{1})^+
    +( -\bb - \lambda^{(k-1)}\rho_1\mathbf{1})^+,
\end{align}
where $\bb=\xb^{(k-1)} -\rho_1 \nabla \bar\Fc(\xb^{(k-1)})$ and $\mathbf{1}$ is an all-one vector.
}

\section{Proposed {Consensus-Based Distributed PDP} Algorithm} \label{sec: proposed}

Our goal is to develop a {distributed} counterpart of the PDP subgradient method in
\eqref{eq: primal-dual central perturbed}. 
{\blue Consider the following saddle-point problem}
\begin{align}\label{saddle point problem}
  \max_{\lambdab\in \Dc}\Bigg\{\min_{\substack{\xb_i\in \Xc_i,\\ i=1,\ldots,N}}~ \Lc(\xb_1,\ldots,\xb_N,\lambdab)\Bigg\} 
\end{align}
where
\begin{align}\label{eq: D set}
  \Dc=\left\{\lambdab\succeq \zerob~|~
  \|\lambdab\|\leq D_\lambda \triangleq \frac{\bar\Fc(\bar\xb)-\tilde q}{\gamma}+\delta\right\}
\end{align} in which {$\bar\xb=(\bar\xb_1^T,\ldots,\bar\xb_N^T)^T$} is a Slater point of \eqref{problem}, $\tilde q=\min_{\substack{\xb_i\in \Xc_i, i=1,\ldots,N}}~ \Lc(\xb_1,\ldots,\xb_N,\tilde \lambdab)$ is the dual function value for some arbitrary $\tilde \lambdab\succeq\zerob$,
$\gamma=\min_{p=1,\ldots,P}\{-\sum_{i=1}^N g_{ip}(\bar\xb_i)\}$, and $\delta>0$ is arbitrary. It has been shown in \cite{Angelia2009_approxprimal} that the optimal dual solution $\hat\lambdab^\star$ of \eqref{dual} satisfies
\begin{align}\label{eq: opt dual}
  \|\hat\lambdab^\star\| \leq \frac{\bar\Fc(\bar\xb)-\tilde q}{\gamma}
\end{align}
and thus {$\hat\lambdab^\star$ lies} in $\Dc$. Here we consider the saddle point problem \eqref{saddle point problem}, instead of the original Lagrange dual problem \eqref{dual},
because $\Dc$ bounds the dual variable $\lambdab$ and thus also bounds the subgradient $ \Lc_{\xb}(\xb^{(k)},\lambdab^{(k)})$ in \eqref{eq: subgradients central a}. This property is important in building the convergence of the distributed algorithm to be {discussed shortly}.
Both \eqref{dual} and \eqref{saddle point problem} have the same optimal dual solution $\hat\lambdab^\star$ and attain the same optimal objective value. One can further verify that any saddle point of \eqref{dual} is also a saddle point of \eqref{saddle point problem}. However, to {\blue relate the saddle points of \eqref{saddle point problem} to solutions of problem~\eqref{problem}}
 some conditions are needed, as given in the following proposition.

\begin{Proposition}\label{prop: opt conditions}{\rm (Primal-dual optimality conditions) \cite{Larsson1999} }
  {\blue Let the Slater condition hold and
  let $(\hat{\xb}_1^\star,\ldots,\hat{\xb}_N^\star,\hat\lambdab^\star)$ be a saddle point of
  \eqref{saddle point problem}.} Then $(\hat{\xb}_1^\star,\ldots,\hat{\xb}_N^\star)$ is {an optimal
  solution for} problem \eqref{problem} if and only if
  \begin{align*}
      \sum_{i=1}^N\gb_i(\hat{\xb}_i^\star) \preceq \zerob~\text{and}~
      (\hat\lambdab^\star)^T\left(\sum_{i=1}^N\gb_i(\hat{\xb}_i^\star)\right)=\zerob.
 \end{align*}
\end{Proposition}


To have a distributed optimization algorithm for solving \eqref{saddle point problem}, in addition to $\xb_i^{(k)}$,
we let each agent $i$ have a local copy of the dual iterate $\lambdab^{(k)}$, denoted by $\lambdab_i^{(k)}$. Moreover, each agent $i$ owns two auxiliary variables, denoted by $\yb_{i}^{(k)}$
and $\zb_i^{(k)}$, representing respectively the
local estimates of the average values of the argument function
$\frac{1}{N}\sum_{i=1}^N \fb_i(\xb_i^{(k)})$
and of the inequality constraint function $\frac{1}{N}\sum_{i=1}^N \gb_i(\xb_i^{(k)})$,
for all $i=1,\ldots,N$.
We consider a \emph{time-varying synchronous network} model \cite{Angelia2010}, where
the network of agents at time $k$ is represented by a {weighted} directed graph $\mathcal{G}(k)=(\mathcal{V},\mathcal{E}(k),\Wb(k))$.
Here $(i,j)\in \mathcal{E}(k)$ if and only if agent $i$ {can receive messages from agent $j$},
and $\Wb(k) \in \mathbb{R}^{N\times N}$ is a {weight} matrix with each entry $[\Wb(k)]_{ij}$
{representing a weight that agent $i$ assigns to the incoming message on link $(i,j)$ at time $k$}.
If $(i,j)\in  \mathcal{E}(k)$, then $[\Wb(k)]_{ij}>0$ and $[\Wb(k)]_{ij}=0$ otherwise.
The agents exchange messages with their neighbors (according to the network graph $\mathcal{G}(k)$) in order to achieve consensus on
$\lambdab^{(k)}$, $\sum_{i=1}^N \gb_i(\xb_i^{(k)})$ and $\sum_{i=1}^N \fb_i(\xb_i^{(k)})$; while computing local perturbation points and primal-dual (sub-)gradient updates locally. Specifically, the proposed
distributed consensus-based PDP
method consists of the following steps at each iteration $k$:

{\bf 1) Averaging consensus:} For $i=1,\ldots,N$,
each agent $i$ sends $\yb_{i}^{(k-1)}$, $\zb_{i}^{(k-1)}$ and $\lambdab_i^{(k-1)}$
to all its neighbors $j$ satisfying $(j,i) \in \mathcal{E}(k)$;
it also receives $\yb_{j}^{(k-1)}$, $\zb_{j}^{(k-1)}$ and $\lambdab_j^{(k-1)}$ from its neighbors, and combines
the received estimates, as follows:
\ifconfver
\begin{align}\label{consensus step}
    &\tilde \yb_{i}^{(k)}= \sum_{j=1}^N [\Wb(k)]_{ij} \yb_{j}^{(k-1)},~ \tilde \zb_{i}^{(k)}= \sum_{j=1}^N [\Wb(k)]_{ij} \zb_{j}^{(k-1)},\notag
    \\
     &\tilde \lambdab_{i}^{(k)}= \sum_{j=1}^N [\Wb(k)]_{ij} \lambdab_{j}^{(k-1)}.~
\end{align}
\else
\begin{align}\label{consensus step}
    \tilde \yb_{i}^{(k)}= \sum_{j=1}^N [\Wb(k)]_{ij} \yb_{j}^{(k-1)},~ \tilde \zb_{i}^{(k)}= \sum_{j=1}^N [\Wb(k)]_{ij} \zb_{j}^{(k-1)},~ \tilde \lambdab_{i}^{(k)}= \sum_{j=1}^N [\Wb(k)]_{ij} \lambdab_{j}^{(k-1)}.
\end{align}\fi
%
%
%

{\bf 2) Perturbation point computation:}
For $i=1,\ldots,N,$ if functions $g_{ip}$, $p=1,\ldots,P,$ are smooth, then each agent $i$ computes the local perturbation points by
\ifconfver
\begin{subequations}\label{local perturbation}
\begin{align}\label{local perturbation x}
              \alphab_i^{(k)}&=\mathcal{P}_{\mathcal{X}_i}
              (\xb_i^{(k-1)} - \rho_1 [   \nabla \fb_i^T(\xb_i^{(k-1)})\nabla \Fc(N\tilde\yb_i^{(k)})
                 \notag \\
                 &~~~~~~~~~~~~~~~~~~+ \nabla  \gb^T_i(\xb_i^{(k-1)}) \tilde \lambdab_i^{(k)} ]) ,\\
              \betab_i^{(k)}&=\mathcal{P}_{\Dc}(\tilde\lambdab_i^{(k)}
                         +\rho_2 ~ N\tilde \zb_i^{(k)}).
              \label{local perturbation lambda}
\end{align}\end{subequations}
\else
\begin{subequations}\label{local perturbation}
\begin{align}\label{local perturbation x}
              \alphab_i^{(k)}&=\mathcal{P}_{\mathcal{X}_i}
              (\xb_i^{(k-1)} - \rho_1 [   \nabla \fb_i^T(\xb_i^{(k-1)})\nabla \Fc(N\tilde\yb_i^{(k)})
                 + \nabla  \gb^T_i(\xb_i^{(k-1)}) \tilde \lambdab_i^{(k)} ]) ,\\
              \betab_i^{(k)}&=\mathcal{P}_{\Dc}(\tilde\lambdab_i^{(k)}
                         +\rho_2 ~ N\tilde \zb_i^{(k)}).
              \label{local perturbation lambda}
\end{align}\end{subequations}
\fi
Note that, comparing to \eqref{eq: gradient perturbation} and \eqref{eq: proximal perturbation point}, agent $i$ here uses the most up-to-date estimates
$N\tilde \yb_{i}^{(k)}$, $N\tilde \zb_{i}^{(k)}$ and $\tilde \lambdab_{i}^{(k)}$ in {place} of $\sum_{i=1}^N \fb_i(\xb_i^{(k-1)})$, $\sum_{i=1}^N \gb_i(\xb_i^{(k-1)})$ and $\lambdab^{(k-1)}$.
If $g_{ip}$, $p=1,\ldots,P,$ are non-smooth, agent $i$ instead computes $\alphab_i^{(k)}$ by
\ifconfver
\begin{align}\label{eq: local proximal perturbation point}
    \alphab_i^{(k)}=&\arg\min_{\alphab_i \in \Xc_i}
    \bigg\{\gb_i^T(\alphab_i)\tilde \lambdab_i^{(k)}
    +\frac{1}{2\rho_1} \|\alphab_i - (\xb_i^{(k-1)} \notag \\
    &- \rho_1 \nabla \fb_i^T(\xb_i^{(k-1)})\nabla \Fc(N\tilde\yb_i^{(k)}))\|^2\bigg\},
 \end{align}
\else
\begin{align}\label{eq: local proximal perturbation point}
    \alphab_i^{(k)}=&\arg\min_{\alphab_i \in \Xc_i}
    \left\{\gb_i^T(\alphab_i)\tilde \lambdab_i^{(k)}
    +\frac{1}{2\rho_1} \|\alphab_i - (\xb_i^{(k-1)} - \rho_1 \nabla \fb_i^T(\xb_i^{(k-1)})\nabla \Fc(N\tilde\yb_i^{(k)}))\|^2\right\},
 \end{align} \fi for $i=1,\ldots,N$.

{\bf3) Primal-dual perturbed subgradient update:}
For $i=1,\ldots,N$, each agent $i$ updates its primal and dual variables $(\xb_i^{(k)},\lambdab_i^{(k)})$ based on the local perturbation point $(\alphab_i^{(k)},\betab_i^{(k)})$:
\ifconfver
\begin{align}\label{local update x}
 \xb_i^{(k)}&=\mathcal{P}_{\mathcal{X}_i}
   (\xb_i^{(k-1)} - a_k [ \nabla \fb_i^T(\xb_i^{(k-1)})\nabla \Fc(N\tilde\yb_i^{(k)})
       \notag \\
       &~~~~~~~~~~~~~~~~~~ + \nabla  \gb^T_i(\xb_i^{(k-1)}) \betab_i^{(k)} ]) ,\\
         \lambdab_i^{(k)}&=\mathcal{P}_{\Dc}(\tilde\lambdab_i^{(k)}
                     +a_k ~ \gb_i(\alphab_i^{(k)})). \label{local update lambda}
\end{align}
\else
\begin{align}\label{local update x}
 \xb_i^{(k)}&=\mathcal{P}_{\mathcal{X}_i}
   (\xb_i^{(k-1)} - a_k [ \nabla \fb_i^T(\xb_i^{(k-1)})\nabla \Fc(N\tilde\yb_i^{(k)})
        + \nabla  \gb^T_i(\xb_i^{(k-1)}) \betab_i^{(k)} ]) ,\\
         \lambdab_i^{(k)}&=\mathcal{P}_{\Dc}(\tilde\lambdab_i^{(k)}
                     +a_k ~ \gb_i(\alphab_i^{(k)})). \label{local update lambda}
\end{align}
\fi

{\bf 4) Auxiliary variable update:} For $i=1,\ldots,N$, each agent $i$ updates variable
$\yb_{i}^{(k)}$, $\zb_{i}^{(k)}$ with the changes of the local argument function  $\fb_i(\xb_i^{(k)})$ and
the constraint function $\gb_i(\xb_i^{(k)})$ :
\begin{align}
\yb_{i}^{(k)}&= \tilde \yb_{i}^{(k)} + \fb_i(\xb_i^{(k)}) - \fb_i(\xb_i^{(k-1)}),\label{local update y} \\
\zb_{i}^{(k)}&= \tilde \zb_{i}^{(k)} + \gb_i(\xb_i^{(k)}) - \gb_i(\xb_i^{(k-1)}).\label{local update z}
\end{align}

Algorithm~1 summarizes the above steps. We prove that Algorithm~1 converges under proper problem and network assumptions in the next section. {Readers who are interested more in numerical performance of Algorithm~1 may go directly to Section \ref{sec: simulation}.}

\begin{algorithm}[t!]
\caption{Distributed {Consensus-Based PDP} Algorithm}
\begin{algorithmic}[1]\label{table: ADMM}
\STATE {\bf Given} initial variables
{$\xb_i^{(0)}\in \Xc_i$, $\lambdab_i^{(0)} \succeq \zerob$}, $\yb_{i}^{(0)}=\fb_i(\xb_i^{(0)})$ and $\zb_{i}^{(0)}=\gb_i(\xb_i^{(0)})$ for each agent $i$, $i=1,\ldots,N;$ Set $k=1$.
\REPEAT
\STATE {\bf Averaging consensus:} For $i=1,\ldots,N$,
each agent $i$ computes \eqref{consensus step}.

\STATE {\bf Perturbation point computation:} For $i=1,\ldots,N$, if $\{g_{ip}\}_{p=1}^P$ are smooth, then each agent $i$ computes the local perturbation points by \eqref{local perturbation};
otherwise, each agent $i$ instead computes $\alphab_i^{(k)}$ by \eqref{eq: local proximal perturbation point}.

\STATE {\bf Local variable updates:} For $i=1,\ldots,N$, each agent $i$ updates
\eqref{local update x}, \eqref{local update lambda}, \eqref{local update y} and \eqref{local update z} sequentially.
%
%
%
%

\STATE {\bf Set} $k=k+1.$
\UNTIL {a predefined stopping criterion (e.g., a maximum iteration number) is satisfied.}%
\end{algorithmic}
\end{algorithm}


%

%
%
\vspace{-0.3cm}
\section{Convergence Analysis}\label{sec: analysis}
Next, in Section \ref{subsec: assumption}, we present additional assumptions on problem \eqref{problem} and the network model. The main convergence results are presented in Section IV-B. {\blue The proofs are presented in Section IV-C and Section IV-D.}
\vspace{-0.3cm}
\subsection{Assumptions}\label{subsec: assumption}
{\blue Our results will make use of the following assumption.}
\begin{Assumption}\label{assumption function}{\rm
  (a) The sets $\Xc_i$, $i=1,\ldots,N,$ are compact. 
  In particular, for $i=1,\ldots,N$, there is a constant $D_x>0$ such that
  \begin{align}\label{eq: bounded set}
      \|\xb_i\| \leq D_x~~\forall\xb_i\in \Xc_i;
  \end{align}
  (b) The functions $f_{i1},\ldots,f_{iM}$, $i=1,\ldots,N$, are continuously differentiable.
 }
\end{Assumption}
Note that Assumption \ref{assumption function}(a) and Assumption \ref{assumption function}(b) imply that
$f_{i1},\ldots,f_{iM}$ have uniformly bounded gradients (denoted by $\nabla f_{im}$, $m=1,\ldots,M$) and are Lipschitz continuous, i.e.,
for some $L_f>0$,
\ifconfver
  \begin{align}
  &\max_{1\le m\le M}\|\nabla f_{im}(\xb_i)\|\leq L_f,~\forall \xb_i\in \mathcal{X}_i \label{eq: bounded subgradient f}\\
  &\max_{1\le m\le M}|f_{im}(\xb_i)-f_{im}(\yb_i)|\notag \\
&~~~~~~~~~~~~~~~~~~\leq L_f \|\xb_i-\yb_i\| \forall \xb_i,\yb_i\in \mathcal{X}_i.
\label{eq: lipschitz f}
  \end{align}
\else
  \begin{align}
  &\max_{1\le m\le M}\|\nabla f_{im}(\xb_i)\|\leq L_f,~\forall \xb_i\in \mathcal{X}_i \label{eq: bounded subgradient f}\\
  &\max_{1\le m\le M}|f_{im}(\xb_i)-f_{im}(\yb_i)|\leq L_f \|\xb_i-\yb_i\|\quad \forall \xb_i,\yb_i\in \mathcal{X}_i.
\label{eq: lipschitz f}
  \end{align}
\fi
{\blue Similarly, Assumption \ref{assumption function}(a)
and the convexity of functions $g_{i1},\ldots,g_{iP}$ 
imply that all $g_{ip}$ have uniformly bounded subgradients, which is equivalent to all $g_{ip}$
being Lipschitz continuous. Thus, for some $L_g>0$, we have}
\ifconfver
\begin{align}
&\max_{1\le p\le P}\|\nabla g_{ip}(\xb_i)\|\leq L_g \quad\forall \xb_i\in \mathcal{X}_i, \label{eq: bounded subgradient g}\\
&\max_{1\le p\le P}|g_{ip}(\xb_i)-g_{ip}(\yb_i)|\notag \\
&~~~~~~~~~~~~~~~~~~\leq L_g \|\xb_i-\yb_i\|~\forall \xb_i,\yb_i\in \mathcal{X}_i.
\label{eq: lipschitz g}
\end{align}
\else
\begin{align}
&\max_{1\le p\le P}\|\nabla g_{ip}(\xb_i)\|\leq L_g \quad\forall \xb_i\in \mathcal{X}_i, \label{eq: bounded subgradient g}\\
&\max_{1\le p\le P}|g_{ip}(\xb_i)-g_{ip}(\yb_i)|\leq L_g \|\xb_i-\yb_i\| \quad \forall \xb_i,\yb_i\in \mathcal{X}_i.
\label{eq: lipschitz g}
\end{align}
\fi
{\blue In addition, by Assumption~\ref{assumption function} and the continuity of each $g_{ip}$ (which is implied by the convexity of $g_{ip}$)
each $\fb_i$ and $\gb_i$ are also bounded on $\Xc$, i.e., there exist}
constants $C_f>0$ and $C_g>0$ such that for all $i=1,\ldots,N$,
  \begin{align}\label{eq: function bounded}
      \|\fb_i(\xb_i)\| \leq C_f,\quad \|\gb_i(\xb_i)\| \leq C_g, \quad\forall\xb_i\in \Xc_i,
  \end{align}
where $\|\fb_i(\xb_i)\|=\sqrt{\sum_{m=1}^M f^2_{im}(\xb_i)}$ and $\|\gb_i(\xb_i)\|=\sqrt{\sum_{p=1}^P g^2_{ip}(\xb_i)}$.

We also {\blue make use of} the following assumption on the network utility costs $\Fc$ and $\bar\Fc$:
\begin{Assumption}\label{assumption function 2}{\rm
 (a) The function
 $\Fc$ is continuously differentiable and has bounded and Lipschitz continuous gradients, i.e., for some $G_\Fc>0$ and $L_\Fc>0$, we have
 \begin{align}
 &\| \nabla \Fc(\xb)- \nabla \Fc(\yb)\| \leq G_\Fc \|\xb-\yb\|\quad \forall \xb,\yb\in
  \mathbb{R}^M, \label{eq: lip gradient Fc} \\
 &\|\nabla \Fc\left(\yb\right)\| \leq L_\Fc\quad\forall \yb\in \mathbb{R}^M; \label{eq: gradient of F}
  \end{align}
  (b) {\blue The function $\bar\Fc$ 
  has Lipschitz continuous gradients, i.e.,}
  for some $G_{\bar\Fc}>0$,
  \begin{align}\label{eq: lip gradient bar Fc}
  \| \nabla \bar\Fc(\xb)- \nabla \bar\Fc(\yb)\| &\leq G_{\bar\Fc} \|\xb-\yb\|\quad \forall \xb,\yb\in \Xc.
  \end{align}  }
\end{Assumption}
{
Note that the convexity of $\bar\Fc$ 
Assumption \ref{assumption function}(a) indicate that $\bar\Fc$ is Lipschitz continuous, i.e., for some $L_{\bar\Fc}>0$,
  \begin{align}\label{eq: lip bar Fc}
  \| \bar\Fc(\xb)- \bar\Fc(\yb)\| &\leq L_{\bar\Fc} \|\xb-\yb\|\quad \forall \xb,\yb\in \Xc.
  \end{align}
  }
{\hspace{-0.4cm} \blue Assumptions \ref{assumption function} and \ref{assumption function 2} are used to ensure that the (sub-)gradients of the Lagrangian function $\Lc(\xb,\lambdab)$ with respect to $\xb$ are well behaved for
applying (sub-)gradient-based methods.
}
In cases that $g_{ip}$, $p=1,\ldots,P,$ are smooth, we make use of the following additional assumption:
\begin{Assumption}\label{assumption diff of g}{\rm
The functions $g_{ip}$, $p=1,\ldots,P,$ are continuously differentiable and have Lipschitz continuous gradients, i.e., there exists a constant $G_g>0$ such that
\ifconfver
\begin{align}
  &\max_{1\leq p \leq P}\| \nabla g_{ip}(\xb_i)- \nabla  g_{ip}(\yb_i)\| \notag \\
&~~~~~~~~~~~~~~~~~~\leq G_{g} \|\xb_i-\yb_i\|
  ~\forall \xb_i,\yb_i\in \mathcal{X}_i.
\end{align}
\else
\begin{align}
  \max_{1\leq p \leq P}\| \nabla g_{ip}(\xb_i)- \nabla  g_{ip}(\yb_i)\| &\leq G_{g} \|\xb_i-\yb_i\|\quad \forall \xb_i,\yb_i\in \mathcal{X}_i.
\end{align}
\fi
}
\end{Assumption}

We also have the following assumption on the network model \cite{Angelia2010,Ram2012}:
%
\begin{Assumption}\label{assumption network}{\rm
{The weighted graphs $\mathcal{G}(k)=\!(\mathcal{V},\mathcal{E}(k),\Wb(k))$ satisfy}:
\begin{enumerate}[(a)]
\item  There exists a scalar $0<\eta<1$ such that $[\Wb(k)]_{ii}>\eta$
for all $i,k$ and $[\Wb(k)]_{ij}>\eta$ if $[\Wb(k)]_{ij}>0$.
\item
{$\Wb(k)$ is doubly stochastic:
$\sum_{j=1}^N[\Wb(k)]_{ij}=1$ for all $i,k$ and $\sum_{i=1}^N[\Wb(k)]_{ij}=1$  $\forall j,k$.}
\item
There is an integer $Q$ such that $(\mathcal{V}, \cup_{\ell=1,\ldots,Q} \mathcal{E}(k+\ell))$
is strongly connected for all $k$.
\end{enumerate}
}
\end{Assumption}
Assumption \ref{assumption network} ensures that all the agents can sufficiently and equally {influence} each other in a long run.

\subsection{Main Convergence Results}

Let $A_k=\sum_{\ell=1}^k a_{\ell}$, and let
\begin{align}\label{running average}
  \hat{\xb}_i^{(k-1)}=\frac{1}{A_{k}}\sum_{\ell=1}^{k} a_\ell~{\xb}_i^{(\ell-1)},~ i=1,\ldots,N,
\end{align}
be the {running weighted-averages} of the primal iterates {$\xb_i^{(0)},\ldots,\xb_i^{(k-1)}$
generated by agent $i$ until time $k-1$}.
Our main convergence result for Algorithm~1 is given {in the following theorem:}

\begin{Theorem}\label{thm convergence}
Let {\blue Assumptions \ref{assumption function}-\ref{assumption network} hold},
and let $\rho_1\leq 1/(G_{\bar \Fc} + D_\lambda \sqrt{P} G_g)$. Assume that the step size
 {sequence $\{a_k\}$ is non-increasing and such that
 $a_k >0$ for all $k\ge 1$,} $\sum_{k=1}^\infty a_k =\infty$ and
 $\sum_{k=1}^\infty a_k^2 < \infty$.
 {\blue Let the sequences $\{\hat\xb^{(k)}\}$ and  $\{\lambdab_i^{(k)}\}$, $i=1,\ldots,N$, be generated
 by Algorithm~1
 using the gradient perturbation points in \eqref{local perturbation}. Then,
 $\{\hat{\xb}^{(k)}\}$ and $\{\lambdab_i^{(k)}\}$, $i=1,\ldots,N$,
 converge to an optimal primal solution $\xb^\star\in \Xc$ and an optimal dual
solution $\lambdab^\star$
 of problem~\eqref{problem}, respectively. }
 \end{Theorem}\vspace{-0.3cm}
Theorem \ref{thm convergence} indicates that the proposed distributed primal-dual algorithm asymptotically yields an optimal primal and dual solution pair for the original problem \eqref{problem}.
The same convergence result holds if the constraint functions $g_{ip}$, $p=1,\ldots,P,$ are non-smooth and the perturbation points $\alphab_i^{(k)}$ are computed {\blue according to} \eqref{eq: local proximal perturbation point}.

\begin{Theorem}\label{thm convergence proximal}
 Let  Assumptions \ref{assumption function}, \ref{assumption function 2}, and \ref{assumption network} hold, and let $\rho_1\leq 1/G_{\bar \Fc}$. Assume that the step size
 sequence $\{a_k\}$ is non-increasing and such that
 $a_k >0$ for all $k\ge 1$, $\sum_{k=1}^\infty a_k =\infty$ and
 $\sum_{k=1}^\infty a_k^2 < \infty$.
 {Let the sequences $\{\hat\xb^{(k)}\}$ and  $\{\lambdab_i^{(k)}\}$, $i=1,\ldots,N$, be generated
 by Algorithm~1 using the perturbation points in \eqref{eq: local proximal perturbation point} and \eqref{local perturbation lambda}.
 Then,
 $\{\hat{\xb}^{(k)}\}$ and $\{\lambdab_i^{(k)}\}$, $i=1,\ldots,N$,
 converge to an optimal primal solution $\xb^\star\in \Xc$ and an optimal }dual
solution $\lambdab^\star$
 of problem~\eqref{problem}, respectively.
 \end{Theorem}

%

{\blue The proofs of Theorems \ref{thm convergence} and \ref{thm convergence proximal} are presented in the next two subsections, respectively.}

\subsection{\blue Proof of Theorem \ref{thm convergence}}

{\blue In this subsection, we present the major steps for proving Theorem \ref{thm convergence}.}
Three key lemmas that will be used in the proof are presented first. The first is a deterministic version of the lemma in \cite[Lemma 11, Chapter 2.2]{BK:Polyak1987}:
\begin{Lemma}\label{lemma supermargingale}
Let $\{b_k\}$, $\{d_k\}$ and $\{c_k\}$ be non-negative sequences. Suppose that
$\sum_{{k=1}}^\infty c_k < \infty$ and
$$
b_{k} \leq b_{k-1} -d_{k-1} +c_{k-1}\qquad {\forall~k\ge1},
$$ then the sequence $\{b_k\}$ converges and $\sum_{{k=1}}^\infty d_k < \infty$.
\end{Lemma}

Moreover, by extending the results in \cite[Theorem 4.2]{Ram2012} and \cite[Lemma 8(a)]{Angelia2010},  we
establish the following result on the consensus of $\{\lambdab_i^{(k)}\}$,  $\{\yb_i^{(k)}\}$, and $\{\zb_i^{(k)}\}$ among agents.
\begin{Lemma}\label{lemma: consensus2}
Suppose that Assumptions \ref{assumption function} and \ref{assumption network} hold. If $\{a_k\}$ is a {positive}, non-increasing sequence satisfying {$\sum_{k=1}^\infty a_k^2 < \infty$}, then
\ifconfver
\begin{align}\label{summable lambda}
\!\!\!\!\!  &\sum_{k=1}^\infty a_k \|\lambdab_{i}^{(k)} - \hat{\lambdab}^{(k)}\| < \infty,~
    \lim_{k\rightarrow \infty} \|\lambdab_{i}^{(k)} -  \hat{\lambdab}^{(k)}\| =0,    \\
   &\sum_{k=1}^\infty a_k \|\tilde\lambdab_{i}^{(k)} - \hat{\lambdab}^{(k-1)}\| < \infty,~
    \lim_{k\rightarrow \infty} \|\tilde \lambdab_{i}^{(k)} -  \hat{\lambdab}^{(k-1)}\| =0,
    \label{summable tilde lambda}
    \\
   &\sum_{k=1}^\infty a_k \|\tilde\yb_{i}^{(k)} - \hat{\yb}^{(k-1)}\| < \infty,~
    \lim_{k\rightarrow \infty} \|\tilde\yb_{i}^{(k)} - \hat{\yb}^{(k-1)}\| =0,
    \label{summable tilde y}
\\
   &\sum_{k=1}^\infty a_k \|\tilde\zb_{i}^{(k)} - \hat{\zb}^{(k-1)}\| < \infty,~
    \lim_{k\rightarrow \infty} \|\tilde\zb_{i}^{(k)} - \hat{\zb}^{(k-1)}\| =0,
    \label{summable tilde z}
  \end{align}
\else
  \begin{align}\label{summable lambda}
  &\sum_{k=1}^\infty a_k \|\lambdab_{i}^{(k)} - \hat{\lambdab}^{(k)}\| < \infty,
   \qquad
    \lim_{k\rightarrow \infty} \|\lambdab_{i}^{(k)} -  \hat{\lambdab}^{(k)}\| =0,
    \\
   &\sum_{k=1}^\infty a_k \|\tilde\lambdab_{i}^{(k)} - \hat{\lambdab}^{(k-1)}\| < \infty,\qquad
    \lim_{k\rightarrow \infty} \|\tilde \lambdab_{i}^{(k)} -  \hat{\lambdab}^{(k-1)}\| =0,
    \label{summable tilde lambda}
    \\
   &\sum_{k=1}^\infty a_k \|\tilde\yb_{i}^{(k)} - \hat{\yb}^{(k-1)}\| < \infty,\qquad
    \lim_{k\rightarrow \infty} \|\tilde\yb_{i}^{(k)} - \hat{\yb}^{(k-1)}\| =0,\label{summable tilde y}
\\
   &\sum_{k=1}^\infty a_k \|\tilde\zb_{i}^{(k)} - \hat{\zb}^{(k-1)}\| < \infty,\qquad
    \lim_{k\rightarrow \infty} \|\tilde\zb_{i}^{(k)} - \hat{\zb}^{(k-1)}\| =0,\label{summable tilde z}
  \end{align}
\fi
  for all $i=1,\ldots,N$, where
\ifconfver
\begin{align}\label{hat hat hat}
&\hat{\yb}^{(k)}=\frac{1}{N} \sum_{i=1}^N \fb_i(\xb_i^{(k)}),~~\hat{\zb}^{(k)}=\frac{1}{N} \sum_{i=1}^N \gb_i(\xb_i^{(k)}),\notag\\
&\hat{\lambdab}^{(k)}=\frac{1}{N}\sum_{i=1}^N \lambdab_i^{(k)}.
\end{align}
\else
\begin{align}\label{hat hat hat}
\hat{\yb}^{(k)}=\frac{1}{N} \sum_{i=1}^N \fb_i(\xb_i^{(k)}),~~\hat{\zb}^{(k)}=\frac{1}{N} \sum_{i=1}^N \gb_i(\xb_i^{(k)}),~~\hat{\lambdab}^{(k)}=\frac{1}{N}\sum_{i=1}^N \lambdab_i^{(k)}.
\end{align}
\fi
\end{Lemma}
{\blue The proof is 
omitted here due to the space limitation; interested readers may refer to the electronic companion \cite{ChangTAC2013_companion}.} Lemma \ref{lemma: consensus2} implies that the local variables $\lambdab_i^{(k)} $, $\yb_i^{(k)} $ and $\zb_i^{(k)} $ at distributed agents will eventually achieve consensus on the values of $\hat\lambdab^{(k)} $ $\hat{\yb}^{(k)}$ and $\hat{\zb}^{(k)}$, respectively.

{\blue The local perturbation points $\alphab_i^{(k)}$ and $\betab_i^{(k)}$ in \eqref{local perturbation} and \eqref{eq: local proximal perturbation point} will also achieve consensus asymptotically.} In particular, following \eqref{eq: gradient perturbation}, we define
\ifconfver
\begin{subequations}\label{local perturbation hat}
\begin{align}\label{local perturbation hat x}
              \hat{\alphab}_i^{(k)}&=\mathcal{P}_{\mathcal{X}_i}
              (\xb_i^{(k-1)} - \rho_1 [\nabla \fb_i^T(\xb_i^{(k-1)}) \nabla \Fc(N \hat \yb^{(k)} )
                 \notag \\
                 &~~~~~~~~~~~~~~~~~~+ \nabla  \gb^T_i(\xb_i^{(k-1)}) \hat\lambdab^{(k-1)} ]),\\
              \hat{\betab}^{(k)}&=\mathcal{P}_{\Dc}(\hat\lambdab^{(k-1)}
                         +\rho_2 ~ N \hat \zb^{(k)}).
              \label{local perturbation hat lambda}
\end{align}\end{subequations}
\else
\begin{subequations}\label{local perturbation hat}
\begin{align}\label{local perturbation hat x}
              \hat{\alphab}_i^{(k)}&=\mathcal{P}_{\mathcal{X}_i}
              (\xb_i^{(k-1)} - \rho_1 [\nabla \fb_i^T(\xb_i^{(k-1)}) \nabla \Fc(N \hat \yb^{(k)} )
                 + \nabla  \gb^T_i(\xb_i^{(k-1)}) \hat\lambdab^{(k-1)} ]),\\
              \hat{\betab}^{(k)}&=\mathcal{P}_{\Dc}(\hat\lambdab^{(k-1)}
                         +\rho_2 ~ N \hat \zb^{(k)}),
              \label{local perturbation hat lambda}
\end{align}\end{subequations}
\fi
for $i=1,\ldots,N,$ as the `centralized' counterparts of \eqref{local perturbation}; similarly, following \eqref{eq: proximal perturbation point}, we define
\ifconfver
\begin{align}\label{eq: local proximal hat alpha}
    \hat\alphab_i^{(k)}=
    &\arg\min_{\alphab_i \in \Xc_i} \gb_i^T(\alphab_i)\hat \lambdab^{(k-1)}
    +\frac{1}{2\rho_1}\|\alphab_i-
    \notag \\
    &(\xb_i^{(k-1)} -\rho_1 \nabla \fb_i^T(\xb_i^{(k-1)})\nabla \Fc(N\hat\yb^{(k-1)}))\|^2,
 \end{align}
\else
\begin{align}\label{eq: local proximal hat alpha}
    \hat\alphab_i^{(k)}=
    &\arg\min_{\alphab_i \in \Xc_i} \gb_i^T(\alphab_i)\hat \lambdab^{(k-1)}
    +\frac{1}{2\rho_1}\|\alphab_i -(\xb_i^{(k-1)} -\rho_1 \nabla \fb_i^T(\xb_i^{(k-1)})\nabla \Fc(N\hat\yb^{(k-1)}))\|^2,
 \end{align}\fi
for $i=1,\ldots,N,$ as the centralized counterparts of the proximal perturbation point in \eqref{eq: local proximal perturbation point}.
We show in Appendix \ref{Appendix: proof of alpha beta consensus} the following lemma:
\begin{Lemma}\label{lemma: alpha beta consensus} Let Assumptions \ref{assumption function} and \ref{assumption function 2} hold.
   For $\{\alphab_i^{(k)}, \betab_i^{(k)}\}_{i=1}^N$ in \eqref{local perturbation} and $(\hat\alphab_1^{(k)}, \ldots, \hat\alphab_N^{(k)}, \hat\betab^{(k)})$ in \eqref{local perturbation hat}, it holds that
   \ifconfver
   \begin{align}\label{eq: alpha alpha consensus}
       \|\hat\alphab_i^{(k)} -\alphab_i^{(k)}\| &\leq  \rho_1 L_g \sqrt{P}    \|\tilde\lambdab_i^{(k)} -\hat\lambdab^{(k-1)}\| \notag\\
       &~+  \rho_1 G_\Fc L_f \sqrt{M}N \|\tilde\yb_i^{(k)} -\hat\yb^{(k-1)}\|, \\
       \|\hat\betab^{(k)} -\betab_i^{(k)}\| &\leq  \|\tilde\lambdab_i^{(k)} -\hat\lambdab^{(k-1)}\|\notag \\
       &~~~~~+ \rho_2 N \|\tilde\zb_i^{(k)} -\hat\zb^{(k-1)}\|,
       \label{eq: beta beta consensus}
    \end{align}
   \else
   \begin{align}\label{eq: alpha alpha consensus}
       \|\hat\alphab_i^{(k)} -\alphab_i^{(k)}\| &\leq  \rho_1 L_g \sqrt{P}    \|\tilde\lambdab_i^{(k)} -\hat\lambdab^{(k-1)}\| +  \rho_1 G_\Fc L_f \sqrt{M}N \|\tilde\yb_i^{(k)} -\hat\yb^{(k-1)}\|, \\
       \|\hat\betab^{(k)} -\betab_i^{(k)}\| &\leq  \|\tilde\lambdab_i^{(k)} -\hat\lambdab^{(k-1)}\|+ \rho_2 N \|\tilde\zb_i^{(k)} -\hat\zb^{(k-1)}\|,
       \label{eq: beta beta consensus}
    \end{align}\fi
  $i=1,\ldots,N.$ Equation \eqref{eq: alpha alpha consensus} also holds for the proximal perturbation point $\alphab_i^{(k)}$ in \eqref{eq: local proximal perturbation point}  and $\hat\alphab_i^{(k)}$ in \eqref{eq: local proximal hat alpha}.
\end{Lemma}
Lemma \ref{lemma: alpha beta consensus} says that, when $\tilde\lambdab_i^{(k)} $, $\tilde\yb_i^{(k)} $ and $\tilde\zb_i^{(k)} $ at distributed agents achieve consensus, each $\alphab_i^{(k)}$ converges to $\hat\alphab_i^{(k)}$, and all the $\betab_i^{(k)}$ converge to the common point $\hat\betab^{(k)}$.

{\blue Now we are ready to prove Theorem \ref{thm convergence}. The proof primarily consists of showing two facts:
(a) the primal-dual iterate
pairs $(\hat{\xb}_1^{(k)},\ldots,\hat{\xb}_N^{(k)},\hat{\lambdab}^{(k)})$ will converge
to a saddle point of \eqref{saddle point problem}, and (b) $(\hat{\xb}_1^{(k)},\ldots,\hat{\xb}_N^{(k)},\hat{\lambdab}^{(k)})$ asymptotically satisfies the primal-dual optimality conditions in Proposition~\ref{prop: opt conditions}. Thus, $(\hat{\xb}_1^{(k)},\ldots,\hat{\xb}_N^{(k)},\hat{\lambdab}^{(k)})$  is asymptotically primal-dual optimal to problem \eqref{problem}. To show the first fact, we use \eqref{local update x}, \eqref{local update lambda} and Lemma \ref{lemma: alpha beta consensus} to characterize the basic relations of the primal and dual iterates.}

\begin{Lemma}\label{lemma: basic iterates}
Let Assumptions \ref{assumption function} and \ref{assumption function 2} hold. Then, for any
$\xb=(\xb_1^T,\ldots,\xb_N^T)^T\in \Xc$ and $\lambdab\in \Dc$, the following two inequalities are true:
\ifconfver
\begin{align}\label{eq: basic iterate primal}
\!\!\!\!\!\! &\|\xb^{(k)}-{\xb}\|^2 \leq
  \|\xb^{(k-1)}-{\xb}\|^2
  -2a_k \bigg(\Lc(\xb^{(k-1)},\hat{\betab}^{(k)}) \notag \\
  &-\Lc(\xb,\hat{\betab}^{(k)})\bigg)
    \!+\! a_k^2 N (\sqrt{M}L_fL_\Fc+D_\lambda \sqrt{P}L_g )^2\!+\!2 a_k ND_{x}
  \notag\\
    &\times \sqrt{M}L_f G_\Fc\sum_{i=1}^N \|\tilde\yb_i^{(k)} - \hat\yb^{(k-1)}\|
    +2 a_k D_{x}\sqrt{P}
    \notag \\
    &\times L_g\sum_{i=1}^N ( \|\tilde\lambdab_i^{(k)} -\hat\lambdab^{(k-1)}\| + \rho_2 N \|\tilde\zb_i^{(k)} -\hat\zb_i^{(k-1)}\|),
\end{align}
\begin{align}
&\sum_{i=1}^N\|\lambdab_i^{(k)}-\lambdab\|^2\leq \sum_{i=1}^N\|\lambdab_i^{(k-1)}-\lambdab\|^2
\notag\\
&~~~+ 2\alpha_k (\Lc(\hat\alphab^{(k)},\hat{\lambdab}^{(k-1)})-\Lc(\hat\alphab^{(k)},\lambdab))\notag \\
&~~~+ a_k^2 N C_g^2
 + 2a_k(2 \rho_1D_\lambda PL_g^2 + C_g )
   \| \tilde\lambdab_i^{(k)} - \hat{\lambdab}^{(k-1)}\| \notag\\
&  ~~~+
   4\rho_1 N D_\lambda G_\Fc \sqrt{PM}L_g L_f a_k\|\tilde\yb_i^{(k)} -\hat\yb_i^{(k-1)}\|.
  \label{eq: basic iterate dual}
\end{align}
\else
\begin{align}\label{eq: basic iterate primal}
\!\!\!\!\!\! &\|\xb^{(k)}-{\xb}\|^2 \leq
  \|\xb^{(k-1)}-{\xb}\|^2 -2a_k \bigg(\Lc(\xb^{(k-1)},\hat{\betab}^{(k)})-\Lc(\xb,\hat{\betab}^{(k)})\bigg) \notag \\
  & ~~~~~~~~~~~~~~~
    + a_k^2 N (\sqrt{M}L_fL_\Fc+D_\lambda \sqrt{P}L_g )^2 +2 a_k ND_{x}\sqrt{M}L_f G_\Fc\sum_{i=1}^N \|\tilde\yb_i^{(k)} - \hat\yb^{(k-1)}\| \notag \\
   &  ~~~~~~~~~~~~~~~~~~~~~~~~~~~~+2 a_k D_{x}\sqrt{P}L_g\sum_{i=1}^N \left( \|\tilde\lambdab_i^{(k)} -\hat\lambdab^{(k-1)}\|+ \rho_2 N \|\tilde\zb_i^{(k)} -\hat\zb_i^{(k-1)}\|\right),
   \\
&\sum_{i=1}^N\|\lambdab_i^{(k)}-\lambdab\|^2
\leq \sum_{i=1}^N\|\lambdab_i^{(k-1)}-\lambdab\|^2 + 2\alpha_k \bigg(\Lc(\hat\alphab^{(k)},\hat{\lambdab}^{(k-1)})-\Lc(\hat\alphab^{(k)},\lambdab)\bigg)+ a_k^2 N C_g^2
 \notag \\
  &~~ + 2a_k(2 \rho_1D_\lambda PL_g^2 + C_g )
   \| \tilde\lambdab_i^{(k)} - \hat{\lambdab}^{(k-1)}\| +
   4\rho_1 N D_\lambda G_\Fc \sqrt{PM}L_g L_f a_k\|\tilde\yb_i^{(k)} -\hat\yb_i^{(k-1)}\|.
  \label{eq: basic iterate dual}
\end{align}
\fi
\end{Lemma}

{\blue The detailed proof is given in the electronic companion \cite{ChangTAC2013_companion}.
The second ingredient is
a relation between the primal-dual iterates $(\xb^{(k-1)},\hat\lambdab^{(k-1)})$ and the perturbation points $(\hat\alphab^{(k)},\hat\betab^{(k)})$, as given below.}


\begin{Lemma}\label{lemma: perturbation}
Let Assumptions \ref{assumption function}, \ref{assumption function 2} and \ref{assumption diff of g} hold. For the gradient perturbation points $(\hat\alphab^{(k)},\hat\betab^{(k)})$ in \eqref{local perturbation hat}, it holds true that
\ifconfver
\begin{align}\label{eq: perturbation inequality}
 &\Lc(\xb^{(k-1)},\hat\betab^{(k)})-\Lc(\hat\alphab^{(k)},\hat\lambdab^{(k-1)})
 \notag \\
 &\geq \bigg(\frac{1}{\rho_1}- (G_{\bar \Fc} + D_\lambda \sqrt{P} G_g)\bigg)  \|\xb^{(k-1)}-\hat\alphab^{(k)}\|^2\notag \\
 &~~~~~~~~~~~~~~~~~~~~~~~~~~~~~~+
 \frac{1}{\rho_2} \|\hat{\lambdab}^{(k-1)}-\hat{\betab}^{(k)}\|^2.
\end{align}
\else
\begin{align}\label{eq: perturbation inequality}
\!\!\! \Lc(\xb^{(k-1)},\hat\betab^{(k)})-&\Lc(\hat\alphab^{(k)},\hat\lambdab^{(k-1)})
 \notag \\
 &\geq \bigg(\frac{1}{\rho_1}- (G_{\bar \Fc} + D_\lambda \sqrt{P} G_g)\bigg)  \|\xb^{(k-1)}-\hat\alphab^{(k)}\|^2+
 \frac{1}{\rho_2} \|\hat{\lambdab}^{(k-1)}-\hat{\betab}^{(k)}\|^2.
\end{align}\fi 
{Moreover, let $\rho_1\leq 1/(G_{\bar \Fc} + D_\lambda \sqrt{P} G_g)$, and suppose that
$ \Lc(\xb^{(k-1)},\hat\betab^{(k)})-\Lc(\hat\alphab^{(k)},\hat\lambdab^{(k-1)}) \rightarrow 0$ and $(\xb^{(k-1)}, \lambdab^{(k-1)})$ converges to some limit point $(\hat\xb^\star, \hat\lambdab^\star)\in \Xc \times \Dc$ as $k\rightarrow \infty$. Then $(\hat\xb^\star, \hat\lambdab^\star)$ is a saddle point of \eqref{saddle point problem}.
%
}

\end{Lemma}

The proof is presented in Appendix \ref{Appendix proof of lemma perturbation}.
{\blue Using the preceding lemmas, we show the first key fact, namely, that
$(\hat{\xb}_1^{(k)},\ldots,\hat{\xb}_N^{(k)},\hat{\lambdab}^{(k)})$ converges to a saddle point of
\eqref{saddle point problem}.}

\begin{Lemma}\label{lemma: saddle point}
Let Assumptions \ref{assumption function}-\ref{assumption network} hold, and let $\rho_1\leq 1/(G_{\bar \Fc} + D_\lambda \sqrt{P} G_g)$. Assume that the step size
{$a_k>0$} is a non-increasing sequence satisfying 
$\sum_{k=1}^\infty a_k =\infty$ and $\sum_{k=1}^\infty a_k^2 < \infty$.
Then
\ifconfver
\begin{align}\label{eq converge of x}
  &\lim_{k \rightarrow \infty}
  \|\xb_i^{(k)} - \hat\xb_i^\star\| =0,\quad i=1,\ldots,N,
  \notag
  \\
  &\lim_{k \rightarrow \infty} \| \hat{\lambdab}^{(k)} - \hat\lambdab^\star\| =0,
  \end{align}
  \begin{align}
   &\lim_{k \rightarrow \infty}
  \|\xb^{(k-1)} - \hat\alphab^{(k)}\| =0,~
   \lim_{k \rightarrow \infty} \| \hat{\lambdab}^{(k-1)} - \hat\betab^{(k)}\| =0,\label{converge to each other}
\end{align}
\else
\begin{align}\label{eq converge of x}
  &\lim_{k \rightarrow \infty}
  \|\xb_i^{(k)} - \hat\xb_i^\star\| =0,\quad i=1,\ldots,N,
  \qquad \lim_{k \rightarrow \infty} \| \hat{\lambdab}^{(k)} - \hat\lambdab^\star\| =0, \\
   &\lim_{k \rightarrow \infty}
  \|\xb^{(k-1)} - \hat\alphab^{(k)}\| =0,
  \qquad \lim_{k \rightarrow \infty} \| \hat{\lambdab}^{(k-1)} - \hat\betab^{(k)}\| =0,\label{converge to each other}
\end{align}
\fi
where $\hat\xb^\star=((\hat\xb_1^\star)^T,\ldots,(\hat\xb_N^\star)^T)^T\in\Xc$ and $\hat\lambdab^\star \in \Dc$ form a saddle point of problem \eqref{saddle point problem}.

\end{Lemma}

\emph{Proof:\ }
By the compactness of the set $\Xc$ and the continuity of the functions $\bar\Fc$ and $\gb_i$,
problem~\eqref{problem} has a solution. Due to the Slater condition, the dual problem also has a solution.
By construction of the set $\Dc$ in \eqref{eq: D set}, all dual optimal solutions are contained in the set $\Dc$.
We let $\xb^\star=((\xb_1^\star)^T,\ldots, (\xb_N^\star)^T)^T\in \Xc$ and $\lambdab^\star\in\Dc$
be an arbitrary saddle point of \eqref{saddle point problem}, and we apply Lemma \ref{lemma: basic iterates} with
$\xb=(\xb_1^T,\ldots,\xb_N^T)^T=\xb^\star$ and $\lambdab=\lambdab^\star$.
By summing \eqref{eq: basic iterate primal} and \eqref{eq: basic iterate dual},
we obtain the following inequality
\ifconfver
\begin{align}\label{eq: saddle point proof stop1}
&(\|\xb^{(k)} - \xb^\star\|^2+\sum_{i=1}^N\|\lambdab_i^{(k)}-\lambdab^\star\|^2)
\notag \\
&~~~~~~\leq (\|\xb^{(k-1)} -\xb^\star\|^2
     +\sum_{i=1}^N\|\lambdab_i^{(k-1)} -\lambdab^\star\|^2)\cr
&~~~~~~
+ \tilde{c}_k-2a_k \bigg(\Lc(\xb^{(k-1)},\hat\betab^{(k)})-\Lc(\xb^\star,\betab^{(k)})
\notag \\
&~~~~~~
-\Lc(\hat\alphab^{(k)},\hat\lambdab^{(k-1)})+\Lc(\hat\alphab^{(k)},\lambdab^\star)\bigg),
\end{align}
\else
\begin{align}\label{eq: saddle point proof stop1}
&(\|\xb^{(k)} - \xb^\star\|^2+\sum_{i=1}^N\|\lambdab_i^{(k)}-\lambdab^\star\|^2)
\leq (\|\xb^{(k-1)} -\xb^\star\|^2
     +\sum_{i=1}^N\|\lambdab_i^{(k-1)} -\lambdab^\star\|^2)\cr
&~~~~~~~~~~~
+ \tilde{c}_k-2a_k \bigg(\Lc(\xb^{(k-1)},\hat\betab^{(k)})-\Lc(\xb^\star,\betab^{(k)})
-\Lc(\hat\alphab^{(k)},\hat\lambdab^{(k-1)})+\Lc(\hat\alphab^{(k)},\lambdab^\star)\bigg),
\end{align}\fi
where
\ifconfver
\begin{align}\label{eq: saddle point proof stop2}
&\tilde{c}_k\triangleq a_k^2 N [(\sqrt{M}L_fL_\Fc+D_\lambda \sqrt{P}L_g )^2
+C_g^2]
     \notag
       \end{align}
 \begin{align}
 &+ 2 [D_{x}\sqrt{P}L_g +  C_g+ 2 \rho_1 P D_\lambda L_g^2] \sum_{i=1}^N (a_k \|\tilde\lambdab_i^{(k)}-\hat\lambdab^{(k-1)}\|)
 \notag
 \\
 &+2N  \sqrt{M}L_f G_\Fc (D_{x}+2\rho_1D_\lambda\sqrt{P}L_g) \sum_{i=1}^N ( a_k \|\tilde \yb_i^{(k)}-\hat\yb^{(k-1)}\|)\notag \\
 &+2N \rho_2D_{x}\sqrt{P}L_g \sum_{i=1}^N ( a_k \|\tilde \zb_i^{(k)}-\hat\zb^{(k-1)}\|).
\end{align}
\else
\begin{align}\label{eq: saddle point proof stop2}
&\tilde{c}_k\triangleq a_k^2 N [(\sqrt{M}L_fL_\Fc+D_\lambda \sqrt{P}L_g )^2
+C_g^2]
     \notag \\
 &~~+ 2 [D_{x}\sqrt{P}L_g +  C_g+ 2 \rho_1 P D_\lambda L_g^2] \sum_{i=1}^N (a_k \|\tilde\lambdab_i^{(k)}-\hat\lambdab^{(k-1)}\|)+2N  \sqrt{M}L_f G_\Fc (D_{x}\notag\\
 &~~+2\rho_1D_\lambda\sqrt{P}L_g) \sum_{i=1}^N ( a_k \|\tilde \yb_i^{(k)}-\hat\yb^{(k-1)}\|)+2N \rho_2D_{x}\sqrt{P}L_g \sum_{i=1}^N ( a_k \|\tilde \zb_i^{(k)}-\hat\zb^{(k-1)}\|).
\end{align}\fi
First of all, by Theorem \ref{thm: saddle point thm}, we have
\ifconfver
\begin{align*}
\Lc(\hat\alphab^{(k)},\lambdab^\star) - \Lc(\xb^\star,\lambdab^\star)\geq 0,~
\Lc(\xb^\star,\lambdab^\star) - \Lc(\xb^\star,\hat\betab^{(k)})\geq 0,
\end{align*}
\else
\begin{align*}
\Lc(\hat\alphab^{(k)},\lambdab^\star) - \Lc(\xb^\star,\lambdab^\star)\geq 0,
\qquad
\Lc(\xb^\star,\lambdab^\star) - \Lc(\xb^\star,\hat\betab^{(k)})\geq 0,
\end{align*}
\fi
implying that $\Lc(\hat\alphab^{(k)},\lambdab^\star)-\Lc(\xb^\star,\betab^{(k)})\geq 0$. Hence we deduce from \eqref{eq: saddle point proof stop2} that
\ifconfver
\begin{align}\label{eq: saddle point proof stop2.5}
&(\|\xb^{(k)} - \xb^\star\|^2+\sum_{i=1}^N\|\lambdab_i^{(k)}-\lambdab^\star\|^2)
\notag \\
&\leq (\|\xb^{(k-1)} -\xb^\star\|^2
     +\sum_{i=1}^N\|\lambdab_i^{(k-1)} -\lambdab^\star\|^2)+ \tilde{c}_k\notag \\
&
-2a_k (\Lc(\xb^{(k-1)},\hat\betab^{(k)})
-\Lc(\hat\alphab^{(k)},\hat\lambdab^{(k-1)})).
\end{align}
\else
\begin{align}\label{eq: saddle point proof stop2.5}
(\|\xb^{(k)} - \xb^\star\|^2+\sum_{i=1}^N\|\lambdab_i^{(k)}-\lambdab^\star\|^2)
&\leq (\|\xb^{(k-1)} -\xb^\star\|^2
     +\sum_{i=1}^N\|\lambdab_i^{(k-1)} -\lambdab^\star\|^2)\cr
&
+ \tilde{c}_k-2a_k (\Lc(\xb^{(k-1)},\hat\betab^{(k)})
-\Lc(\hat\alphab^{(k)},\hat\lambdab^{(k-1)})).
\end{align}\fi
Secondly, by $\sum_{k=1}^\infty a_k^2 < \infty$ and
by Lemma \ref{lemma: consensus2}, we see that all the four terms in $\tilde{c}_k$ are summable over $k$, and thus $\sum_{k=1}^\infty \tilde{c}_k < \infty$.
Thirdly, by Lemma \ref{lemma: perturbation} and under the premise of $\rho_1\leq 1/(G_{\bar \Fc} + D_\lambda \sqrt{P} G_g)$, 
we have $\Lc(\xb^{(k-1)},\hat\betab^{(k)})
-\Lc(\hat\alphab^{(k)},\hat\lambdab^{(k-1)})\geq 0$.
Therefore, by applying Lemma \ref{lemma supermargingale} to
{relation~\eqref{eq: saddle point proof stop2.5},
we conclude that
the sequence $\{\|\xb^{(k)} - \xb^\star\|^2
+\sum_{i=1}^N\|\lambdab_i^{(k)} - \lambdab^\star\|^2\}$ converges for any saddle point $(\xb^\star, \lambdab^\star)$, and it holds that}
$\sum_{k=1}^\infty a_k
\left(\Lc(\xb^{(k-1)},\hat\betab^{(k)})
-\Lc(\hat\alphab^{(k)},\hat\lambdab^{(k-1)})\right)<\infty.
$ 
Because $\sum_{k=1}^\infty a_k =\infty$, the preceding relation
implies that
\begin{align}\label{eq: saddle point proof stop4}
  \liminf_{k \rightarrow \infty}
  \Lc(\xb^{(k-1)},\hat\betab^{(k)})
-\Lc(\hat\alphab^{(k)},\hat\lambdab^{(k-1)})  = 0.
\end{align}
Equation \eqref{eq: saddle point proof stop4} implies that there exists a subsequence $\ell_1,\ell_2,\ldots$ such that
\ifconfver
\begin{align}\label{eq: saddle point proof stop4p5}
\!\!\!\!\!\!  \Lc(\xb^{(\ell_k-1)},\hat\betab^{(\ell_k)})
-\Lc(\hat\alphab^{(\ell_k)},\hat\lambdab^{(\ell_k-1)})\rightarrow 0~\text{as}~k\rightarrow \infty.
\end{align}
\else
\begin{align}\label{eq: saddle point proof stop4p5}
  \Lc(\xb^{(\ell_k-1)},\hat\betab^{(\ell_k)})
-\Lc(\hat\alphab^{(\ell_k)},\hat\lambdab^{(\ell_k-1)})\rightarrow 0~\text{as}~k\rightarrow \infty.
\end{align}
\fi
According to Lemma \ref{lemma: perturbation}, the above equation indicates that
\ifconfver
\begin{align}\label{eq: saddle point proof stop5}
  {\lim_{k\to\infty}\|\xb^{(\ell_k-1)}  - \hat\alphab^{(\ell_k)}\|=0,~
 \lim_{k\to\infty}\|\hat{\lambdab}^{(\ell_k-1)} - \hat{\betab}^{(\ell_k)}\|=0.
 }
\end{align}
\else
\begin{align}\label{eq: saddle point proof stop5}
  {\lim_{k\to\infty}\|\xb^{(\ell_k-1)}  - \hat\alphab^{(\ell_k)}\|=0,\qquad
 \lim_{k\to\infty}\|\hat{\lambdab}^{(\ell_k-1)} - \hat{\betab}^{(\ell_k)}\|=0.
 }
\end{align}\fi
Moreover, because $\{(\xb^{(\ell_k-1)},\hat{\lambdab}^{(\ell_k-1)})\}\subset\Xc\times\Dc$ is a bounded sequence, there must exist a limit point, say $(\hat\xb^\star,\hat\lambdab^\star)\in\Xc\times\Dc$, such that
\begin{align}\label{eq: saddle point proof stop6}
  \xb^{(\ell_k-1)} \rightarrow \hat\xb^\star,\qquad
 \hat{\lambdab}^{(\ell_k-1)}\rightarrow \hat\lambdab^\star,~\text{as}~k\rightarrow \infty.
\end{align}
Under the premise of $\rho_1\leq 1/(G_{\bar \Fc} + D_\lambda \sqrt{P} G_g)$, and by \eqref{eq: saddle point proof stop4p5} and \eqref{eq: saddle point proof stop6}, we obtain from Lemma \ref{lemma: perturbation} that $(\hat\xb^\star,\hat\lambdab^\star)\in\Xc\times\Dc$ is a saddle point of \eqref{saddle point problem}.
Moreover, because
\ifconfver
\begin{align*}
 &\|\xb^{(\ell_k)} - \hat\xb^\star\|^2+\sum_{i=1}^N\|\lambdab_i^{(\ell_k)} - \hat\lambdab^\star\|^2 \leq \|\xb^{(\ell_k)} - \hat\xb^\star\|^2
 \notag \\
 &~~~~~~~~~~~~~+ \sum_{i=1}^N  ( \|\lambdab^{(\ell_k)}_i - \hat\lambdab^{(\ell_k)}\| + \|\hat\lambdab^{(\ell_k)}-\hat\lambdab^\star\|)^2,
\end{align*}
\else
$$
 \|\xb^{(\ell_k)} - \hat\xb^\star\|^2+\sum_{i=1}^N\|\lambdab_i^{(\ell_k)} - \hat\lambdab^\star\|^2 \leq \|\xb^{(\ell_k)} - \hat\xb^\star\|^2
 + \sum_{i=1}^N  ( \|\lambdab^{(\ell_k)}_i - \hat\lambdab^{(\ell_k)}\| + \|\hat\lambdab^{(\ell_k)}-\hat\lambdab^\star\|)^2,
$$\fi we obtain from Lemma \ref{lemma: consensus2} and \eqref{eq: saddle point proof stop6} that the sequence
$\{\|\xb^{(k)} - \hat\xb^\star\|^2+\sum_{i=1}^N\|\lambdab_i^{(k)} - \hat\lambdab^\star\|^2\}$ has a limit value equal to zero.
Since
the sequence $\{\|\xb^{(k)} - \xb^\star\|^2
+\sum_{i=1}^N\|\lambdab_i^{(k)} - \lambdab^\star\|^2\}$ converges for any saddle point of \eqref{saddle point problem},
we conclude that $\{\|\xb^{(k)} - \hat\xb^\star\|^2
+\sum_{i=1}^N\|\lambdab_i^{(k)} - \hat\lambdab^\star\|^2\}$ in fact converges to zero, and therefore \eqref{eq converge of x} is proved.
Finally, {relation} \eqref{converge to each other} can also be obtained by  \eqref{eq converge of x}, \eqref{eq: saddle point proof stop2.5} and \eqref{eq: perturbation inequality}, provided that $\rho_1\leq 1/(G_{\bar \Fc} + D_\lambda \sqrt{P} G_g)$.
\hfill $\blacksquare$

According to \cite[Lemma 3]{Larsson1999}, if $\xb^{(k)} \rightarrow \xb^\star$ as
$k \rightarrow \infty$, then its weighted running average $\xb^{(k)}$ defined in \eqref{running average}
also converges to $\xb^\star$ as $k \rightarrow \infty$. {\blue What remains is to show the second fact that
$(\hat{\xb}^{(k)},\hat{\lambdab}^{(k)})$ asymptotically
satisfies the optimality conditions given by Proposition \ref{prop: opt conditions}.
We prove in Appendix \ref{Appendix proof of asympototically feasible} that the following lemma holds.}

\begin{Lemma}\label{lemma: asympototically feasible}
Under the {assumptions} of Lemma \ref{lemma: saddle point}, it holds
\ifconfver
\begin{align}\label{eq: asymp opt conditions}
  &\lim_{k \rightarrow \infty} \bigg\|\bigg(\sum_{i=1}^N \gb_i(\hat{\xb}_i^{(k)})\bigg)^+\bigg\| =0, \notag \\
  &\lim_{k \rightarrow \infty} (\hat{\lambdab}^{(k)})^T \left(\sum_{i=1}^N \gb_i(\hat\xb_i^{(k)})\right)=0.
\end{align}
\else
\begin{align}\label{eq: asymp opt conditions}
  &\lim_{k \rightarrow \infty} \bigg\|\bigg(\sum_{i=1}^N \gb_i(\hat{\xb}_i^{(k)})\bigg)^+\bigg\| =0,
  \qquad\lim_{k \rightarrow \infty} (\hat{\lambdab}^{(k)})^T \left(\sum_{i=1}^N \gb_i(\hat\xb_i^{(k)})\right)=0.
\end{align}\fi
\end{Lemma}

By Lemma~\ref{lemma: saddle point}, Lemma~\ref{lemma: asympototically feasible} and
Proposition~\ref{prop: opt conditions}, we conclude that Theorem~\ref{thm convergence} is true.
{\blue Finally, we remark that when the step size $a_k$ has the form of $a/(b+k)$ where {$a>0,b\geq 0$}, one can simply consider the running average below \cite{Larsson1999}
\begin{align}
   \bar \xb^{(k)}= \frac{1}{k}\sum_{\ell=0}^{k-1} ~{\xb}^{(\ell)}=
   \bigg(1-\frac{1}{k}\bigg)\bar \xb^{(k-1)} + \frac{1}{k}\, {\xb}^{(k-1)},
\end{align}
instead of the running weighted-average in \eqref{running average} while Lemma~\ref{lemma: asympototically feasible} still holds true.} 

\subsection{\blue Proof of Theorem \ref{thm convergence proximal}}

Theorem \ref{thm convergence proximal} essentially can be obtained in the same line as the proof of Theorem \ref{thm convergence}, except for Lemma \ref{lemma: perturbation}. What we need to show here is that
the centralized proximal perturbation point $\hat\alphab^{(k)}$ in \eqref{eq: local proximal hat alpha} and $\hat\betab^{(k)}$ in \eqref{local perturbation hat lambda}
and
the primal-dual iterates $(\xb^{(k-1)},\lambdab^{(k-1)})$ satisfy a result similar to Lemma \ref{lemma: perturbation}. The lemma below is proved in Appendix \ref{Appendix proof of lemma perturbation proximal}:

\begin{Lemma}\label{lemma: perturbation proximal}
Let Assumptions \ref{assumption function} and \ref{assumption function 2} hold. For the centralized perturbation points $\hat\alphab^{(k)}$ in \eqref{eq: local proximal hat alpha} and $\hat\betab^{(k)}$ in \eqref{local perturbation hat lambda}, it holds true that
\ifconfver
\begin{align}\label{eq: perturbation inequality2}
 &\Lc(\xb^{(k-1)},\hat\betab^{(k)})-\Lc(\hat\alphab^{(k)},\hat\lambdab^{(k-1)})
 \geq \bigg(\frac{1}{2\rho_1}- \frac{G_{\bar \Fc}}{2}\bigg)
 \notag \\
 &\times  \|\xb^{(k-1)}-\hat\alphab^{(k)}\|^2+
 \frac{1}{\rho_2} \|\hat{\lambdab}^{(k-1)}-\hat{\betab}^{(k)}\|^2.
\end{align}
\else
\begin{align}\label{eq: perturbation inequality2}
 \Lc(\xb^{(k-1)},\hat\betab^{(k)})-&\Lc(\hat\alphab^{(k)},\hat\lambdab^{(k-1)})
 \notag \\
 &\geq \bigg(\frac{1}{2\rho_1}- \frac{G_{\bar \Fc}}{2}\bigg)  \|\xb^{(k-1)}-\hat\alphab^{(k)}\|^2+
 \frac{1}{\rho_2} \|\hat{\lambdab}^{(k-1)}-\hat{\betab}^{(k)}\|^2.
\end{align}\fi 
Moreover, {let $\rho_1\leq 1/G_{\bar\Fc}$,  and let $ \Lc(\xb^{(k-1)},\hat\betab^{(k)})-\Lc(\hat\alphab^{(k)},\hat\lambdab^{(k-1)}) \rightarrow 0$ and
$(\xb^{(k-1)}, \lambdab^{(k-1)})\to (\hat\xb^\star, \hat\lambdab^\star)$
as $k\rightarrow \infty$, where $(\hat\xb^\star, \hat\lambdab^\star)\in \Xc \times \Dc$}. Then $(\hat\xb^\star, \hat\lambdab^\star)$ is a saddle point of \eqref{saddle point problem}.
\end{Lemma}
\ifconfver
\vspace{-0.4cm}
\section{Simulation Results}\label{sec: simulation}
\else
\vspace{-0.4cm}
\section{Simulation Results}\label{sec: simulation}
\vspace{-0.0cm}
\fi
{\blue In this section, we examine the efficacy of the proposed distributed PDP method (Algorithm~1) by considering 
the DSM problem discussed in Section \ref{subsec: applications}.}
\vspace{-0.0cm}
We consider the DSM problem presented in \eqref{DSM problem} and \eqref{DSM problem 2}. The cost functions were set to $C_{\rm p}(\cdot)=\pi_{\rm p}\|\cdot\|^2$ and $C_{\rm s}(\cdot)=\pi_{\rm s}\|\cdot\|^2$, respectively, where $\pi_{\rm p}$ and $\pi_{\rm s}$ are some price parameters.
The load profile function $\psib_i(\xb_i)$ is based on the load model in \cite{ChangPES2012}, which were proposed to model deferrable, non-interruptible loads such as electrical vehicle, washing machine and tumble dryer et. al. According to \cite{ChangPES2012}, $\psib_i(\xb_i)$ can be modeled as a linear function, i.e., $\psib_i(\xb_i)=\Psib_i \xb_i$, where $\Psib_i \in \mathbb{R}^{T \times T}$ is a coefficient matrix composed of load profiles of appliances of customer $i$. The control variable $\xb_i \in \mathbb{R}^T$
determines the operation scheduling of appliances of customer $i$. Due to some physical conditions and quality of service constraints, each $\xb_i$ is subject to a local constraint set
$
 \Xc_i = \{\xb_i \in \mathbb{R}^T~|~\Ab_i \db_i \preceq \bb_i,~ {\bm l}_i \leq \db_i \leq \ub_i \}
$ where $\Ab_i \in \mathbb{R}^{T \times T}$ and ${\bm l}_i,\ub_i \in \mathbb{R}^T$ \cite{ChangPES2012}. The problem formulation corresponding to \eqref{DSM problem} is thus given by
\begin{align}\label{eq: demo}
\min_{\substack{\xb_i\in  \mathcal{X}_i,\\i=1,\ldots,N} } &\pi_{\rm p} \bigg\|\bigg(\sum_{i=1}^N \Psib_i \xb_i- \pb\bigg)^+ \bigg\|^2\!\! + \pi_{\rm s} \bigg\|\bigg(\pb-\sum_{i=1}^N \Psib_i \xb_i\bigg)^+ \bigg\|^2.
\end{align}
Analogous to \eqref{DSM problem 2}, problem \eqref{eq: demo} can be reformulated as
\begin{subequations}\label{eq: demo2}
\begin{align}
\min_{\substack{\xb_i \in \mathcal{X}_i,i=1,\ldots,N, \\\zb\succeq \zerob}} &~~\pi_{\rm p} \|\zb \|^2 + \pi_{\rm s} \bigg\|\zb -\sum_{i=1}^N \Psib_i \xb_i +\pb \bigg\|^2 \\
\text{s.t.}~& \sum_{i=1}^N \Psib_i \xb_i-\pb -\zb \preceq \zerob,\label{eq: demo2 C1}
\end{align}
\end{subequations} to which the proposed distributed PDP method can be applied.
We consider a scenario with 400 customers ($N=400$), and follow the same methods as in \cite{ChangTSG2013} to generate the power bidding {$\pb$ and coefficients $\Psib_i$, $\Ab_i$, $\bb_i$, ${\bm l}_i$, $\ub_i$, $i=1,\ldots,N$.}
The network graph $\mathcal{G}$ was randomly generated. The price parameters $\pi_{\rm p}$ and $\pi_{\rm s}$ were simply set to $1/N$ and $0.8/N$, respectively.
In addition to the distributed PD method in \cite{MZhu2012}, we also compare the proposed distributed PDP method with the distributed dual subgradient (DDS) method\footnote{One can utilize the linear structure to show that \eqref{eq: demo} is equivalent to the following saddle point problem (by Lagrange dual)
\begin{align}\label{inner LP}
  \max_{\substack{\lambdab\succeq \zerob,\\ \etab\succeq \zerob} }~ \bigg\{\min_{\substack{\xb_i\in \Xc_i\\i=1,\ldots,N }} & -\frac{1}{4\pi_{\rm p}}\|\lambdab\|^2
  -\frac{1}{4\pi_{\rm s}}\|\etab\|^2 + (\lambdab-\etab)^T(\sum_{i=1}^N \Psib_i \xb_i - \pb) \bigg\}
\end{align} to which the method in \cite{MZhu2012} and the DDS method \cite{YangJohansson2010} can be applied.} \cite{YangJohansson2010,ChangPES2012}. This method is based on the same idea as the dual decomposition technique \cite{YangJohansson2010}, where, given the dual variables, each customer globally solves the corresponding inner minimization problem. The average consensus subgradient technique \cite{Angelia2009_multiagent} is applied to the dual domain for distributed dual optimization.

Figure \ref{fig: fig_TAC_cohem}(a) shows the convergence curves of the three methods under test. The curves shown in this figure are the corresponding objective values in \eqref{eq: demo} of the running average iterates of the three methods.
The step size of the distributed PD method in \cite{MZhu2012} was set to $a_k=\frac{15}{10+k}$ and that of the DDS method was set to $a_k=\frac{0.05}{10+k}$. For the proposed distributed PDP method, $a_k$, $\rho_1$ and $\rho_2$ were respectively set to $a_k=\frac{0.1}{10+k}$ and $\rho_1=\rho_2=0.001$.
From this figure, we observe that the proposed distributed PDP method and the DDS method exhibit comparable convergence behavior; both methods converge within 100 iterations and
outperform the distributed PD method in \cite{MZhu2012}. One should note that the DDS method is computational more expensive than the proposed distributed PDP method since, {\blue in each iteration, the former requires to globally solve the inner minimization problem while the latter takes two primal gradient updates only.
For the proposed PDP Algorithm~1, the complexity order per iteration per customer is given by $\mathcal{O}(4T)$ [see \eqref{local perturbation}, \eqref{local update x} and \eqref{local update lambda}]. For the DDS method, each customer has to solve the inner linear programming (LP) in \eqref{inner LP}
$\min_{\substack{\xb_i\in \Xc_i}} (\lambdab-\etab)^T\Psib_i \xb_i $
per iteration. According to \cite{Lustig1994}, the worst-case complexity of interior point methods for solving an LP is given by $\mathcal{O}(T^{0.5}(3T^2+T^3)) \approx \mathcal{O}(T^{3.5})$.

{\blue In Figure \ref{fig: fig_TAC_cohem}(b), we display the load profiles of the power supply and the unscheduled load (without DSM), while, in Figure \ref{fig: fig_TAC_cohem}(c), we show the load profiles scheduled by the three optimization methods under consideration. The results were obtained by respectively combining each of the optimization method with the certainty equivalent control (CEC) approach in \cite[Algorithm~1]{ChangPES2012} to handle a stochastic counterpart of problem \eqref{eq: demo}.
The stopping criterion was set to the maximum iteration number of 500.
We can observe from this figure that, for all the three methods, the power balancing can be much improved compared to that without DSM control. However, we still can observe from Figure \ref{fig: fig_TAC_cohem}(c) that the proposed PDP method and the DDS method exhibit better results than the distributed PD method in \cite{MZhu2012}. Specifically, the cost in \eqref{eq: demo} is $4.49\times 10^4$ KW for the unscheduled load whereas that of the load scheduled by the proposed distributed PDP method is $2.44 \times 10^4$ KW ($45.65\%$ reduction).
The cost for the load scheduled by the distributed DDS method is slightly lower which is $2.38 \times 10^4$ KW; whereas that scheduled by the distributed PD method in \cite{MZhu2012} has a higher cost of $3.81 \times 10^4$ KW.}

As discussed in Section \ref{subsec: applications}, problem \eqref{problem} also incorporates the important regression problems. In \cite{Changglobalsip13}, we have applied the proposed PDP method to solving a distributed sparse regression problem (with a non-smooth constraint function). The simulation results can be found in \cite{Changglobalsip13}.}
\ifconfver
\begin{figure}[!t]
\begin{center}
{\subfigure[][Convergence curve]{\resizebox{.45\textwidth}{!}
{\includegraphics{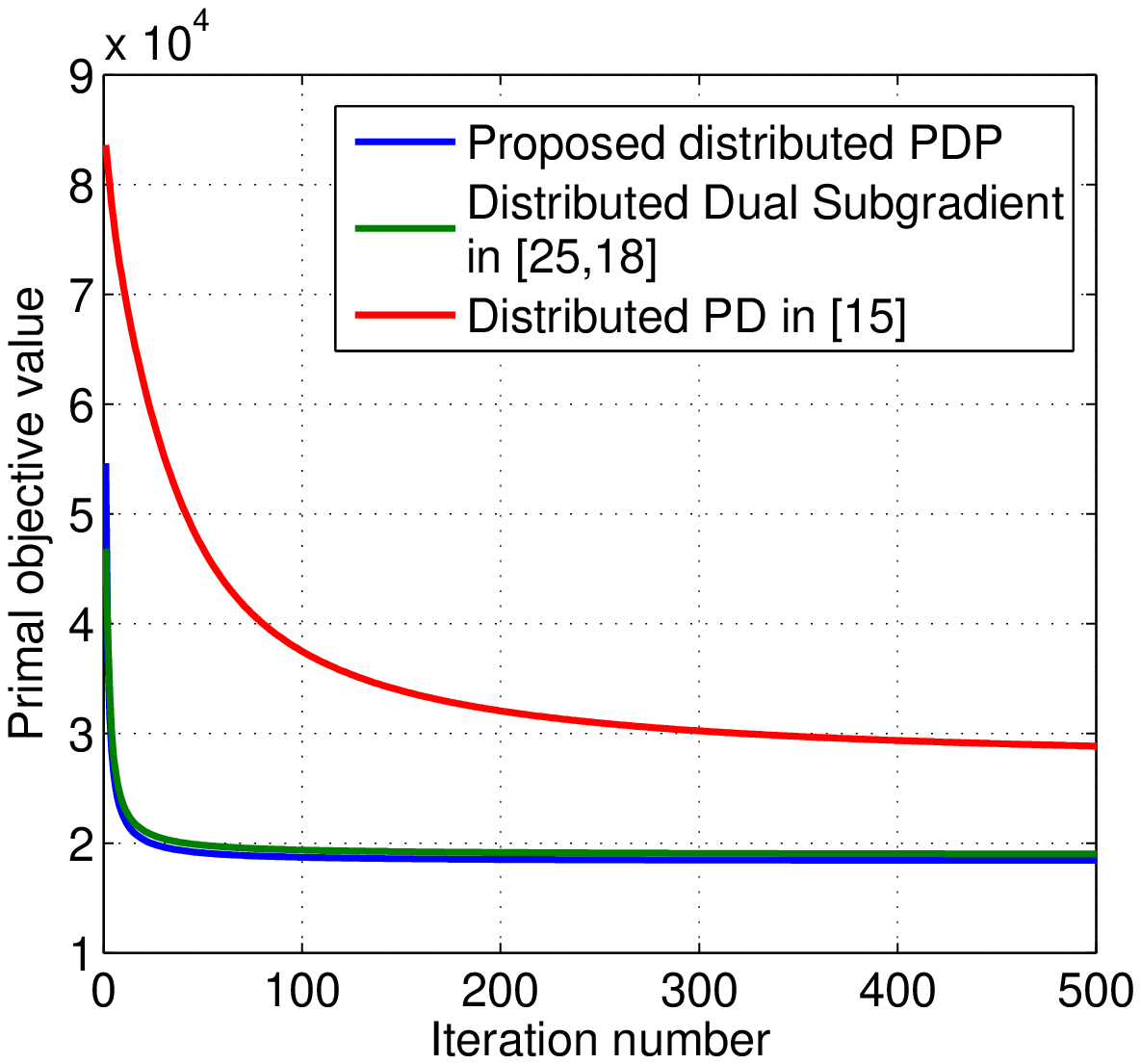}}}
}
\hspace{-0.1pc}
{\subfigure[][Unscheduled load profile]{\resizebox{.45\textwidth}{!}{\includegraphics{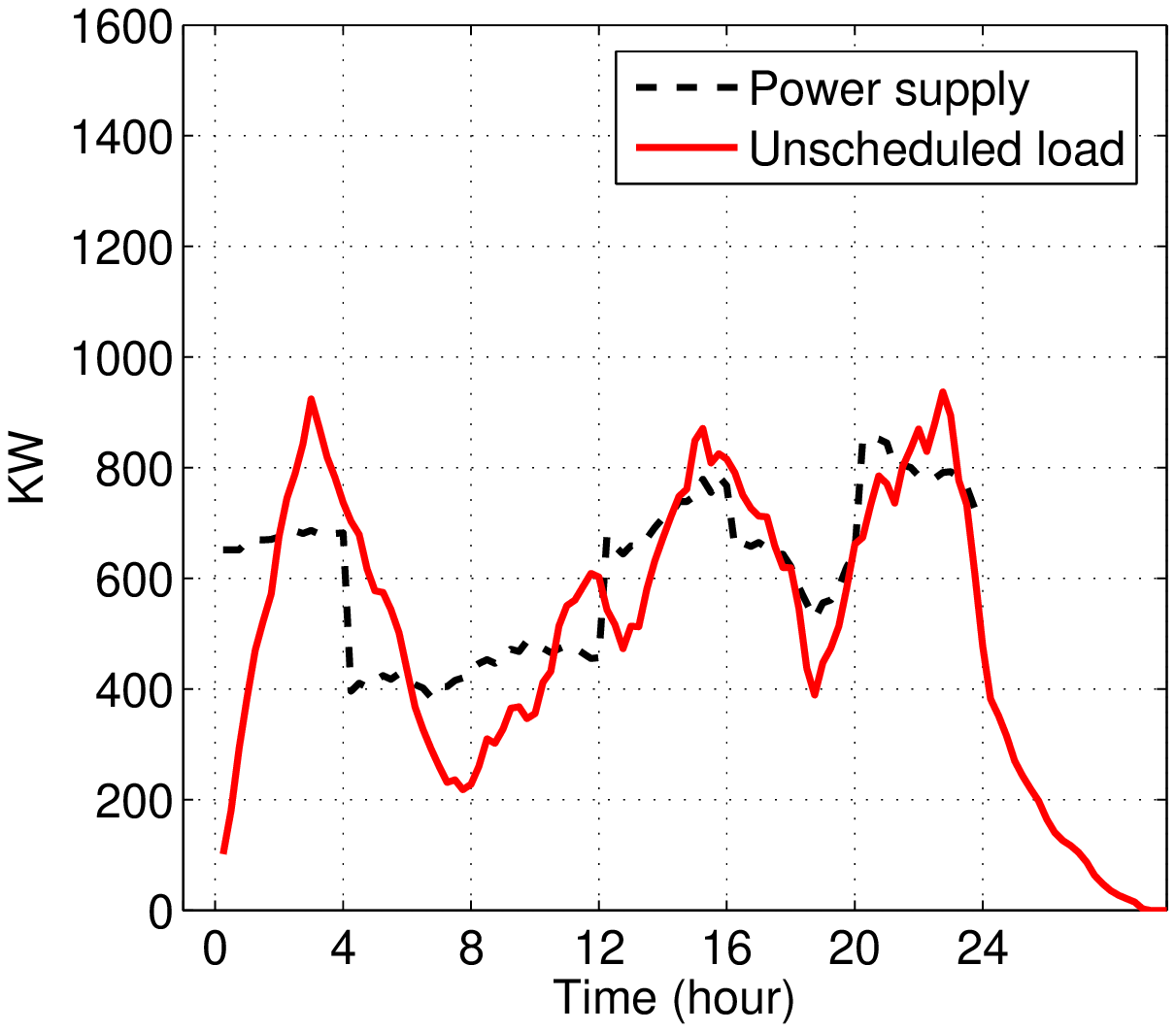}}}
 }
\hspace{-0.1pc}
{\subfigure[][Scheduled load profiles]{\resizebox{.45\textwidth}{!}{\includegraphics{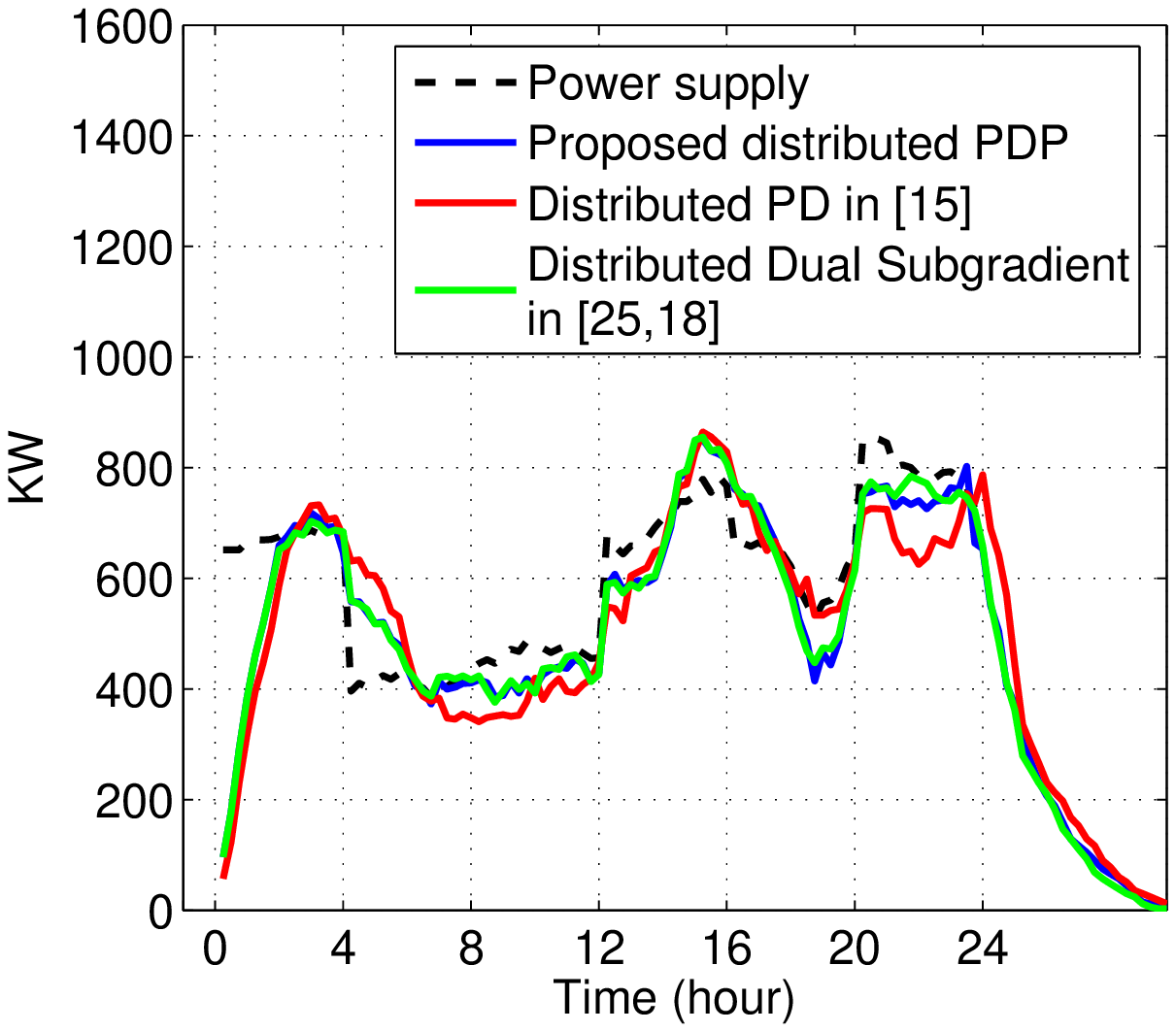}}}
 }
\end{center}\vspace{-0.0cm}
\caption{\blue Numerical results for the smart grid DSM problem \eqref{eq: demo} with $400$ customers.}
\vspace{-0.3cm}\label{fig: fig_TAC_cohem}
\end{figure}
\else
\begin{figure}[!t]
\begin{center}
{\subfigure[][Convergence curve]{\resizebox{.48\textwidth}{!}
{\includegraphics{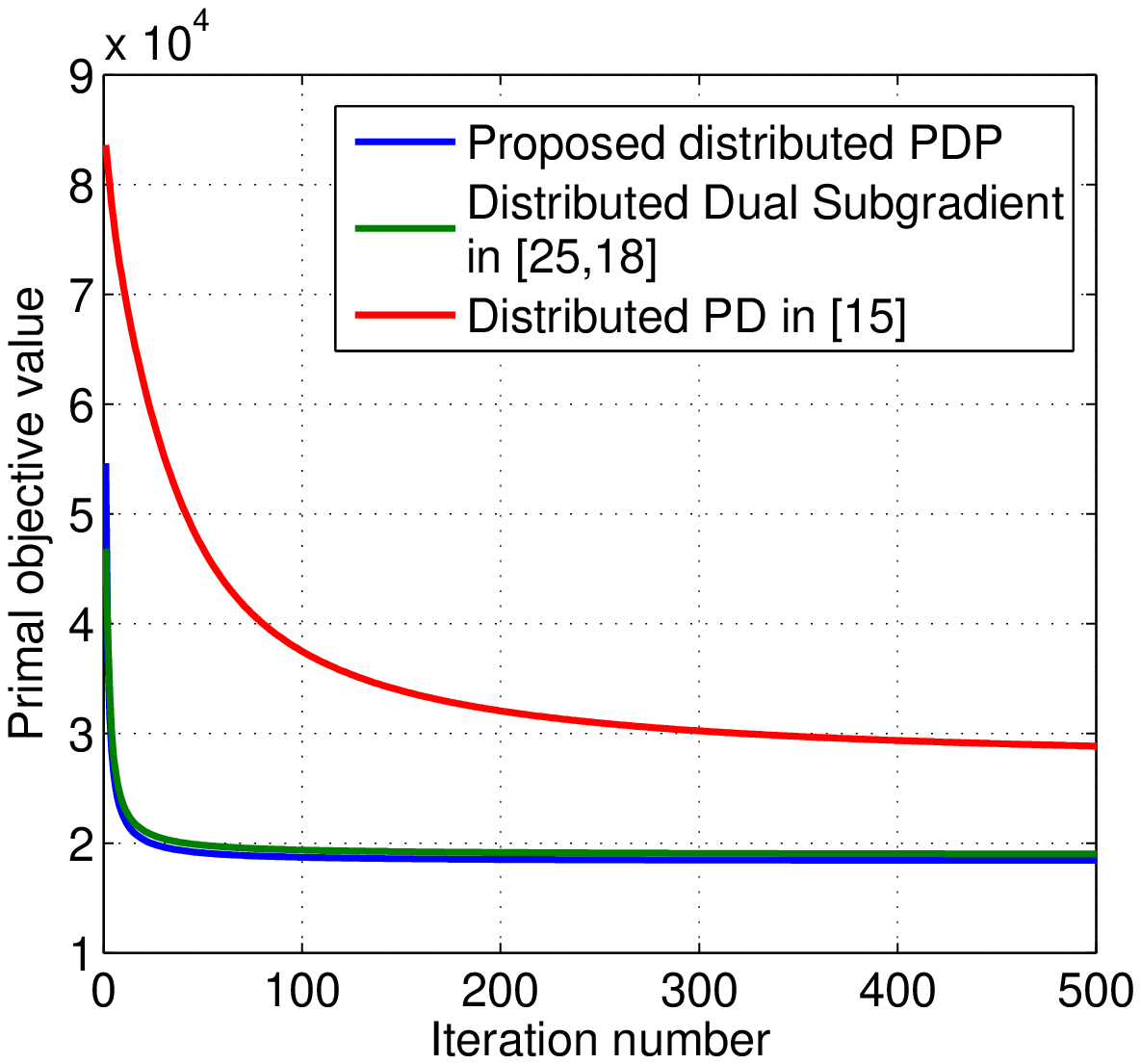}}}
}
\hspace{-0pc}
{\subfigure[][Unscheduled load profile]{\resizebox{.48\textwidth}{!}{\includegraphics{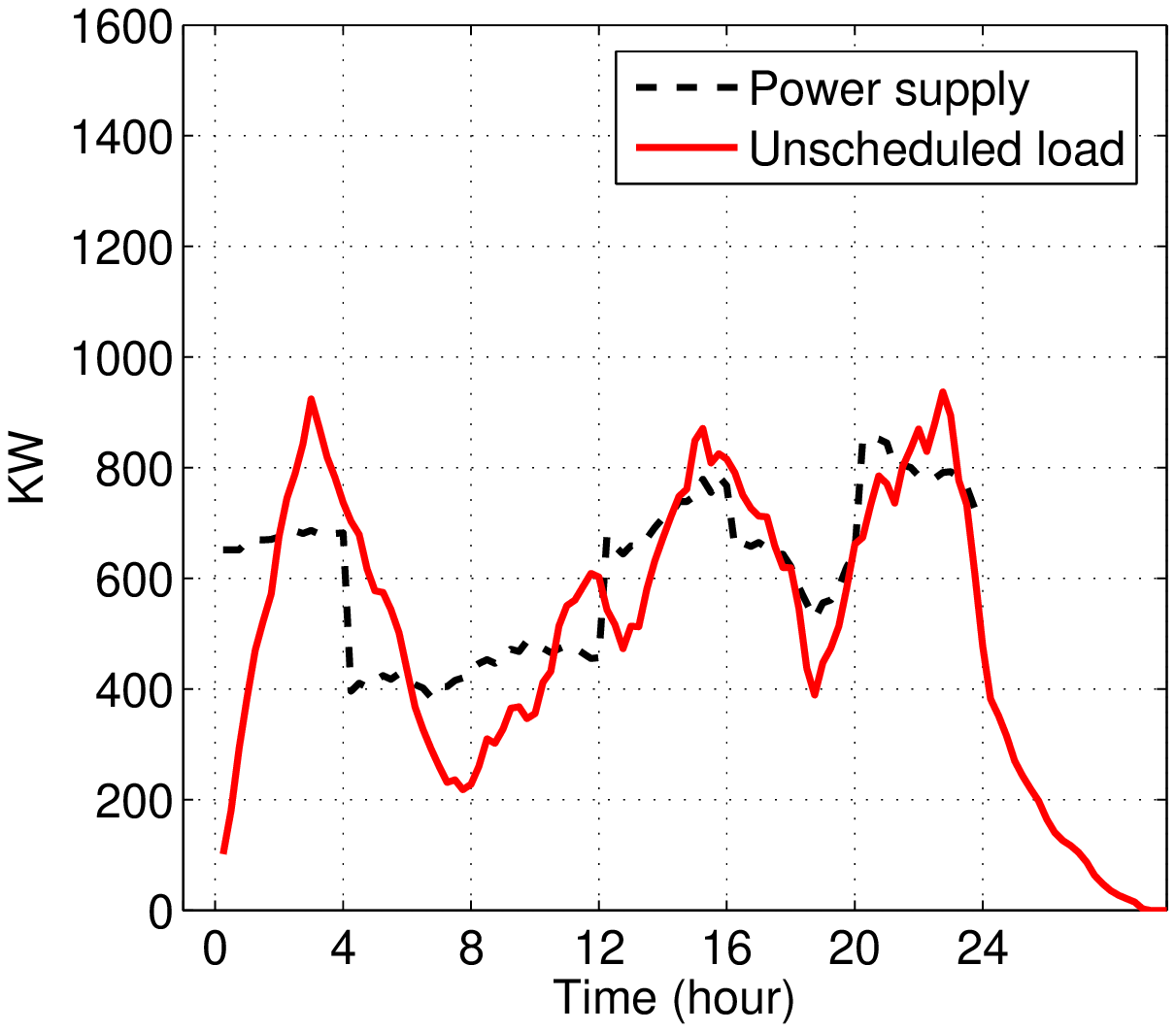}}}
 }
\hspace{-0pc}
{\subfigure[][Scheduled load profiles]{\resizebox{.5\textwidth}{!}{\includegraphics{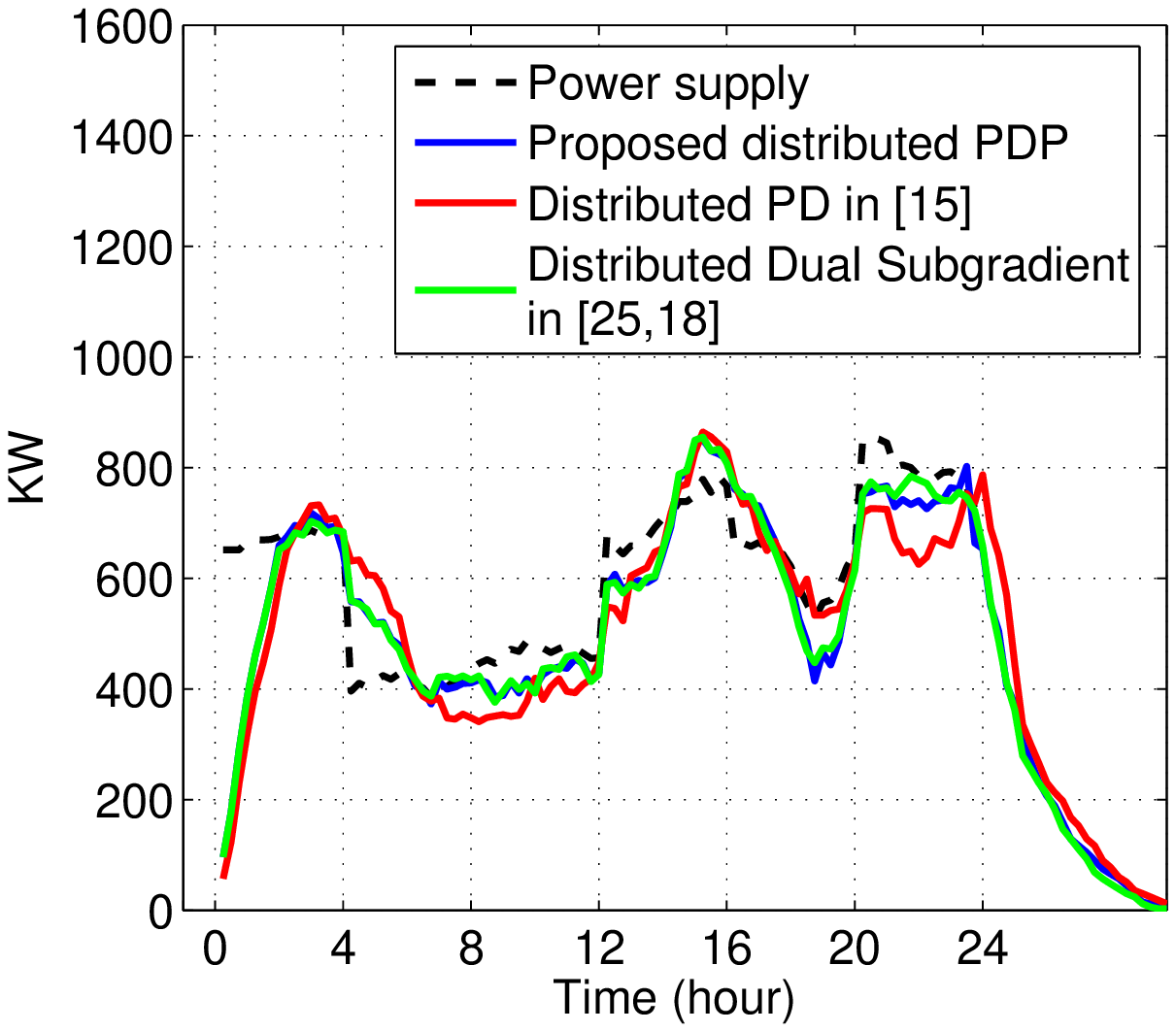}}}
 }
\end{center}\vspace{-0.5cm}
\caption{\blue Numerical results for the smart grid DSM problem \eqref{eq: demo} with $400$ customers.}
\vspace{-0.8cm}\label{fig: fig_TAC_cohem}
\end{figure}
\fi

\vspace{-0.2cm}
\section{Conclusions}\label{sec: conclusion}
{\blue We have presented a distributed consensus-based PDP algorithm for solving problem of the
form~\eqref{problem}, which has a globally coupled cost function and inequality constraints. The algorithm
employs the average consensus technique and the primal-dual perturbed
(sub-) gradient method.
We have provided a convergence analysis showing that the proposed algorithm enables the agents across the network to achieve a global optimal primal-dual solution of the considered problem in a distributed manner.
The effectiveness of the proposed algorithm has been demonstrated by applying it to a smart grid demand response control problem and a sparse linear regression problem \cite{Changglobalsip13}. In particular, the proposed algorithm is shown to have better convergence property than the distributed PD method in \cite{MZhu2012} which does not have perturbation. In addition, the proposed algorithm performs comparably with the distributed dual subgradient method \cite{YangJohansson2010} for the demand response control problem, even though the former is computationally cheaper.}


\ifconfver
\vspace{-0.4cm}\else \fi
\appendices {\setcounter{equation}{0}
\renewcommand{\theequation}{A.\arabic{equation}}

\section{Proof of Lemma \ref{lemma: alpha beta consensus}}
  \label{Appendix: proof of alpha beta consensus}

We first show \eqref{eq: beta beta consensus}. By definitions in \eqref{local perturbation hat lambda} and \eqref{local perturbation lambda}, and by the non-expansiveness of projection, we readily obtain
\ifconfver
\begin{align*}
&\|\hat\betab^{(k)} -\betab_i^{(k)}\| \notag \\
&\leq \bigg\|\mathcal{P}_{\Dc}\bigg(\tilde\lambdab_i^{(k)}
                         +\rho_2 ~ N\tilde \zb_i^{(k)}\bigg)-
                         \mathcal{P}_{\Dc}\bigg(\hat\lambdab^{(k-1)}
                         +\rho_2 ~ N\hat\zb^{(k-1)}\bigg)\bigg\| \notag \\
                         & \leq  \|\tilde\lambdab_i^{(k)} -\hat\lambdab^{(k-1)}\|+ \rho_2 N \|\tilde\zb_i^{(k)} -\hat\zb^{(k-1)}\|.
\end{align*}
\else
\begin{align*}\|\hat\betab^{(k)} -\betab_i^{(k)}\| &\leq \bigg\|\mathcal{P}_{\Dc}\bigg(\tilde\lambdab_i^{(k)}
                         +\rho_2 ~ N\tilde \zb_i^{(k)}\bigg)-
                         \mathcal{P}_{\Dc}\bigg(\hat\lambdab^{(k-1)}
                         +\rho_2 ~ N\hat\zb^{(k-1)}\bigg)\bigg\| \notag \\
                         & \leq  \|\tilde\lambdab_i^{(k)} -\hat\lambdab^{(k-1)}\|+ \rho_2 N \|\tilde\zb_i^{(k)} -\hat\zb^{(k-1)}\|.
\end{align*}\fi
Equation \eqref{eq: alpha alpha consensus} for the $\alphab_i^{(k)}$ in \eqref{local perturbation x} and  $\hat\alphab_i^{(k)}$ in \eqref{local perturbation hat x}
\ifconfver
can be shown in a similar line, as shown in \eqref{iteration lamb stop3} at the top of the next page,
\begin{figure*}[t]
\begin{align}\label{iteration lamb stop3}
 & \|\alphab_i^{(k)}-\hat{\alphab}_i^{(k)})\|
=\bigg\|\mathcal{P}_{\mathcal{X}_i}
              \bigg(\xb_i^{(k-1)} - \rho_1 \bigg[\nabla \fb_i^T(\xb_i^{(k-1)}) \nabla \Fc(N\tilde \yb_i^{(k)})
                 + \nabla  \gb^T_i(\xb_i^{(k-1)}) \tilde \lambdab_i^{(k)} \bigg]\bigg)
                 \notag \\
                 &~~~~~~~~~~~~\qquad-
              \mathcal{P}_{\mathcal{X}_i}
              \bigg(\xb_i^{(k-1)} - \rho_1 \bigg[\nabla \fb_i^T(\xb_i^{(k-1)})  \nabla \Fc(N\hat \yb^{(k-1)})
                 + \nabla  \gb^T_i(\xb_i^{(k-1)}) \hat\lambdab^{(k-1)} \bigg]\bigg)\bigg\|
                 \notag \\
              &~~~~~~\leq  \rho_1  \|\nabla \fb_i^T(\xb_i^{(k-1)})   \|  \|  \nabla \Fc(N\tilde \yb_i^{(k)})- \nabla \Fc(N\hat \yb^{(k-1)}) \|
              +\rho_1 \|\nabla\gb_i(\xb_i^{(k-1)})\|  \|\tilde \lambdab_i^{(k)}-\hat\lambdab^{(k-1)}\|       \notag       \\
              &~~~~~~ \leq  \rho_1 L_g \sqrt{P}    \|\tilde\lambdab_i^{(k)} -\hat\lambdab^{(k-1)}\| +  \rho_1 G_\Fc L_f \sqrt{M}N \|\tilde\yb_i^{(k)} -\hat\yb^{(k-1)}\|,
\end{align}\normalsize\hrulefill\vspace{-5pt}
\end{figure*}
\else
can be shown in a similar line:
\begin{align}\label{iteration lamb stop3}
 & \|\alphab_i^{(k)}-\hat{\alphab}_i^{(k)})\|
=\bigg\|\mathcal{P}_{\mathcal{X}_i}
              \bigg(\xb_i^{(k-1)} - \rho_1 \bigg[\nabla \fb_i^T(\xb_i^{(k-1)}) \nabla \Fc(N\tilde \yb_i^{(k)})
                 + \nabla  \gb^T_i(\xb_i^{(k-1)}) \tilde \lambdab_i^{(k)} \bigg]\bigg)
                 \notag \\
                 &~~~~~~~~~~~~\qquad-
              \mathcal{P}_{\mathcal{X}_i}
              \bigg(\xb_i^{(k-1)} - \rho_1 \bigg[\nabla \fb_i^T(\xb_i^{(k-1)})  \nabla \Fc(N\hat \yb^{(k-1)})
                 + \nabla  \gb^T_i(\xb_i^{(k-1)}) \hat\lambdab^{(k-1)} \bigg]\bigg)\bigg\|
                 \notag \\
              &~~~~~~\leq  \rho_1  \|\nabla \fb_i^T(\xb_i^{(k-1)})   \|  \|  \nabla \Fc(N\tilde \yb_i^{(k)})- \nabla \Fc(N\hat \yb^{(k-1)}) \|
              +\rho_1 \|\nabla\gb_i(\xb_i^{(k-1)})\|  \|\tilde \lambdab_i^{(k)}-\hat\lambdab^{(k-1)}\|       \notag       \\
              &~~~~~~ \leq  \rho_1 L_g \sqrt{P}    \|\tilde\lambdab_i^{(k)} -\hat\lambdab^{(k-1)}\| +  \rho_1 G_\Fc L_f \sqrt{M}N \|\tilde\yb_i^{(k)} -\hat\yb^{(k-1)}\|,
\end{align}\fi where, in the second inequality, we have used the boundedness of gradients (cf.  \eqref{eq: bounded subgradient f}, \eqref{eq: bounded subgradient g}) and the Lipschitz continuity of $\nabla \Fc$ (Assumption \ref{assumption function 2}).

To show that \eqref{eq: alpha alpha consensus} holds for $\alphab_i^{(k)}$ in \eqref{eq: local proximal perturbation point}  and $\hat\alphab_i^{(k)}$ in \eqref{eq: local proximal hat alpha}, we use the following lemma:

\begin{Lemma} {\rm \cite[Lemma 4.1]{BertsekasADMM}} \label{lemma 4.1} If $\yb^\star=\arg \min_{\yb \in \mathcal{Y}} J_1(\yb)+J_2(\yb)$, where $J_1: \mathbb{R}^n \rightarrow \mathbb{R} $ and
$J_2: \mathbb{R}^n \rightarrow \mathbb{R} $ are convex functions and $\mathcal{Y}$ is a closed convex set. Moreover, $J_2$ is continuously differentiable. Then
$
    \yb^\star=\arg \min_{\yb \in \mathcal{Y}} \{J_1(\yb)+\nabla J_2^T(\yb^\star)\yb \}.
$
\end{Lemma}

By applying the above lemma to \eqref{eq: local proximal perturbation point} using $J_1(\alphab_1)= \gb_i^T(\alphab_i)\tilde \lambdab_i^{(k)}$
and $$J_2(\alphab_i)=
\frac{1}{2\rho_1} \|\alphab_i - (\xb_i^{(k-1)} - \rho_1 \nabla \fb_i^T(\xb_i^{(k-1)})\nabla \Fc(N\tilde\yb_i^{(k)}))\|^2,$$ we obtain
\ifconfver
\begin{align}
    &\alphab_i^{(k)}=\arg\min_{\alphab_i \in \Xc_i} \gb_i^T(\alphab_i)\tilde \lambdab_i^{(k)} +
        ( \nabla \fb_i^T(\xb_i^{(k-1)})
        \notag \\
    &~~~\times\nabla \Fc(N\tilde\yb_i^{(k)})
        +\frac{1}{\rho_1} (\alphab_i^{(k)}-\xb_i^{(k-1)}) )^T\alphab_i.\label{eq: local proximal perturbation point transform}
 \end{align}
\else
\begin{align} \label{eq: local proximal perturbation point transform}
    &\alphab_i^{(k)}=\arg\min_{\alphab_i \in \Xc_i} \gb_i^T(\alphab_i)\tilde \lambdab_i^{(k)}+
        ( \nabla \fb_i^T(\xb_i^{(k-1)})\nabla \Fc(N\tilde\yb_i^{(k)})
        +\frac{1}{\rho_1} (\alphab_i^{(k)}-\xb_i^{(k-1)}) )^T\alphab_i.
 \end{align}\fi
Similarly, applying Lemma \ref{lemma 4.1} to  \eqref{eq: local proximal hat alpha}, we obtain
\ifconfver
\begin{align}
    &\hat\alphab_i^{(k)}=\arg\min_{\alphab_i \in \Xc_i} \gb_i^T(\alphab_i)\hat \lambdab^{(k-1)}+
        ( \nabla \fb_i^T(\xb_i^{(k-1)})
        \notag \\
        &~~~\times \nabla \Fc(N\hat\yb^{(k-1)})+\frac{1}{\rho_1} (\hat\alphab_i^{(k)}-\xb_i^{(k-1)}) )^T\alphab_i .\label{eq: local proximal perturbation point transform hat}
 \end{align}
\else
\begin{align}\label{eq: local proximal perturbation point transform hat}
    &\hat\alphab_i^{(k)}=\arg\min_{\alphab_i \in \Xc_i} \gb_i^T(\alphab_i)\hat \lambdab^{(k-1)}+
        ( \nabla \fb_i^T(\xb_i^{(k-1)})\nabla \Fc(N\hat\yb^{(k-1)})
        +\frac{1}{\rho_1} (\hat\alphab_i^{(k)}-\xb_i^{(k-1)}) )^T\alphab_i .
 \end{align}\fi
 {From \eqref{eq: local proximal perturbation point transform}
 it follows that}
\ifconfver
\begin{align}
  &\gb_i^T(\alphab_i^{(k)})\tilde \lambdab_i^{(k)}\notag \\
  &+
        ( \nabla \fb_i^T(\xb_i^{(k-1)})\nabla \Fc(N\tilde\yb_i^{(k)})
        +\frac{1}{\rho_1} (\alphab_i^{(k)}-\xb_i^{(k-1)}) )^T\alphab_i^{(k)} \notag \\
   &\leq \gb_i^T(\hat\alphab_i^{(k)})\tilde \lambdab_i^{(k)} \notag \\
   &+
        ( \nabla \fb_i^T(\xb_i^{(k-1)})\nabla \Fc(N\tilde\yb_i^{(k)})
        +\frac{1}{\rho_1} (\alphab_i^{(k)}-\xb_i^{(k-1)}) )^T\hat\alphab_i^{(k)}, \notag
\end{align}
\else
\begin{align}
  &\gb_i^T(\alphab_i^{(k)})\tilde \lambdab_i^{(k)}+
        ( \nabla \fb_i^T(\xb_i^{(k-1)})\nabla \Fc(N\tilde\yb_i^{(k)})
        +\frac{1}{\rho_1} (\alphab_i^{(k)}-\xb_i^{(k-1)}) )^T\alphab_i^{(k)} \notag \\
   &~~~\leq \gb_i^T(\hat\alphab_i^{(k)})\tilde \lambdab_i^{(k)}+
        ( \nabla \fb_i^T(\xb_i^{(k-1)})\nabla \Fc(N\tilde\yb_i^{(k)})
        +\frac{1}{\rho_1} (\alphab_i^{(k)}-\xb_i^{(k-1)}) )^T\hat\alphab_i^{(k)}, \notag
\end{align}\fi
which is equivalent to
\ifconfver
\begin{align}\label{lemma alpha beta c stop 1}
&0\leq (\gb_i^T(\hat\alphab_i^{(k)})-\gb_i^T(\alphab_i^{(k)}))\tilde \lambdab_i^{(k)}
\!+\! \nabla \fb_i^T(\xb_i^{(k-1)})\nabla \Fc(N\tilde\yb_i^{(k)})\notag \\
 \!\!&\times (\hat\alphab_i^{(k)}  \!-\! \alphab_i^{(k)}  )
\!+\!\frac{1}{\rho_1} (\alphab_i^{(k)}-\xb_i^{(k-1)})  ( \hat\alphab_i^{(k)}  - \alphab_i^{(k)} ).
\end{align}
\else
\begin{align}\label{lemma alpha beta c stop 1}
0&\leq (\gb_i^T(\hat\alphab_i^{(k)})-\gb_i^T(\alphab_i^{(k)}))\tilde \lambdab_i^{(k)}
\notag \\
&~~~~~~~~~~~~+ \nabla \fb_i^T(\xb_i^{(k-1)})\nabla \Fc(N\tilde\yb_i^{(k)}) (\hat\alphab_i^{(k)}  - \alphab_i^{(k)}  )
+\frac{1}{\rho_1} (\alphab_i^{(k)}-\xb_i^{(k-1)})  ( \hat\alphab_i^{(k)}  - \alphab_i^{(k)} ).
\end{align}\fi
Similarly, equation \eqref{eq: local proximal perturbation point transform hat} implies that
\ifconfver
\begin{align}\label{lemma alpha beta c stop 2}
0&\leq (\gb_i^T(\alphab_i^{(k)})-\gb_i^T(\hat\alphab_i^{(k)}))\hat \lambdab^{(k-1)}
\notag \\
&~~~~~~~+ \nabla \fb_i^T(\xb_i^{(k-1)})\nabla \Fc(N\hat \yb^{(k-1)}) (\alphab_i^{(k)}  - \hat\alphab_i^{(k)}  )
\notag \\
&~~~~~~~+\frac{1}{\rho_1} (\hat\alphab_i^{(k)}-\xb_i^{(k-1)})  (\alphab_i^{(k)}  -  \hat\alphab_i^{(k)} ).
\end{align}
\else
\begin{align}\label{lemma alpha beta c stop 2}
0&\leq (\gb_i^T(\alphab_i^{(k)})-\gb_i^T(\hat\alphab_i^{(k)}))\hat \lambdab^{(k-1)}
\notag \\
&~~~~~~~~+ \nabla \fb_i^T(\xb_i^{(k-1)})\nabla \Fc(N\hat \yb^{(k-1)}) (\alphab_i^{(k)}  - \hat\alphab_i^{(k)}  )
+\frac{1}{\rho_1} (\hat\alphab_i^{(k)}-\xb_i^{(k-1)})  (\alphab_i^{(k)}  -  \hat\alphab_i^{(k)} ).
\end{align}\fi
By combining \eqref{lemma alpha beta c stop 1} and \eqref{lemma alpha beta c stop 2}, we obtain
\ifconfver
\begin{align}
  &\frac{1}{\rho_1} \|\hat \alphab_i^{(k)}  -  \alphab_i^{(k)}  \|^2 \leq
  (\gb_i^T(\hat\alphab_i^{(k)})-\gb_i^T(\alphab_i^{(k)}))(\tilde \lambdab_i^{(k)}-\hat \lambdab^{(k-1)})
  \notag \\
  &+ \nabla \fb_i^T(\xb_i^{(k-1)}) (\nabla \Fc(N\tilde\yb_i^{(k)})\!-\!\! \nabla \Fc(N\hat \yb^{(k-1)})) ( \hat\alphab_i^{(k)}  - \alphab_i^{(k)} )
  \notag \\
  &\leq \big(\sqrt{P}L_g \|\tilde \lambdab_i^{(k)}-\hat \lambdab^{(k-1)}\|
   \notag \\
   &~~~~~~~~~~~~~+G_\Fc L_f \sqrt{M}N \|\tilde\yb_i^{(k)} -\hat\yb^{(k-1)}\|\big) \| \hat\alphab_i^{(k)}  - \alphab_i^{(k)}\|, \notag
\end{align}
\else
\begin{align}
  \frac{1}{\rho_1} \|\hat \alphab_i^{(k)}  -  \alphab_i^{(k)}  \|^2 &\leq
  (\gb_i^T(\hat\alphab_i^{(k)})-\gb_i^T(\alphab_i^{(k)}))(\tilde \lambdab_i^{(k)}-\hat \lambdab^{(k-1)})
  \notag \\
  &~~~~~+ \nabla \fb_i^T(\xb_i^{(k-1)}) (\nabla \Fc(N\tilde\yb_i^{(k)})- \nabla \Fc(N\hat \yb^{(k-1)})) ( \hat\alphab_i^{(k)}  - \alphab_i^{(k)} )
  \notag \\
  &\leq \left(\sqrt{P}L_g \|\tilde \lambdab_i^{(k)}-\hat \lambdab^{(k-1)}\|
  +G_\Fc L_f \sqrt{M}N \|\tilde\yb_i^{(k)} -\hat\yb^{(k-1)}\|\right) \| \hat\alphab_i^{(k)}  - \alphab_i^{(k)}\|, \notag
\end{align}\fi where we have used the boundedness of gradients (cf. \eqref{eq: bounded subgradient f}, \eqref{eq: bounded subgradient g}), the Lipschitz continuity of $\nabla \Fc$ (Assumption \ref{assumption function 2}) as well as the Lipschitz continuity of $\gb_i$ (in \eqref{eq: lipschitz g}).
{The desired result in \eqref{eq: alpha alpha consensus} follows from the preceding relation}. \hfill $\blacksquare$

\section{Proof of Lemma \ref{lemma: perturbation}} \label{Appendix proof of lemma perturbation}

{We first prove that relation~\eqref{eq: perturbation inequality} holds
for the perturbation points $\hat\alphab_i^{(k)}$ and $\hat\betab^{(k)}$ in \eqref{local perturbation hat} assuming that Assumption \ref{assumption diff of g} is satisfied.}
Note that \eqref{local perturbation hat x} is equivalent to
\ifconfver
\begin{align*}
   &\hat\alphab^{(k)}_i=\arg \min_{\alphab_i \in \mathcal{X}_i} \|
   \alphab_i- \xb_i^{(k-1)} + \rho_1 \Lc_{\xb_i}(\xb^{(k-1)},\hat\lambdab^{(k-1)})
   \|^2,
\end{align*}
$i=1,\ldots,N,$ where $\Lc_{\xb_i}(\xb^{(k-1)},\hat\lambdab^{(k-1)})=\nabla \fb_i^T(\xb_i^{(k-1)}) \\\nabla \Fc(N \hat \yb^{(k)} )
                 + \nabla  \gb^T_i(\xb_i^{(k-1)}) \hat\lambdab^{(k-1)}$.
\else
\begin{align*}
   \hat\alphab^{(k)}_i=\arg \min_{\alphab_i \in \mathcal{X}_i} \|
   \alphab_i- \xb_i^{(k-1)} + \rho_1 \Lc_{\xb_i}(\xb^{(k-1)},\hat\lambdab^{(k-1)})
   \|^2,~i=1,\ldots,N,
\end{align*}
where $\Lc_{\xb_i}(\xb^{(k-1)},\hat\lambdab^{(k-1)})=\nabla \fb_i^T(\xb_i^{(k-1)}) \nabla \Fc(N \hat \yb^{(k)} )
                 + \nabla  \gb^T_i(\xb_i^{(k-1)}) \hat\lambdab^{(k-1)}$. \fi By the optimality condition, we have that, for all $\xb_i \in \mathcal{X}_i,$
\begin{align*}
  (\xb_i - \hat\alphab^{(k)}_i)^T (\hat\alphab^{(k)}_i- \xb_i^{(k-1)} + \rho_1 \Lc_{\xb_i}(\xb^{(k-1)},\hat\lambdab^{(k-1)})) \geq 0.
\end{align*}
By choosing $\xb_i=\xb_i^{(k-1)},$ one obtains
\begin{align*}
  (\xb_i^{(k-1)} - \hat\alphab^{(k)}_i)^T \Lc_{\xb_i}(\xb^{(k-1)},\hat\lambdab^{(k-1)})
   \geq \frac{1}{\rho_1} \| \xb_i^{(k-1)}-\hat\alphab^{(k)}_i\|^2,
\end{align*}
which, by summing over $i=1,\ldots,N$, gives rise to
\begin{align*}
  (\xb^{(k-1)} - \hat\alphab^{(k)})^T \Lc_{\xb}(\xb^{(k-1)},\hat\lambdab^{(k-1)})
   \geq \frac{1}{\rho_1} \| \xb^{(k-1)}-\hat\alphab^{(k)}\|^2.
\end{align*}
Further write the above equation as follows
\ifconfver
\begin{align}\label{eq: lemma 5 stop 2}
  &(\xb^{(k-1)} - \hat\alphab^{(k)})^T \Lc_{\xb}(\hat\alphab^{(k)},\hat\lambdab^{(k-1)})
   \notag \\
  &\geq \frac{1}{\rho_1} \| \xb^{(k-1)}-\hat\alphab^{(k)}\|^2
   -(\xb^{(k-1)} - \hat\alphab^{(k)})^T \notag \\
   &~~~~~~~~~~~\times(\Lc_{\xb}(\xb^{(k-1)},\hat\lambdab^{(k-1)})-
   \Lc_{\xb}(\hat\alphab^{(k)},\hat\lambdab^{(k-1)})) \notag \\
  &\geq  \frac{1}{\rho_1} \| \xb^{(k-1)}-\hat\alphab^{(k)}\|^2
   -\|\xb^{(k-1)} - \hat\alphab^{(k)}\| \notag \\
   &~~~~~~~\times\|\Lc_{\xb}(\xb^{(k-1)},\hat\lambdab^{(k-1)})-
   \Lc_{\xb}(\hat\alphab^{(k)},\hat\lambdab^{(k-1)})\|.
\end{align}
\else
\begin{align}\label{eq: lemma 5 stop 2}
  &(\xb^{(k-1)} - \hat\alphab^{(k)})^T \Lc_{\xb}(\hat\alphab^{(k)},\hat\lambdab^{(k-1)})
   \notag \\
  &~\geq \frac{1}{\rho_1} \| \xb^{(k-1)}-\hat\alphab^{(k)}\|^2
   -(\xb^{(k-1)} - \hat\alphab^{(k)})^T(\Lc_{\xb}(\xb^{(k-1)},\hat\lambdab^{(k-1)})-
   \Lc_{\xb}(\hat\alphab^{(k)},\hat\lambdab^{(k-1)})) \notag \\
  &~\geq  \frac{1}{\rho_1} \| \xb^{(k-1)}-\hat\alphab^{(k)}\|^2
   -\|\xb^{(k-1)} - \hat\alphab^{(k)}\|\times\|\Lc_{\xb}(\xb^{(k-1)},\hat\lambdab^{(k-1)})-
   \Lc_{\xb}(\hat\alphab^{(k)},\hat\lambdab^{(k-1)})\|.
\end{align}\fi By \eqref{eq: subgradients central}, Assumption \ref{assumption function 2}, Assumption \ref{assumption diff of g} and the boundedness of $\hat\lambdab^{(k-1)}\in \mathcal{D}$, we can bound the second term in \eqref{eq: lemma 5 stop 2} as
\ifconfver
\begin{align}\label{eq: lemma 5 stop 3}
 &\|\Lc_{\xb}(\xb^{(k-1)},\hat\lambdab^{(k-1)})-
   \Lc_{\xb}(\hat\alphab^{(k)},\hat\lambdab^{(k-1)})\| \notag \\
 &\leq  \|\nabla \bar\Fc(\xb^{(k-1)})- \nabla \bar\Fc(\hat\alphab^{(k)})\|
                 \notag \\
 &+\| \hat\lambdab^{(k-1)}\|~ \bigg\|
                 \begin{bmatrix}
                 \nabla \gb^T_1(\xb_1^{(k-1)})
                 -\nabla  \gb^T_1(\hat\alphab_1^{(k)}) \\
                 \vdots \\
                 \nabla \gb^T_N(\xb_N^{(k-1)})
                 -\nabla  \gb^T_N(\hat\alphab_N^{(k)})
                 \end{bmatrix}\bigg\|_F\notag \\
 &\leq (G_{\bar \Fc} + D_\lambda \sqrt{P} G_g) \|\xb^{(k-1)}-\hat\alphab^{(k)}\|,
\end{align}
\else
\begin{align}\label{eq: lemma 5 stop 3}
 &\|\Lc_{\xb}(\xb^{(k-1)},\hat\lambdab^{(k-1)})-
   \Lc_{\xb}(\hat\alphab^{(k)},\hat\lambdab^{(k-1)})\| \notag \\
 &\leq  \|\nabla \bar\Fc(\xb^{(k-1)})- \nabla \bar\Fc(\hat\alphab^{(k)})\|
                 +\| \hat\lambdab^{(k-1)}\|~ \bigg\|
                 \begin{bmatrix}
                 \nabla \gb^T_1(\xb_1^{(k-1)})
                 -\nabla  \gb^T_1(\hat\alphab_1^{(k)}) \\
                 \vdots \\
                 \nabla \gb^T_N(\xb_N^{(k-1)})
                 -\nabla  \gb^T_N(\hat\alphab_N^{(k)})
                 \end{bmatrix}\bigg\|_F\notag \\
 &\leq (G_{\bar \Fc} + D_\lambda \sqrt{P} G_g) \|\xb^{(k-1)}-\hat\alphab^{(k)}\|,
\end{align}\fi where $\|\cdot\|_F$ denotes the Frobenious norm. By combining \eqref{eq: lemma 5 stop 2} and \eqref{eq: lemma 5 stop 3}, we obtain
\ifconfver
\begin{align}\label{eq: lemma 5 stop 4}
  &(\xb^{(k-1)} - \hat\alphab^{(k)})^T \Lc_{\xb}(\hat\alphab^{(k)},\hat\lambdab^{(k-1)})
  \notag \\
  &\geq \bigg(\frac{1}{\rho_1}- (G_{\bar \Fc} + D_\lambda \sqrt{P} G_g)\bigg)\|\xb^{(k-1)} - \hat\alphab^{(k)}\|^2.
\end{align}
\else
\begin{align}\label{eq: lemma 5 stop 4}
  &(\xb^{(k-1)} - \hat\alphab^{(k)})^T \Lc_{\xb}(\hat\alphab^{(k)},\hat\lambdab^{(k-1)})
  \geq \bigg(\frac{1}{\rho_1}- (G_{\bar \Fc} + D_\lambda \sqrt{P} G_g)\bigg)\|\xb^{(k-1)} - \hat\alphab^{(k)}\|^2.
\end{align}\fi
Since
$
  \Lc(\xb^{(k-1)},\hat\lambdab^{(k-1)})-\Lc(\hat\alphab^{(k)},\hat\lambdab^{(k-1)})
  \geq (\xb^{(k-1)} - \hat\alphab^{(k)})^T \Lc_{\xb}(\hat\alphab^{(k)},\hat\lambdab^{(k-1)})
$ by the convexity of $\Lc$ in $\xb$, we further obtain
\ifconfver
\begin{align}\label{eq: lemma 5 stop 5}
  &\Lc(\xb^{(k-1)},\hat\lambdab^{(k-1)})-\Lc(\hat\alphab^{(k)},\hat\lambdab^{(k-1)})
  \notag \\
  &\geq \bigg(\frac{1}{\rho_1}- (G_{\bar \Fc} + D_\lambda \sqrt{P} G_g)\bigg)\|\xb^{(k-1)} - \hat\alphab^{(k)}\|^2.
\end{align}
\else
\begin{align}\label{eq: lemma 5 stop 5}
  &\Lc(\xb^{(k-1)},\hat\lambdab^{(k-1)})-\Lc(\hat\alphab^{(k)},\hat\lambdab^{(k-1)})
  \geq \bigg(\frac{1}{\rho_1}- (G_{\bar \Fc} + D_\lambda \sqrt{P} G_g)\bigg)\|\xb^{(k-1)} - \hat\alphab^{(k)}\|^2.
\end{align}\fi

On the other hand, by \eqref{local perturbation hat lambda}, we know that
$
   \hat\betab^{(k)}=\arg \min_{\beta \in \mathcal{D}} \|
   \betab- \hat\lambdab^{(k-1)} - \rho_2 \sum_{i=1}^N\gb_i(\xb^{(k-1)}_i)
   \|^2.$
By the optimality condition and the linearity of $\Lc$ in $\lambdab$, we have
\ifconfver
\begin{align}\label{eq: lemma 5 stop 6}
  &\Lc(\xb^{(k-1)},\hat\betab^{(k)})-\Lc(\xb^{(k-1)},\hat\lambdab^{(k-1)})=
  -(\hat\lambdab^{(k-1)}-\hat\betab^{(k)})^T \notag \\
  &\times\left(\sum_{i=1}^N\gb_i(\xb^{(k-1)}_i)\right)
  \geq \frac{1}{\rho_2} \|\hat\lambdab^{(k-1)}-\hat\betab^{(k)}\|^2.
\end{align}
\else
\begin{align}\label{eq: lemma 5 stop 6}
  \Lc(\xb^{(k-1)},\hat\betab^{(k)})-\Lc(\xb^{(k-1)},\hat\lambdab^{(k-1)})&=
  -(\hat\lambdab^{(k-1)}-\hat\betab^{(k)})^T \left(\sum_{i=1}^N\gb_i(\xb^{(k-1)}_i)\right)
  \notag \\
  &\geq \frac{1}{\rho_2} \|\hat\lambdab^{(k-1)}-\hat\betab^{(k)}\|^2.
\end{align}\fi
Combining \eqref{eq: lemma 5 stop 5} and \eqref{eq: lemma 5 stop 6} yields \eqref{eq: perturbation inequality}.

Suppose that
 $ \Lc(\xb^{(k-1)},\hat\betab^{(k)})-\Lc(\hat\alphab^{(k)},\hat\lambdab^{(k-1)}) \rightarrow 0$ and $(\xb^{(k-1)}, \hat\lambdab^{(k-1)})$ converges to some limit point $(\hat\xb^\star, \hat\lambdab^\star)$ as $k\rightarrow \infty$. Since $\rho_1\leq 1/(G_{\bar \Fc} + D_\lambda \sqrt{P} G_g)$, we infer from \eqref{eq: perturbation inequality} that
 {$\|\xb^{(k-1)} - \hat\alphab^{(k)}\|\to0$ and $\|\hat\lambdab^{(k-1)} - \hat\betab^{(k)}\|\to0$,
 as $k\to\infty$.}
It then follows from \eqref{local perturbation hat} and the fact that projection is a continuous mapping \cite{BertsekasADMM} that $(\hat\xb^\star,\hat\lambdab^\star)\in\Xc\times\Dc$ satisfies
\ifconfver
{\small\begin{align*}
              \hat\xb^\star_i&=\mathcal{P}_{\mathcal{X}_i}
              \bigg(\hat\xb^\star_i - \rho_1 [\nabla \fb_i^T(\hat\xb^\star_i)\nabla \Fc\bigg(\sum_{i=1}^M \fb_i(\hat{\xb}_i^\star)\bigg)
                \notag \\
                &~~~~~~~~~~~~~~~~~~~~ + \nabla  \gb^T_i(\hat\xb^\star_i) \hat\lambdab^\star ]\bigg),~i=1,\ldots,N,\\
             \hat\lambdab^\star&=\mathcal{P}_{\Dc}(\hat\lambdab^\star
                       +\rho_2 ~ \sum_{i=1}^N \gb_i(\hat\xb^\star_i))
\end{align*}}
\else
\begin{align*}
              \hat\xb^\star_i&=\mathcal{P}_{\mathcal{X}_i}
              \bigg(\hat\xb^\star_i - \rho_1 [\nabla \fb_i^T(\hat\xb^\star_i)\nabla \Fc\bigg(\sum_{i=1}^M \fb_i(\hat{\xb}_i^\star)\bigg)
                 + \nabla  \gb^T_i(\hat\xb^\star_i) \hat\lambdab^\star ]\bigg),~i=1,\ldots,N,\\
             \hat\lambdab^\star&=\mathcal{P}_{\Dc}(\hat\lambdab^\star
                       +\rho_2 ~ \sum_{i=1}^N \gb_i(\hat\xb^\star_i))
\end{align*}\fi
which, respectively, imply that
$\hat\xb^\star =\arg \min_{\xb \in \Xc} \Lc(\xb,\hat\lambdab^\star)$ and
$\hat\lambdab^\star =\arg \max_{\lambdab \in \Dc} \Lc(\hat\xb^\star,\lambdab)$
i.e., $(\hat\xb^\star,\hat\lambdab^\star)$ is a saddle point of problem \eqref{saddle point problem}.
\hfill  $\blacksquare$

\section{Proof of Lemma \ref{lemma: asympototically feasible}} \label{Appendix proof of asympototically feasible}
By \eqref{eq: basic iterate dual} in Lemma \ref{lemma: basic iterates}  and the fact of
$ \Lc(\hat\alphab^{(k)},\lambdab) = \Lc(\hat\alphab^{(k)},\hat{\lambdab}^{(k-1)}) + (\lambdab-\hat{\lambdab}^{(k-1)} )^T \Lc_{\lambdab}(\hat\alphab^{(k)},\hat{\lambdab}^{(k-1)})$,
we have
\ifconfver
\begin{align}\label{eq: lemma 6 stop 1}
&(\lambdab-\hat{\lambdab}^{(k-1)})^T \gb(\hat\alphab^{(k)})
\leq  \frac{\bar c_k}{2 a_k} \notag \\
&~+ \frac{1}{2 a_k}\left(\sum_{j=1}^N\|\lambdab_j^{(k-1)}-\lambdab\|^2 - \sum_{i=1}^N\|\lambdab_i^{(k)}-\lambdab\|^2\right),
\end{align}
\else
\begin{align}\label{eq: lemma 6 stop 1}
 (\lambdab-\hat{\lambdab}^{(k-1)})^T \gb(\hat\alphab^{(k)})
\leq  \frac{\bar c_k}{2 a_k}+ \frac{1}{2 a_k}\left(\sum_{j=1}^N\|\lambdab_j^{(k-1)}-\lambdab\|^2 - \sum_{i=1}^N\|\lambdab_i^{(k)}-\lambdab\|^2\right),
\end{align}\fi
where $\gb(\hat\alphab^{(k)})=\sum_{i=1}^N \gb_i(\hat\alphab^{(k)}_i)$ and
\ifconfver
\begin{align*}
\bar c_k &\triangleq  a_k^2 N C_g^2+ 2a_k(2 \rho_1D_\lambda PL_g^2 + C_g )
   \| \tilde\lambdab_i^{(k)} - \hat{\lambdab}^{(k-1)}\| \notag \\
   &+4\rho_1 N D_\lambda G_\Fc \sqrt{PM}L_g L_f a_k\|\tilde\yb_i^{(k)} -\hat\yb_i^{(k-1)}\|.
\end{align*}
\else
\begin{align*}
\bar c_k \triangleq  a_k^2 N C_g^2+ 2a_k(2 \rho_1D_\lambda PL_g^2 + C_g )
   \| \tilde\lambdab_i^{(k)} - \hat{\lambdab}^{(k-1)}\| +
   4\rho_1 N D_\lambda G_\Fc \sqrt{PM}L_g L_f a_k\|\tilde\yb_i^{(k)} -\hat\yb_i^{(k-1)}\|.
\end{align*}\fi
%
%
\ifconfver
By following a similar argument as in \cite[Proposition 5.1]{Angelia2009_saddle} and by
\eqref{eq: lemma 6 stop 1}, \eqref{eq: D set}, \eqref{eq: lipschitz g} and  \eqref{eq: function bounded}, one can show that
\begin{align}
& (\lambdab-\hat{\lambdab}^\star)^T \gb(\xb^{(k-1)})
 \leq  \frac{\bar c_k}{2 a_k} \notag \\
&+ \frac{1}{2 a_k}\left(\sum_{j=1}^N\|\lambdab_j^{(k-1)}-\lambdab\|^2 - \sum_{i=1}^N\|\lambdab_i^{(k)}-\lambdab\|^2\right) \notag \\
&+2N\sqrt{P}D_\lambda L_g \|\xb^{(k-1)}-\hat\alphab^{(k)}\|+NC_g\|\hat{\lambdab}^{(k-1)}-\hat{\lambdab}^\star\|.
\label{eq: lemma 6 stop 2}
\end{align}
\else
{By following a similar argument as in \cite[Proposition 5.1]{Angelia2009_saddle} and by
\eqref{eq: lemma 6 stop 1}, \eqref{eq: D set}, \eqref{eq: lipschitz g} and  \eqref{eq: function bounded}, one can show that
\begin{align}
 (\lambdab-\hat{\lambdab}^\star)^T \gb(\xb^{(k-1)})
& \leq  \frac{\bar c_k}{2 a_k}+ \frac{1}{2 a_k}\left(\sum_{j=1}^N\|\lambdab_j^{(k-1)}-\lambdab\|^2 - \sum_{i=1}^N\|\lambdab_i^{(k)}-\lambdab\|^2\right) \notag \\
&~~~~~~~~~+2N\sqrt{P}D_\lambda L_g \|\xb^{(k-1)}-\hat\alphab^{(k)}\|+NC_g\|\hat{\lambdab}^{(k-1)}-\hat{\lambdab}^\star\|.
\label{eq: lemma 6 stop 2}
\end{align} }\fi%
%
%
%
%
By taking the weighted running average of \eqref{eq: lemma 6 stop 2}, we obtain
\ifconfver
\begin{align}
&(\lambdab-\hat{\lambdab}^\star)^T \gb(\hat{\xb}^{(k-1)}) \leq \frac{1}{A_k}\sum_{\ell=1}^{k} a_\ell(\lambdab-\hat{\lambdab}^\star)^T \gb(\xb^{(\ell-1)}) \notag \\
& \leq  \frac{1}{2A_k}\sum_{\ell=1}^{k} \bar c_\ell \!+
\! \frac{1}{2A_k}\! \left(\!\sum_{j=1}^N\|\lambdab_j^{(0)}\!-\!\lambdab\|^2 \!-\! \sum_{i=1}^N\!\|\lambdab_i^{(k)}\!-\!\lambdab\|^2\!\right)\!\notag
\end{align}
\begin{align}
&+ \frac{2N\sqrt{P}D_\lambda L_g}{A_k}\sum_{\ell=1}^{k} a_\ell \|\xb^{(\ell-1)}-\hat\alphab^{(\ell)}\|
\notag\\
&+\!\frac{NC_g}{A_k}\sum_{\ell=1}^{k} a_\ell \|\hat{\lambdab}^{(\ell-1)}-\hat{\lambdab}^\star\|
\notag
\\
& \leq  \frac{1}{2A_k}\sum_{\ell=1}^{k} \bar c_\ell \!+\! \frac{2ND_\lambda^2}{A_k}
\!+\!\frac{2N\sqrt{P}D_\lambda L_g}{A_k}\sum_{\ell=1}^{k} a_\ell \|\xb^{(\ell-1)}\!-\!\hat\alphab^{(\ell)}\|
 \notag
 \\
 &+\!\frac{NC_g}{A_k}\sum_{\ell=1}^k a_\ell \|\hat{\lambdab}^{(\ell-1)}-\hat{\lambdab}^\star\|
\label{eq: lemma 6 stop 3}
\triangleq \xi^{(k-1)}
\end{align}
\else
\begin{align}
&(\lambdab-\hat{\lambdab}^\star)^T \gb(\hat{\xb}^{(k-1)}) \leq \frac{1}{A_k}\sum_{\ell=1}^{k} a_\ell(\lambdab-\hat{\lambdab}^\star)^T \gb(\xb^{(\ell-1)}) \notag \\
& \leq  \frac{1}{2A_k}\sum_{\ell=1}^{k} \bar c_\ell \!+
\! \frac{1}{2A_k}\! \left(\!\sum_{j=1}^N\|\lambdab_j^{(0)}\!-\!\lambdab\|^2 \!-\! \sum_{i=1}^N\!\|\lambdab_i^{(k)}\!-\!\lambdab\|^2\!\right)\!\notag \\
&~~~~~~~~~~~~~~~~+ \frac{2N\sqrt{P}D_\lambda L_g}{A_k}\sum_{\ell=1}^{k} a_\ell \|\xb^{(\ell-1)}-\hat\alphab^{(\ell)}\|
+\!\frac{NC_g}{A_k}\sum_{\ell=1}^{k} a_\ell \|\hat{\lambdab}^{(\ell-1)}-\hat{\lambdab}^\star\| \notag
\\
& \leq  \frac{1}{2A_k}\sum_{\ell=1}^{k} \bar c_\ell \!+\! \frac{2ND_\lambda^2}{A_k}
 +\frac{2N\sqrt{P}D_\lambda L_g}{A_k}\sum_{\ell=1}^{k} a_\ell \|\xb^{(\ell-1)}-\hat\alphab^{(\ell)}\| +\!\frac{NC_g}{A_k}\sum_{\ell=1}^k a_\ell \|\hat{\lambdab}^{(\ell-1)}-\hat{\lambdab}^\star\|
\label{eq: lemma 6 stop 3} \\
&
\triangleq \xi^{(k-1)}, \notag
\end{align}\fi where the first inequality is owing to the fact that $\gb(\xb)$ is convex, and the last inequality is obtained by dropping $-\! \sum_{i=1}^N\!\|\lambdab_i^{(k)}\!-\!\lambdab\|^2$ followed by applying \eqref{eq: D set}.
{We claim} that
\begin{align}\label{xi converge zero}
 \lim_{k\rightarrow \infty}\xi^{(k-1)} =0.
\end{align}
To see this, {note that
the first and second terms in $\xi^{(k-1)}$
converge} to zero as $k \rightarrow \infty$ {since $\lim_{k\to\infty} A_k =\infty$} and
$\sum_{\ell=1}^\infty \bar c_\ell<\infty$. The term
$\frac{1}{A_k}\sum_{\ell=1}^k {a_\ell} \|\hat{\lambdab}^{(\ell-1)}-\hat{\lambdab}^\star\|$
also converges to zero since, by Lemma~\ref{lemma: saddle point},
$\lim_{k \rightarrow \infty}\|\hat{\lambdab}^{(k)}-\hat{\lambdab}^\star\|=0$
and so does its weighted running average by~\cite[Lemma 3]{Larsson1999}.
Similarly, the term $\frac{1}{A_k}\sum_{\ell=1}^k {a_\ell} \|\xb^{(\ell-1)}-\hat\alphab^{(\ell)}\|$ also converges to zero since $\lim_{k \rightarrow \infty}\|\xb^{(k-1)}-\hat\alphab^{(k)}\|=0$ due to \eqref{converge to each other}.

Now let
$\lambdab=\hat{\lambdab}^\star + \delta~\frac{\left(\gb(\hat{\xb}^{(k-1)})\right)^+}{\|\left(\gb(\hat{\xb}^{(k-1)})\right)^+\|}$
which lies in $\Dc$, since $\|\lambdab\| \leq \|\hat{\lambdab}^\star\| +\delta
\leq D_\lambda$ by \eqref{eq: opt dual}. Substituting 
$\lambdab$ into \eqref{eq: lemma 6 stop 3} gives rise to
\begin{align}
\delta \|\left(\gb(\hat{\xb}^{(k-1)})\right)^+\|
& \leq  \xi^{(k-1)}.
\label{eq: lemma 6 stop 4}
\end{align}
As a result, the first term in \eqref{eq: asymp opt conditions} is obtained by taking
$k \rightarrow \infty$ in~\eqref{eq: lemma 6 stop 4} and by \eqref{xi converge zero}.

To show that the second limit in \eqref{eq: asymp opt conditions} holds true, we first let
$\lambdab=\hat{\lambdab}^\star + \delta~\frac{\hat{\lambdab}^{(k-1)}}{\|\hat{\lambdab}^{(k-1)}\|} \in \Dc.$
By substituting it 
into \eqref{eq: lemma 6 stop 3} and by \eqref{eq: D set}, we obtain
$ (\hat{\lambdab}^{(k-1)})^T\gb(\hat{\xb}^{(k-1)})
 \leq  \left(\frac{D_\lambda}{\delta}\right) \xi^{(k-1)} $
which, by taking $k \rightarrow \infty$, leads to
\begin{align}\label{sup term}
  \limsup_{k\rightarrow \infty}~(\hat{\lambdab}^{(k-1)})^T\gb(\hat{\xb}^{(k-1)}) \leq 0.
\end{align}
On the other hand,  {by letting $\lambdab=\zerob \in \Dc$, from \eqref{eq: lemma 6 stop 3} we have}
$-(\hat{\lambdab}^{(k-1)})^T \gb(\hat{\xb}^{(k-1)})\leq
\xi^{(k-1)}+ (\hat{\lambdab}^{\star}-\hat{\lambdab}^{(k-1)})^T \gb(\hat{\xb}^{(k-1)})\leq \xi^{(k-1)}+ N C_g \|\hat{\lambdab}^{(k-1)}-\hat{\lambdab}^{\star}\|.
$  
{Since $\lim_{k\rightarrow \infty}\xi^{(k-1)} =0$ and $\lim_{k \rightarrow \infty} \| \hat{\lambdab}^{(k)} - \hat\lambdab^\star\| =0$ by Lemma \ref{lemma: saddle point}, it follows that
$\liminf_{k\rightarrow \infty}~(\hat{\lambdab}^{(k-1)})^T\gb(\hat{\xb}^{(k-1)}) \geq 0,$ 
which along with \eqref{sup term} 
yields the second term in~\eqref{eq: asymp opt conditions}.}
\hfill $\blacksquare$

\section{Proof of Lemma \ref{lemma: perturbation proximal}} \label{Appendix proof of lemma perturbation proximal}

The definition of $\hat\alphab^{(k)}$ in \eqref{eq: local proximal hat alpha} implies that
\ifconfver
\begin{align}\notag 
    &\gb_i^T(\hat\alphab_i^{(k)})\hat \lambdab^{(k-1)}\notag\\
    &~~~~~~~~~~+
        (\hat\alphab_i^{(k)}-\xb_i^{(k-1)})^T
        \nabla \fb_i^T(\xb_i^{(k-1)})\nabla \Fc(N\hat\yb^{(k-1)})
        \notag \\
        &~~~~~~~~~~+\frac{1}{2\rho_1} \|\hat\alphab_i^{(k)}-\xb_i^{(k-1)}\|^2~ \leq ~\gb_i^T(\xb_i^{(k-1)})\hat \lambdab^{(k-1)}, \notag
 \end{align}
\else
\begin{align}\notag 
    &\gb_i^T(\hat\alphab_i^{(k)})\hat \lambdab^{(k-1)}+
        (\hat\alphab_i^{(k)}-\xb_i^{(k-1)})^T
        \nabla \fb_i^T(\xb_i^{(k-1)})\nabla \Fc(N\hat\yb^{(k-1)})
        \notag \\
        &~~~~~~~~~~~~~~~~~~~~~~~~~~~~~~+\frac{1}{2\rho_1} \|\hat\alphab_i^{(k)}-\xb_i^{(k-1)}\|^2~ \leq ~\gb_i^T(\xb_i^{(k-1)})\hat \lambdab^{(k-1)}, \notag
 \end{align} \fi which, by summing over $i=1,\ldots,N,$ yields
\ifconfver
\begin{align}\label{eq: lemma 5 stop 8}
    &\gb^T(\hat\alphab^{(k)})\hat \lambdab^{(k-1)}+
        (\hat\alphab^{(k)}-\xb^{(k-1)})^T
        \nabla  \bar\Fc(\xb^{(k-1)})
        \notag \\
        &~~~~+\frac{1}{2\rho_1} \|\hat\alphab^{(k)}-\xb^{(k-1)}\|^2\leq ~\gb^T(\xb^{(k)})\hat \lambdab^{(k-1)},
 \end{align}
\else
\begin{align}\label{eq: lemma 5 stop 8}
    &\gb^T(\hat\alphab^{(k)})\hat \lambdab^{(k-1)}+
        (\hat\alphab^{(k)}-\xb^{(k-1)})^T
        \nabla  \bar\Fc(\xb^{(k-1)})
        \notag \\
        &~~~~~~~~~~~~~~~~~~~~~~~~~~~~~~+\frac{1}{2\rho_1} \|\hat\alphab^{(k)}-\xb^{(k-1)}\|^2~ \leq ~\gb^T(\xb^{(k)})\hat \lambdab^{(k-1)},
 \end{align}\fi where $\gb(\hat\alphab^{(k)})=\sum_{i=1}^N\gb_i^T(\hat\alphab_i^{(k)})$.
 By substituting the decent lemma in \cite[Lemma 2.1]{BertsekasADMM}
 \ifconfver
 \begin{align}\label{eq: decent lemma}
  \bar\Fc(\hat\alphab^{(k)})\leq \bar\Fc(\xb^{(k-1)})&+(\hat\alphab^{(k)}-\xb^{(k-1)})^T
        \nabla  \bar\Fc(\xb^{(k-1)})\notag \\
        &+\frac{G_{\bar\Fc}}{2} \|\hat\alphab^{(k)}-\xb^{(k-1)}\|^2
 \end{align}
 \else
 \begin{align}\label{eq: decent lemma}
  \bar\Fc(\hat\alphab^{(k)})\leq \bar\Fc(\xb^{(k-1)})+(\hat\alphab^{(k)}-\xb^{(k-1)})^T
        \nabla  \bar\Fc(\xb^{(k-1)})+\frac{G_{\bar\Fc}}{2} \|\hat\alphab^{(k)}-\xb^{(k-1)}\|^2
 \end{align}\fi
 into \eqref{eq: lemma 5 stop 8}, we then obtain
 \ifconfver
 \begin{align}\notag
   &\left(\frac{1}{2\rho_1}-\frac{G_{\bar\Fc}}{2}\right) \|\hat\alphab^{(k)}-\xb^{(k-1)}\|^2
   \notag \\
    &~~~~~~~~~~~~~\leq \Lc(\xb^{(k-1)},\hat\lambdab^{(k-1)})-\Lc(\hat\alphab^{(k)},\hat\lambdab^{(k-1)})
 \end{align}
 \else
 \begin{align}\notag
   \left(\frac{1}{2\rho_1}-\frac{G_{\bar\Fc}}{2}\right) \|\hat\alphab^{(k)}-\xb^{(k-1)}\|^2
   \leq \Lc(\xb^{(k-1)},\hat\lambdab^{(k-1)})-\Lc(\hat\alphab^{(k)},\hat\lambdab^{(k-1)})
 \end{align}\fi
 which, after combining with \eqref{eq: lemma 5 stop 6}, yields \eqref{eq: perturbation inequality2}.

 To show the second part of this lemma, let us recall \eqref{eq: local proximal perturbation point transform hat} that
 $\hat\alphab_i^{(k)}$ in \eqref{eq: local proximal hat alpha} can be alternatively written as
 \ifconfver
 \begin{align}\notag 
    &\hat\alphab_i^{(k)}=\arg\min_{\alphab_i \in \Xc_i} \gb_i^T(\alphab_i)\hat \lambdab^{(k-1)}+
        ( \nabla \fb_i^T(\xb_i^{(k-1)}) \notag \\
        &~~~~~~~~~~~~~~~~\times \nabla \Fc(N\hat\yb^{(k-1)})
        +\frac{1}{\rho_1} (\hat\alphab_i^{(k)}-\xb_i^{(k-1)}) )^T\alphab_i,\notag
 \end{align}
 \else
 \begin{align}\notag 
    &\hat\alphab_i^{(k)}=\arg\min_{\alphab_i \in \Xc_i} \gb_i^T(\alphab_i)\hat \lambdab^{(k-1)}+
        ( \nabla \fb_i^T(\xb_i^{(k-1)})\nabla \Fc(N\hat\yb^{(k-1)})
        +\frac{1}{\rho_1} (\hat\alphab_i^{(k)}-\xb_i^{(k-1)}) )^T\alphab_i,\notag
 \end{align}\fi
 which implies that, for all $\xb_i \in \Xc_i$, we have
 \ifconfver
 \begin{align}
    &\gb_i^T(\hat\alphab_i^{(k)})\hat \lambdab^{(k-1)}+
        ( \nabla \fb_i^T(\xb_i^{(k-1)})\nabla \Fc(N\hat\yb^{(k-1)})
        \notag \\
        &~~~~~~~~~~~~~~~~~~~~~~~+\frac{1}{\rho_1} (\hat\alphab_i^{(k)}-\xb_i^{(k-1)}) )^T(\hat\alphab_i^{(k)}-\xb_i^{(k-1)}) \notag \\
    &\leq    \gb_i^T(\xb_i)\hat \lambdab^{(k-1)}+
        ( \nabla \fb_i^T(\xb_i^{(k-1)})\nabla \Fc(N\hat\yb^{(k-1)})
        \notag \\
        &~~~~~~~~~~~~~~~~~~~~~~~+\frac{1}{\rho_1} (\hat\alphab_i^{(k)}-\xb_i^{(k-1)}) )^T(\xb_i-\xb_i^{(k-1)}). \notag
 \end{align}
 \else
 \begin{align}
    &\gb_i^T(\hat\alphab_i^{(k)})\hat \lambdab^{(k-1)}+
        ( \nabla \fb_i^T(\xb_i^{(k-1)})\nabla \Fc(N\hat\yb^{(k-1)})
        +\frac{1}{\rho_1} (\hat\alphab_i^{(k)}-\xb_i^{(k-1)}) )^T(\hat\alphab_i^{(k)}-\xb_i^{(k-1)}) \notag \\
    &~~~~\leq~    \gb_i^T(\xb_i)\hat \lambdab^{(k-1)}+
        ( \nabla \fb_i^T(\xb_i^{(k-1)})\nabla \Fc(N\hat\yb^{(k-1)})
        +\frac{1}{\rho_1} (\hat\alphab_i^{(k)}-\xb_i^{(k-1)}) )^T(\xb_i-\xb_i^{(k-1)}). \notag
 \end{align}\fi
By summing the above inequality over $i=1,\ldots,N,$ one obtains, for all $\xb\in \Xc,$
\ifconfver
\begin{align*}
    &\gb^T(\hat\alphab^{(k)})\hat \lambdab^\star+
         \nabla \bar \Fc^T(\xb^{(k-1)})(\hat\alphab^{(k)}-\xb^{(k-1)})\notag \\
        &~~~~~~~~~~~~~~~~~~~~~~~~~~~~~~~~~
        +\frac{1}{\rho_1} \|\hat\alphab^{(k)}-\xb^{(k-1)}\|^2\notag \\
    &\leq    \gb^T(\xb)\hat \lambdab^\star+   \nabla \bar\Fc^T(\xb^{(k-1)})(\xb-\xb^{(k-1)})
    +\frac{1}{\rho_1} \sum_{i=1}^N(\hat\alphab_i^{(k)}\notag \\
    &-\xb_i^{(k-1)}) (\xb_i-\xb_i^{(k-1)})+(\hat \lambdab^\star-\hat \lambdab^{(k-1)}) (\gb(\hat\alphab^{(k)})-\gb(\xb))\notag
\\
     &\leq   \gb^T(\xb)\hat \lambdab^\star +\bar\Fc(\xb) -\bar\Fc(\xb^{(k-1)}) \notag \\
     &~~~~+ \frac{2D_x}{\rho_1} \sum_{i=1}^N\|\hat\alphab_i^{(k)}-\xb_i^{(k-1)}\| + 2C_g \|\hat \lambdab^\star-\hat \lambdab^{(k-1)} \|,
\end{align*}
\else
\begin{align*}
    &\gb^T(\hat\alphab^{(k)})\hat \lambdab^\star+
         \nabla \bar \Fc^T(\xb^{(k-1)})(\hat\alphab^{(k)}-\xb^{(k-1)})
        +\frac{1}{\rho_1} \|\hat\alphab^{(k)}-\xb^{(k-1)}\|^2\notag \\
    &~~~~\leq~    \gb^T(\xb)\hat \lambdab^\star+   \nabla \bar\Fc^T(\xb^{(k-1)})(\xb-\xb^{(k-1)})
      \notag \\
      &~~~~~~~~  +\frac{1}{\rho_1} \sum_{i=1}^N(\hat\alphab_i^{(k)}-\xb_i^{(k-1)}) (\xb_i-\xb_i^{(k-1)})+(\hat \lambdab^\star-\hat \lambdab^{(k-1)}) (\gb(\hat\alphab^{(k)})-\gb(\xb))\notag
\\
     &~~~~ \leq   \gb^T(\xb)\hat \lambdab^\star +\bar\Fc(\xb) -\bar\Fc(\xb^{(k-1)}) + \frac{2D_x}{\rho_1} \sum_{i=1}^N\|\hat\alphab_i^{(k)}-\xb_i^{(k-1)}\| + 2C_g \|\hat \lambdab^\star-\hat \lambdab^{(k-1)} \|,
\end{align*}\fi where we have utilized the convexity of $\bar \Fc$,
boundedness of $\Xc_i$ and the constraint functions
(cf. Assumption \ref{assumption function} and \eqref{eq: function bounded}) in obtaining the last inequality. By applying \eqref{eq: decent lemma} to the above inequality and by the premise of $1/\rho_1\geq G_{\bar\Fc}>G_{\bar\Fc}/2$,
we further obtain, for all $\xb\in \Xc,$
\ifconfver
\begin{align}\label{eq: lemma 8 stop11}
  &\Lc(\hat\xb^\star,\hat \lambdab^\star)\leq \Lc(\xb,\hat \lambdab^\star)
  + \frac{2D_x}{\rho_1} \sum_{i=1}^N\|\hat\alphab_i^{(k)}-\xb_i^{(k-1)}\| \notag \\
  &+2C_g \|\hat \lambdab^\star-\hat \lambdab^{(k-1)}\| + |\Lc(\hat\xb^\star,\hat \lambdab^\star)-\Lc(\hat\alphab^{(k)},\hat \lambdab^\star)|,
\end{align}
\else
\begin{align}\label{eq: lemma 8 stop11}
  \Lc(\hat\xb^\star,\hat \lambdab^\star)\leq \Lc(\xb,\hat \lambdab^\star)
  &+ \frac{2D_x}{\rho_1} \sum_{i=1}^N\|\hat\alphab_i^{(k)}-\xb_i^{(k-1)}\| \notag \\
  &+2C_g \|\hat \lambdab^\star-\hat \lambdab^{(k-1)}\| + |\Lc(\hat\xb^\star,\hat \lambdab^\star)-\Lc(\hat\alphab^{(k)},\hat \lambdab^\star)|,
\end{align}\fi in which one can bound the last term, using \eqref{eq: lip bar Fc}, \eqref{eq: lipschitz f}, \eqref{eq: lipschitz g} and \eqref{eq: D set}, by
\ifconfver
\begin{align}\label{eq: lemma 8 stop12}
 &|\Lc(\hat\xb^\star,\hat \lambdab^\star)-\Lc(\hat\alphab^{(k)},\hat \lambdab^\star)|\notag \\
 &~~~~~~~~~~~~~\leq {(L_{\bar\Fc}                                                              +ND_\lambda \sqrt{P}L_g)} \|\hat\xb^\star-\hat\alphab^{(k)}\|.
\end{align}
\else
\begin{align}\label{eq: lemma 8 stop12}
 |\Lc(\hat\xb^\star,\hat \lambdab^\star)-\Lc(\hat\alphab^{(k)},\hat \lambdab^\star)|\leq {(L_{\bar\Fc}                                                              +ND_\lambda \sqrt{P}L_g)} \|\hat\xb^\star-\hat\alphab^{(k)}\|.
\end{align}\fi

Suppose that
 $ \Lc(\xb^{(k-1)},\hat\betab^{(k)})-\Lc(\hat\alphab^{(k)},\hat\lambdab^{(k-1)}) \rightarrow 0$ and $(\xb^{(k-1)}, \hat\lambdab^{(k-1)})$ converges to some limit point $(\hat\xb^\star, \hat\lambdab^\star)$ as $k\rightarrow \infty$. Then, by \eqref{eq: perturbation inequality2} and since $1/\rho_1\geq G_{\bar\Fc}$, we have
{$\|(\xb^{(k-1)},\hat\lambdab^{(k-1)} ) - (\hat\alphab^{(k)},\hat\betab^{(k)})\|\to0$,
 as $k\rightarrow \infty.$
 } Therefore,
\ifconfver
{\small \begin{align*}
 \lim_{k\rightarrow \infty}
 &\bigg(\frac{2D_x}{\rho_1} \sum_{i=1}^N\|\hat\alphab_i^{(k)}-\xb_i^{(k-1)}\|+2C_g \|\hat \lambdab^\star-\hat \lambdab^{(k-1)}\|
  \notag \\
  &~~~~~~~~~~~~~~~~~~+ |\Lc(\hat\alphab^{(k)},\hat \lambdab^\star)-\Lc(\hat\xb^\star,\hat \lambdab^\star)|\bigg) =0.
\end{align*}}
\else
{\small $$
 \lim_{k\rightarrow \infty}
 \left(\frac{2D_x}{\rho_1} \sum_{i=1}^N\|\hat\alphab_i^{(k)}-\xb_i^{(k-1)}\|+2C_g \|\hat \lambdab^\star-\hat \lambdab^{(k-1)}\| + |\Lc(\hat\alphab^{(k)},\hat \lambdab^\star)-\Lc(\hat\xb^\star,\hat \lambdab^\star)|\right) =0.
$$}\fi
Thus, it follows from \eqref{eq: lemma 8 stop11}, \eqref{eq: lemma 8 stop12} and the above equation that
{$ \Lc(\hat\xb^\star,\hat \lambdab^\star)\leq \Lc(\xb,\hat \lambdab^\star)$
for all $\xb\in \Xc$.
The rest of the proof is similar to that of Lemma~\ref{lemma: perturbation}.
}
\hfill  $\blacksquare$

\vspace{-0.0cm}
\vspace{-0.0cm}
\vspace{-0.0cm}


\footnotesize
\bibliography{distributed_opt,smart_grid}

\begin{thebibliography}{10}
\providecommand{\url}[1]{#1}
\csname url@samestyle\endcsname
\providecommand{\newblock}{\relax}
\providecommand{\bibinfo}[2]{#2}
\providecommand{\BIBentrySTDinterwordspacing}{\spaceskip=0pt\relax}
\providecommand{\BIBentryALTinterwordstretchfactor}{4}
\providecommand{\BIBentryALTinterwordspacing}{\spaceskip=\fontdimen2\font plus
\BIBentryALTinterwordstretchfactor\fontdimen3\font minus
  \fontdimen4\font\relax}
\providecommand{\BIBforeignlanguage}[2]{{%
\expandafter\ifx\csname l@#1\endcsname\relax
\typeout{** WARNING: IEEEtran.bst: No hyphenation pattern has been}%
\typeout{** loaded for the language `#1'. Using the pattern for}%
\typeout{** the default language instead.}%
\else
\language=\csname l@#1\endcsname
\fi
#2}}
\providecommand{\BIBdecl}{\relax}
\BIBdecl

\bibitem{Lesser2003}
V.~Lesser, C.~Ortiz, and M.~Tambe, \emph{Distributed Sensor Networks: A
  Multiagent Perspective}.\hskip 1em plus 0.5em minus 0.4em\relax Kluwer
  Academic Publishers, 2003.

\bibitem{Rabbat2004}
M.~Rabbat and R.~Nowak, ``Distributed optimization in sensor networks,'' in
  \emph{Proc. ACM IPSN}, Berkeley, CA, USA, April 26-27, 2004, pp. 20--27.

\bibitem{Chiang2008}
M.~Chiang, P.~Hande, T.~Lan, and W.~C. Tan, ``Power control in wireless
  cellular networks,'' \emph{Foundations and Trends in Networking}, vol.~2,
  no.~4, pp. 381--533, 2008.

\bibitem{ChaoTSP2012}
S.~Chao, T.-H. Chang, K.-Y. Wang, Z.~Qiu, and C.-Y. Chi, ``Distributed robust
  multicell coordianted beamforming with imperfect csi: {An ADMM} approach,''
  \emph{IEEE Trans. Signal Processing}, vol.~60, no.~6, pp. 2988--3003, 2012.

\bibitem{Belomestny2010}
D.~Belomestny, A.~Kolodko, and J.~Schoenmakers, ``Regression methods for
  stochastic control problems and their convergence analysis,'' \emph{SIAM J.
  on Control and Optimization}, vol.~48, no.~5, pp. 3562--3588, 2010.

\bibitem{Hastie2001Book}
T.~Hastie, R.~Tibshirani, and J.~Friedman, \emph{The Elements of Statistical
  Learning: Data Mining, Inference, and Prediction}.\hskip 1em plus 0.5em minus
  0.4em\relax New York, NY, USA: Springer-Verlag, 2001.

\bibitem{Elad2009Book}
M.~Elad, \emph{Sparse and Redundant Rerpesentations}.\hskip 1em plus 0.5em
  minus 0.4em\relax New York, NY, USA: Springer Science + Business Media, 2010.

\bibitem{Yamada2009}
R.~Cavalcante, I.~Yamada, and B.~Mulgrew, ``An adaptive projected subgradient
  approach to learning in diffusion networks,'' \emph{IEEE Trans. Signal
  Processing}, vol.~57, no.~7, pp. 2762--2774, Aug. 2009.

\bibitem{Johansson08}
B.~Johansson, T.~Keviczky, M.~Johansson, and K.~Johansson, ``Subgradient
  methods and consensus algorithms for solving convex optimization problems,''
  in \emph{Proc. IEEE CDC}, Cancun, Mexica, Dec. 9-11, 2008, pp. 4185--4190.

\bibitem{Angelia2009_multiagent}
A.~Nedi\'{c}, A.~Ozdaglar, , and A.~Parrilo, ``Distributed subgradient methods
  for multi-agent optimization,'' \emph{IEEE Trans. Automatic Control},
  vol.~54, no.~1, pp. 48--61, Jan. 2009.

\bibitem{Angelia2010}
------, ``Constrained consensus and optimization in multi-agent networks,''
  \emph{IEEE Trans. Automatic Control}, vol.~55, no.~4, pp. 922--938, April
  2010.

\bibitem{Lobel2011}
I.~Lobel and A.~Ozdaglar, ``Distributed subgradient methods for convex
  optimization over random networks,'' \emph{IEEE Trans. Automatic Control},
  vol.~56, no.~6, pp. 1291--1306, June 2011.

\bibitem{Ram2010_stoc}
S.~S. Ram, A.~Nedi\'{c}, and V.~V. Veeravalli, ``Distributed stochastic
  subgradeint projection algorithms for convex optimization,'' \emph{J. Optim.
  Theory Appl.}, vol. 147, pp. 516--545, 2010.

\bibitem{Chen_Sayed2012}
J.~Chen and A.~H. Sayed, ``Diffusion adaption strategies for distributed
  optimization and learning networks,'' \emph{IEEE. Trans. Signal Process.},
  vol.~60, no.~8, pp. 4289--4305, Aug. 2012.

\bibitem{MZhu2012}
M.~Zhu and S.~Mart\'{i}nez, ``On distributed convex optimization under
  inequality and equality constraints,'' \emph{IEEE Trans. Automatic Control},
  vol.~57, no.~1, pp. 151--164, Jan. 2012.

\bibitem{Yuan2011}
D.~Yuan, S.~Xu, and H.~Zhao, ``Distributed primal-dual subgradient method for
  multiagent optimization via consensus algorithms,'' \emph{IEEE Trans.
  Systems, Man, and Cybernetics- Part B}, vol.~41, no.~6, pp. 1715--1724, Dec.
  2011.

\bibitem{Ram2012}
S.~S. Ram, A.~Nedi\'{c}, and V.~V. Veeravalli, ``A new class of distributed
  optimization algorithm: {A}pplication of regression of distributed data,''
  \emph{Optimization Methods and Software}, vol.~27, no.~1, pp. 71--88, 2012.

\bibitem{ChangPES2012}
T.-H. Chang, M.~Alizadeh, and A.~Scaglione, ``Coordinated home energy
  management for real-time power balancing,'' in \emph{Proc. IEEE PES General
  Meeting}, San Diego, CA, July 22-26, 2012, pp. 1--8.

\bibitem{LiLow2011PES}
N.~Li, L.~Chen, and S.~H. Low, ``Optimal demand response based on utility
  maximization in power networks,'' in \emph{IEEE PES General Meeting},
  Detroit, MI, USA, July 24-29, 2011, pp. 1--8.

\bibitem{Juan2009}
J.~C.~V. Quintero, ``Decentralized control techniques applied to electric power
  distributed generation in microgrids,'' {MS} thesis, {D}epartament
  d'{E}nginyeria de {S}istemes, {Autom\'{a}tica i Inform\'{a}tica Industrial,
  Universitat Polit\'{e}cnica de Catalunya}, 2009.

\bibitem{Kallio1994}
M.~Kallio and A.~Ruszczy\'{n}ski, ``Perturbation methods for saddle point
  computation,'' 1994, report No. WP-94- 38, International Institute for
  Applied Systems Analysis.

\bibitem{Kallio1999}
M.~Kallio and C.~H. Rosa, ``Large-scale convex optimization via saddle-point
  computation,'' \emph{Oper. Res.}, pp. 93--101, 1999.

\bibitem{Olshevsky2006}
A.~Olshevsky and J.~N. Tsitsiklis, ``Convergence rates in distributed consensus
  averaging,'' in \emph{Proc. IEEE CDC}, San Diego, CA, USA, Dec. 13-15, 2006,
  pp. 3387--3392.

\bibitem{Xiao2003}
L.~Xiao and S.~Boyd, ``Fast linear iterations for distributed averaging,''
  \emph{Systems and Control Letters}, vol.~53, pp. 65--78, 2004.

\bibitem{YangJohansson2010}
B.~Yang and M.~Johansson, ``Distributed optimization and games: {A} tutorial
  overview,'' {C}hapter 4 of \emph{Networked Control Systems}, A. Bemporad, M.
  Heemels and M. Johansson (eds.), LNCIS 406, Springer-Verlag, 2010.

\bibitem{Uzawa1958}
H.~Uzawa, ``Iterative methods in concave programming,'' 1958, in Arrow, K.,
  Hurwicz, L., Uzawa, H. (eds.) Studies in Linear and Nonlinear Programming,
  pp. 154-165. Stanford University Press, Stanford.

\bibitem{Angelia2009_saddle}
A.~Nedi\'{c} and A.~Ozdaglar, ``Subgradeint methods for saddle-point
  problems,'' \emph{J. OPtim. Theory Appl.}, vol. 142, pp. 205--228, 2009.

\bibitem{Srivastava12_minmax}
K.~Srivastava, A.~Nedi\'{c}, and D.~Stipanovi\'{c}, ``Distributed
  {Bregman}-distance algorithms for min-max optimization,'' in book
  \emph{Agent-Based Optimization}, I. Czarnowski, P. Jedrzejowicz and J.
  Kacprzyk (Eds.), Springer Studies in Computational Intelligence (SCI), 2012.

\bibitem{dsm2}
M.~Alizadeh, X.~Li, Z.~Wang, A.~Scaglione, and R.~Melton, ``Demand side
  management in the smart grid: {I}nformation processing for the power
  switch,'' \emph{IEEE Signal Process. Mag.}, vol.~59, no.~5, pp. 55--67, Sept.
  2012.

\bibitem{GuanTSG2010}
X.~Guan, Z.~Xu, and Q.-S. Jia, ``Energy-efficient buildings facilitated by
  microgrid,'' \emph{IEEE Trans. Smart Grid}, vol.~1, no.~3, pp. 243--252, Dec.
  2010.

\bibitem{Cai2010}
N.~Cai and J.~Mitra, ``A decentralized control architecture for a microgrid
  with power electronic interfaces,'' in \emph{Proc. North American Power
  Symposium (NAPS)}, Sept. 26-28 2010, pp. 1--8.

\bibitem{Hershberger2001}
D.~Hershberger and H.~Kargupta, ``Distributed multivariate regression using
  wavelet based collective data mining,'' \emph{J. Parallel Distrib. Comput.},
  vol.~61, no.~3, pp. 372--400, March 2001.

\bibitem{Kargupta1999}
H.~Kargupta, B.-H. Park, D.~Hershberger, and E.~Johnson, ``Collective data
  mining: {A} new perspective toward distributed data mining,'' in
  \emph{Advances in Distributed Data Mining}, H. Kargupta and P. Chan (eds.),
  AAAI/MIT Press, 1999.

\bibitem{BK:Bersekas_netowork}
D.~P. Bertsekas, \emph{{Network Optimization : Continuous and Discrete
  Models}}.\hskip 1em plus 0.5em minus 0.4em\relax Athena Scientific, 1998.

\bibitem{Madan2006}
R.~Madan and S.~Lall, ``Distributed algorithms for maximum lifetime routing in
  wireless sensor networks,'' \emph{IEEE. Trans. Wireless Commun.}, vol.~5,
  no.~8, pp. 2185--2193, Aug. 2006.

\bibitem{Changglobalsip13}
T.-H. Chang, A.~Nedich, and A.~Scaglione, ``Distributed sparse regression by
  consensus-based primal-dual perturbation optimization,'' accepted by
  \emph{IEEE Global Conf. on Signal and Info. Process. (GlobalSIP)}, Austin,
  Texas, Dec. 3-5, 2013.

\bibitem{BK:BoydV04}
S.~Boyd and L.~Vandenberghe, \emph{Convex {O}ptimization}.\hskip 1em plus 0.5em
  minus 0.4em\relax Cambridge, UK: Cambridge University Press, 2004.

\bibitem{BK:Bertsekas2003_analysis}
D.~P. Bertsekas, A.~Nedi\'{c}, and A.~E. Ozdaglar, \emph{Convex analysis and
  optimization}.\hskip 1em plus 0.5em minus 0.4em\relax Cambridge,
  Massachusetts: Athena Scientific, 2003.

\bibitem{Zabotin88}
I.~Y. Zabotin, ``A subgradient method for finding a saddle point of a
  convex-concave function,'' \emph{Issled. Prikl. Mat.}, vol.~15, pp. 6--12,
  1988.

\bibitem{Nesterov2005}
Y.~Nesterov, ``Smooth minimization of nonsmooth functions,'' \emph{Math.
  Program.}, vol. 103, no.~1, pp. 127--152, 2005.

\bibitem{Mateos2010}
G.~Mateos, J.~A. Bazerque, and G.~B. Giannakis, ``Distributed sparse linear
  regression,'' \emph{IEEE Trans. Signal Processing}, vol.~58, no.~10, pp.
  5262--5276, Dec. 2010.

\bibitem{Mota2012}
J.~F.~C. Mota, J.~M.~F. Xavier, P.~M.~Q. Aguiar, and M.~Puschel, ``Distributed
  basis pursuit,'' \emph{IEEE. Trans. Signal Process.}, vol.~60, no.~4, pp.
  1942--1956, April 2012.

\bibitem{Angelia2009_approxprimal}
A.~Nedi\'{c} and A.~Ozdaglar, ``Approximate primal solutions and rate analysis
  for dual subgradient methods,'' \emph{SIAM J. OPtim.}, vol.~19, no.~4, pp.
  1757--1780, 2009.

\bibitem{Larsson1999}
T.~Larsson, .~Patriksson, and A.-B. Str\"{o}mberg, ``Ergodic, primal
  convergence in dual subgradient schemes for convex programming,'' \emph{Math.
  Program.}, vol.~86, pp. 238--312, 1999.

\bibitem{BK:Polyak1987}
B.~T. Polyak, \emph{Introduction to Optimization}.\hskip 1em plus 0.5em minus
  0.4em\relax New Yoir: Optimization Software Inc., 1987.

\bibitem{ChangTAC2013_companion}
T.-H. Chang, A.~Nedich, and A.~Scaglione, ``Electronic companion for
  {D}istributed constrained optimization by consensus-based primal-dual
  perturbation method,'' available on \url{http://arxiv.org}.

\bibitem{ChangTSG2013}
T.-H. Chang, M.~Alizadeh, and A.~Scaglione, ``Real-time power balancing via
  decentralized coordinated home energy scheduling,'' \emph{IEEE Trans. Smart
  Grid}, vol.~4, no.~3, pp. 1490--1504, Sept. 2013.

\bibitem{Lustig1994}
I.~J. Lustig and a.~D. F.~S. R.~E.~Marsten, ``Interior point methods for linear
  programming: {C}omputational state of the art,'' \emph{ORSA J. Comput.},
  vol.~6, no.~1, pp. 1--14, 1994.

\bibitem{BertsekasADMM}
D.~P. Bertsekas and J.~N. Tsitsiklis, \emph{Parallel and distributed
  computation: {Numerical} methods}.\hskip 1em plus 0.5em minus 0.4em\relax
  Upper Saddle River, NJ, USA: Prentice-Hall, Inc., 1989.

\end{thebibliography}
\end{document}